\documentclass{aa}  

\usepackage{graphics,graphicx}
\usepackage{xcolor}
\usepackage{enumitem}

\usepackage{txfonts}
\usepackage{longtable}
\usepackage[unicode=true,pdfusetitle,
 bookmarks=true,bookmarksnumbered=false,bookmarksopen=false,
 breaklinks=false,pdfborder={0 0 0},pdfborderstyle={},backref=false,colorlinks=true]
 {hyperref}
\hypersetup{
 urlcolor=blue,citecolor=blue}

\newcommand{\xmm}{\textit{XMM}}

\newcommand{\nustar}{\textit{NuSTAR}}

\newcommand{\xii}{~ergs~cm~s$^{-1}$~}

\newcommand{\kms}{km~s$^{-1}$}
\newcommand{\fluxcgs}{ergs~s$^{-1}$~cm$^{-2}$}

\newcommand{\ks}{\rm\thinspace ks}
\newcommand{\kev}{{\rm\thinspace keV}}
\def\rg{{\thinspace r_{\rm g}}}
\newcommand{\et}{{et al.\ }}

\def\cm{{\rm\thinspace cm}}
\def\pscm{\hbox{$\cm^{-2}\,$}}
\def\feka{{Fe~K$\alpha$}}

\begin{document} 

   \title{The eROSITA Final Equatorial Depth Survey (eFEDS): Complex absorption and soft excesses in hard X-ray--selected active galactic nuclei}

   \author{S. G. H. Waddell\inst{1}, 
          K. Nandra\inst{1},
          J. Buchner\inst{1},
          Q. Wu\inst{2},
          Y. Shen\inst{2,3},
          R. Arcodia\inst{1},
          A. Merloni\inst{1},
          M. Salvato\inst{1},
          T. Dauser\inst{4},
          Th. Boller\inst{1},
          T. Liu\inst{1},
          J. Comparat\inst{1},
          J. Wolf\inst{1},
          T. Dwelly\inst{1},
          C. Ricci\inst{5},
          J. R. Brownstein\inst{6}, and
          M. Brusa\inst{7,8}
          }

   \institute{Max-Planck-Institut f\"{u}r extraterrestrische Physik, Giessenbachstrasse 1, 85748 Garching, Germany\\
              \email{swaddell@mpe.mpg.de}
         \and
             Department of Astronomy, University of Illinois at Urbana-Champaign, Urbana, IL 61801, USA
         \and
             National Center for Supercomputing Applications, University of Illinois at Urbana-Champaign, Urbana, IL 61801, USA
         \and
             Dr. Karl Remeis-Observatory \& ECAP, University of Erlangen-Nuremberg, Sternwartstr. 7, 96049 Bamberg, Germany
         \and
             N\'{u}cleo de Astronom\'{i}a de la Facultad de Ingenier\'{i}a, Universidad Diego Portales, Av. E\'{j}ercito Libertador 441, Santiago, Chile
         \and 
             Department of Physics and Astronomy, University of Utah, 115 S. 1400 E., Salt Lake City, UT 84112, USA
         \and
             Dipartimento di Fisica e Astronomia ``Augusto Righi'', Universit\'{a} di Bologna, via Gobetti 93/2, 40129 Bologna, Italy
         \and
             INAF – Osservatorio di Astrofisica e Scienza dello Spazio di Bologna, via Gobetti 93/3, 40129 Bologna, Italy
             }

   \date{Received XXX; accepted YYY}

 
  \abstract
   {The soft excess, a surplus of X-ray photons below $2\kev$ with respect to a power law, is a feature of debated physical origin found in the X-ray spectra of many type-1 active galactic nuclei (AGN). The eROSITA instrument aboard the Spectrum-Roentgen-Gamma (SRG) mission will provide an all-sky census of AGN. Spectral fitting of these sources can help identify the physical origin of the soft excess.}
   {The eROSITA Final Equatorial Depth Survey (eFEDS) field, designed to mimic the expected average equatorial depth of the all-sky survey, provides the ideal sample to test the power of eROSITA. The primary goal of this work is to test a variety of models for the soft X-ray emission of AGN (thermal emission, non-thermal emission, ionised absorption, or neutral partial covering absorption) to help identify the physical origin of the soft X-ray spectral complexity. Differences between these models are examined in the context of this sample to understand the physical properties. }
   {We used Bayesian X-ray analysis to fit a sample of 200 AGN from the eFEDS hard X-ray--selected sample with a variety of phenomenological and physically motivated models. Model selection is performed using the Bayes factor to compare the applicability of each model for individual sources as well as for the full sample, and source properties are compared and discussed. Black hole masses and Eddington ratios were estimated from optical spectroscopy.}
   {We find that 29 sources have evidence for a soft excess at a confidence level $>97.5$\%, all of which are better modelled by an additional soft power-law, as opposed to thermal blackbody emission. Applying more physically motivated soft excess emission models, we find that 23 sources prefer a warm corona model, while only six sources are best fit with relativistic blurred reflection. Sources with a soft excess show a significantly higher Eddington ratio than the remainder of the sample. Of the remainder of the sample, many sources show evidence for complex absorption, with 29 preferring a warm absorber, and 25 a partial covering absorber. Many (18/26) sources that show significant neutral absorption when modelled with an absorbed power law, in fact show evidence that the absorber is ionised, which has important implications on the understanding of obscured AGN. In contrast to the soft excesses, warm absorber sources show significantly lower Eddington ratios than the remainder of the sample. We discuss the implications of these results for the physical processes in the central regions of AGN.}
   {Spectral fitting with Bayesian statistics is ideal for the identification of complex absorption and soft excesses in the X-ray spectra of AGN, and can allow one to distinguish between different physical interpretations. Applying the techniques from this work to the eROSITA all-sky survey will provide a more complete picture of the prevalence and origin of soft excesses and warm absorbers in type-1 AGN in the local Universe. }

    \titlerunning{Soft excesses and complex absorbers in eFEDS AGN}
    \authorrunning{Waddell et al.}
   \maketitle
%
\section{Introduction}
eROSITA  \citep[extended ROentgen Survey with an Imaging Telescope Array:][]{Merloniero,Predehlero} is the soft X-ray instrument aboard the Spectrum-Roentgen-Gamma \citep[SRG;][]{Sunyaevero} mission, which successfully launched in July 2019. The primary operation mode of SRG/eROSITA is continuous scanning, and the mission was designed to create an eight-pass map of the entire X-ray sky, providing X-ray variability information as well as spectroscopy in the $0.2-8\kev$ band. The most numerous class of objects to be detected will be millions of Active Galactic Nuclei (AGN), powered by accreting supermassive black holes at the centres of galaxies. According to the AGN unification model, the presence or lack of broad lines in the optical spectra of AGN can be explained according to their viewing angle, where type-1 (or Seyfert 1) galaxies offer a direct view of the central accretion disc and broad emission line region, while type-2 (or Seyfert 2) galaxies are viewed through an obscuring, dusty torus \citep{1993Antonucci,1995Urry}. Observing type-1 AGN at X-ray energies allows for the direct study of the innermost regions of the central engine, where extreme relativistic effects can occur, while observing type-2 AGN can probe the physical properties of the torus. 

The primary source of X-ray emission in AGN is the corona, which consists of hot and/or relativistic electrons located at some height (few $\rg$ to 100s $\rg$) above the inner accretion disc \citep[e.g.][]{1991Haardt,1993Haardt,2000Merloni,2004Fabian}. The resulting coronal emission arises from Compton upscattering of lower energy photons, and takes the form of a power law that dominates the X-ray spectra of AGN above energies of $\sim2\kev$. In the soft X-ray band, the spectra of many AGN show evidence for a soft excess \citep{1981Pravdo,1985Arnaud,1985Singh}, a surplus of photons over the primary power-law continuum below $1-2\kev$. The origin of this component is highly debated, and a variety of physical mechanisms have been proposed to explain this feature. 

Initially, it was proposed that the tail of the disc blackbody from the hottest, innermost regions of the accretion disc may be responsible, as the soft excess shape is well fitted with a blackbody with temperatures of $\sim0.1\kev$ \citep[e.g.][]{2004Gierlinski}. The temperature of the disc, however, should scale with black hole mass as T $\propto$ M$_{\rm BH}^{-1/4}$ \citep{1973discmod}. Given this relationship, the blackbody photons even from the innermost accretion discs in AGN with typical masses of $10^7$ -- $10^9$ solar masses are highly unlikely to be visible in the X-ray regime. Furthermore, the expected trend between the fitted blackbody temperature and the black hole mass has not been found \citep[e.g.][]{2004Gierlinski,2006Crummy}. 

Instead, it has been proposed that the soft excess may be due to a blurred reflection component \citep[e.g.][]{1999Ross,2005Ross}. Some of the photons from the corona will be incident upon the accretion disc. These can be Compton back-scattered from the disc, or excite fluorescence or recombination processes, thus producing a multitude of absorption and emission features. If the accretion disc is highly ionised, these features are concentrated primarily at low energies below $2\kev$. Due to the proximity of the inner accretion disc to the black hole, these features are relativistically broadened. This produces a smooth soft excess in excess of a power law at low energies that can have a form similar to a blackbody, as well as a broad iron line and absorption edge in the hard X-ray. Blurred reflection modelling has been used to successfully explain the spectral shape and variability of numerous type-1 AGN \citep[e.g.][]{2004Fabian, 2008Zoghbi, 2017Wilkins, 2019Gallo, 2019Jiang, 2019Waddell, 2021Boller}. 

In another interpretation, the soft excess can be produced by a secondary, warm corona \citep[e.g.][]{2012Done,2018Petrucci,2020Petrucci} which is optically thick \citep[$\tau \sim 10$;][]{2020Petrucci} and cooler than the primary hot X-ray corona. Blackbody seed photons from the disc undergo Comptonisation in the warm corona. Due to the lower temperature ($\sim0.1-1\kev$) and higher optical depth of the secondary corona, compared to that producing the hard power law, the tail of the Comptonisation spectrum can be seen in the soft X-rays, which can then produce a soft excess over the power law continuum. In general, the warm corona is interpreted to be a slab above the accretion disc \citep[e.g.][]{2012Done}. The warm corona may only be stable under certain restrictive conditions \citep[e.g. the gas cannot be too hot or too cold; see][]{2020Ballantyne} and may also sometimes produce significant absorption features for temperatures below $10^7$~K \citep{2019Garcia}. This model has been shown to fit the spectral shape of numerous type-1 AGN \citep[e.g.][]{2012Done, 2018Ehler, 2019Tripathi, 2018Petrucci, 2020Petrucci}.

In contrast to these emission mechanisms, it has been proposed that the soft excess is an artefact of improperly modelled absorption features \citep[e.g.][]{2004Gierlinski, 2004Tanaka, 2014Parkerabs, 2015Fang, 2015Gallo, 2021Boller, 2021Parker}. In a neutral partial covering absorption scenario, X-rays from the corona pass through an absorber with moderate column density and covering fraction, producing significant absorption in the soft X-ray. This produces a flat spectrum that appears to have some excess emission. This model also produces a deep iron absorption edge at $\sim7\kev$ \citep{2004Tanaka, 2021Parker}, which can explain the hard X-ray curvature observed in some AGN spectra. Rapid variability observed in the spectra of some type-1 AGN has also been attributed to changes in the column density or covering fraction of the absorber. 

Instead of an excess of soft photons, many AGN show evidence for complex ionised (warm) absorption which produces features concentrated in the soft X-ray. In particular, partially ionised neon, oxygen and iron absorption lines and edges in the $0.5-2\kev$ band are observed in many sources \citep[e.g.][]{1998George, 2000Kaastra, 2000Kaspi, 2004Blustin, 2005Blustin, 2004Gierlinski, 2007Mckernan, 2014Laha, 2019Mizumoto}. It has been proposed that these warm absorbers can be physically connected to disc winds launched in some AGN systems, or even to large-scale outflows \citep[e.g.][]{2004Blustin, 2005Blustin, 2019Kallman}. These low-velocity winds are well-studied in AGN, especially using high resolution (e.g. using gratings) spectroscopy \citep[e.g.][]{2004Blustin}. Proper characterisation of warm absorption or partial covering absorption is essential for characterising not only the soft excess, but also the hard X-ray continuum, as absorption can create an apparent soft excess at low energies.

Distinguishing between these various models for the soft excess is not a straightforward task, and many previous attempts to do so have shown that all models can sometimes reproduce the observed X-ray spectral shape \citep[e.g.][]{2018Ehler, 2019Tripathi, 2019Waddell, 2022Chalise}. Each model typically has caveats, and it is often difficult to simultaneously explain the spectral shape as well as the short and long term variability. It is also likely that more than one soft excess component exists simultaneously, often complicated by the superposition of multiple absorption components \citep{2021Boller}. Since each of the different soft excess models have very different physical interpretations, consequences for the understanding of X-ray emission from AGN differ. X-ray reverberation mapping (see \citealp[for a review]{2014Uttley}), where a search is performed for time lags between different X-ray energy bands, can also be used to probe the geometry and height of the corona above the black hole \citep[e.g.][]{2013Zoghbi, 2015Wilkins, 2016Kara}, and the lags can be interpreted as light travel time between two coronas \citep{2017Chainakun} or between the hot X-ray corona and the accretion disc \citep{2013Demarco}. Other timing methods including fractional variability analysis \citep{2003Vaughan} or principal component analysis \citep{2015Parkerpca} can also be compared with simulations to distinguish between models.

In this work, the X-ray spectra of 200 AGN from the hard X-ray--selected sample (Nandra \et in prep.) of the eROSITA Final Equatorial Depth Survey (eFEDS) field \citep{Brunnerefeds, 2022Liu, 2022Salvato} are fit with a variety of phenomenological and physically motivated models to search for the presence of soft excesses and attempt to determine their physical origin.  Therefore, the eROSITA bandpass of $0.2-8\kev$ is ideal for such measurements, as it provides excellent coverage and resolution as well as high effective area for energies below $1\kev$. In Sect. 2, the data reduction and techniques used in this work are described. In Sect. 3, the preliminary models used for spectral fitting are described, and the absorption and soft excess samples are constructed. Section 4 describes the physical models applied to the sample of sources with soft excess. In Sect. 5, the spectral properties identified in this work are presented, and in Sect. 6, a further discussion of results is given. Finally, conclusions are drawn in Sect. 7.

\section{Data reduction and fitting}
\subsection{The eFEDS hard X-ray--selected sample}
The eFEDS field was observed in the eROSITA calibration and performance verification phase and covers $\sim140$ deg$^2$ \citep{Brunnerefeds}. This equatorial survey overlaps with a plethora of multi-wavelength data, facilitating source characterisation and classification, as well as redshift measurements \citep{2022Salvato}. The eFEDS field is slightly deeper than, but comparable to, the average equatorial exposure of the originally planned eROSITA All Sky Survey (eRASS:8; eight passes of the entire sky), with an average exposure per pixel of $2.2\ks$. The eFEDS data are therefore representative of and can be used to predict the all-sky survey performance. More details of this survey and the resulting data products, including the source detection algorithm and data reduction techniques, are presented in \citet{Brunnerefeds}. All eFEDS data have been made public in June 2021 with the Early Data Release (EDR) of the eROSITA German consortium\footnote{https://erosita.mpe.mpg.de/edr/}.

The main X-ray source catalogue in the eFEDS field is assembled from sources detected in the $0.2-2.3\kev$ band \citep{Brunnerefeds, 2022Liu}. For the current work, proper characterisation of the hard X-ray emission is crucial for understanding the strength and shape of the soft excess component, so sources which only have detections below $2.3\kev$ are less suitable for our analysis. Therefore, this work makes use of the hard X-ray--selected catalogue, based on the detection likelihood in the $2.3-5\kev$ band (DET\_LIKE\_3 > 10; see Nandra \et in prep.). This sample is nearly spectroscopically complete, with 197/246 sources having a spectroscopic redshift from a variety of sources. Since most objects which are significantly detected in the hard X-ray also have significant soft emission, the total of 246 sources in the hard X-ray--selected catalogue largely overlap with the 27,910 sources presented in the main catalogue, with 20 sources being present only in the hard catalogue. These sources are classified according to the process outlined in \citet{2022Salvato}. This classification combined the results of three independent methods and considers the multi-wavelength properties of the sources' optical/IR counterparts. The best matches are determined based on a combination of the astrometric and photometric information (see \cite{2022Salvato}). We then apply the selection methods presented in Nandra \et (in prep) and \citet{2022Salvato} wherein sources with secure counterparts, which are likely extragalactic based on their redshift and colour-colour diagnostics, and which have secure photo-z or spectroscopic redshift, are considered to be AGN.

Applying these classifications and cuts, a final sample of 200 hard X-ray--selected AGN is obtained. Of the remaining 46 sources, around one-third are likely galactic, one-third do not have sufficient data quality for counterpart identification, and around one-third do not have sufficiently secure redshifts for spectral modelling. For completeness, all AGN from the sample, including those only detected in the hard band, are included. All sources are listed using their eROSITA name and source ID in Appendix B.

Most AGN in this sample are high flux and low redshift (median $z \sim 0.35$) sources (see Nandra \et in prep), and targeting of bright eFEDS sources as part of several SDSS follow-up programmes has resulted in a high level of spectroscopic coverage (Nandra \et in prep., Merloni \et in prep.). A total of 156 sources have usable optical spectra obtained as part of SDSS-IV \citep{2006Gunnsdss, 2013Smeeboss, 2016Dawsoneboss, 2017Blantonsdssiv, 2020sdssiv}, or from SDSS-V \citep{1973Bowensdss, 2006Gunnsdss, 2013Smeeboss, 2017sdssv, 2019Wilsonsdss, 2023sdssdr18}, see also Anderson \et (in prep.) and Kollmeier \et (in prep.). For sources without a spectroscopic redshift, photometric redshifts were computed according to the method outlined in \citet{2022Salvato}. To ensure accurate photometric redshifts are obtained, IR, optical, and UV data are used in order to construct and SED, and this is fit to measure the redshift. Independent methods (LePhare; \citet{2006lephare} and DNNZ; Nishizawa \et in prep) are compared, and the most reliable redshifts are those which agree between the two methods. Only these sources are considered in spectral fitting (for more details on redshift measurements, see \citealp{2022Salvato}). Since we have very high spectral coverage for the AGN sample, for sources which rely on a photo-z, the peak of the probability density function for each redshift is taken, and associated errors are not considered. Redshift and luminosity distributions are presented in Sect. 5.1, and also in Nandra \et (in prep). 

\subsection{X-ray spectral analysis}
X-ray spectral extraction was performed in the manner described in \citet{2022Liu} and Nandra \et (in prep), using the eROSITA standard analaysis software system (eSASS) version c001 \citep{Brunnerefeds}. Spectral fits were also performed in a manner similar to those described in \citet{2022Liu} and Nandra \et (in prep), with the exception that here the spectra were not rebinned before modelling. This maximises the spectral information at the expense of computational speed. 

Following \citet{2018Simmonds}, a parametric spectral model for the eROSITA background \citep[see][]{2020Freyberg} is learned empirically using all eFEDS background spectra \citep{Brunnerefeds, 2022Liu}. First, the parametric model is determined using principle component analysis (PCA) in logarithmic count space. Next, the background spectrum of each source is iteratively fitted by adding principle components in logarithmic space, and further adding Gaussian lines in linear space, as required by the data according to the Akaike information criterion (AIC). In this way, when the addition of further Gaussian lines no longer changes the AIC, the background model is considered satisfactory. During the joint source and background fit, the normalisation of the background shape is a free parameter, while the shape parameters are kept fixed; however when the relative areas of the source and background regions are accounted for, this value is almost always 1. This technique ensures that improper subtraction of the background does not affect the spectral fits, particularly at higher energies, which is highly relevant for this sample.

After the background model has been applied, spectral fitting is performed using Bayesian X-ray analysis \citep[BXA;][]{2014Buchnerbxa, 2019Buchnerbxa}. BXA combines the X-ray spectral fitting packages and models used in XSPEC \citep{xspec} with UltraNest \citep{2021Buchnerultranest}, a nested sampling algorithm. By using BXA, the full range of parameter space can be explored to ensure that the best fit is found. Input priors on parameters are used to constrain the values to a reasonable parameter space, and posterior distributions can be examined after fitting to better understand the constraints that can be placed on parameters. 

Using BXA for spectral fitting, we can also perform model comparison. The Bayesian evidence (Z) is computed for each spectral fit. This value encompasses both the available parameter space and the fit quality, so it can be used to compare models. This is done using the Bayes factor, K;

\begin{equation}
    K = \frac{Z_{M1}}{Z_{M2}}
    \label{eqn:kbayes}
\end{equation}

\noindent where M1 and M2 are the models to be compared. While this value cannot directly be linked to a confidence interval or significance, Bayes factors can still be used to identify the best fitting spectral models. This method will be used in the following sections in order to robustly compare spectral models, and select sources with soft excesses and warm absorbers on a sound statistical basis.

\section{Preliminary spectral modelling}
Before applying more physically motivated spectral models to each source, we seek a simple characterisation of the spectral shape. This baseline model can then be rejected if the data show statistical evidence in favour of a more complex model including a soft excess or absorption component. The continuum can be modelled using a power law, modified by absorption from the Milky Way (taken from \citet{2013Willingale}) as well as the host galaxy, and the results of this fit are discussed in Sect. 3.1. The method for performing model comparison is summarised in Sect. 3.2, with more details in Appendix A. Next, the complex absorption modelling is presented in Sect. 3.3; here, neutral partial covering absorption and warm (ionised) absorption are compared. In Sect. 3.4, two different toy models are presented and compared to characterise the shape of the soft excess; a second power law component, and a blackbody component. Sect. 3.5 discusses the model comparison in more detail and present a final sample of sources with complex absorption and soft excesses. A full list of the {\sc xspec} models is given in Table~\ref{tab:models}, and an example source (ID 00011) with all models applied along with the residuals for each of the best fits is shown in Fig.~\ref{fig:allmodex}. All sources are listed in Appendix B along with the PL model fit parameters, information on which model provided the best fit to the source, and complex model parameters where relevant.  

\begin{figure}
   \centering
    \includegraphics[width=0.98\columnwidth]{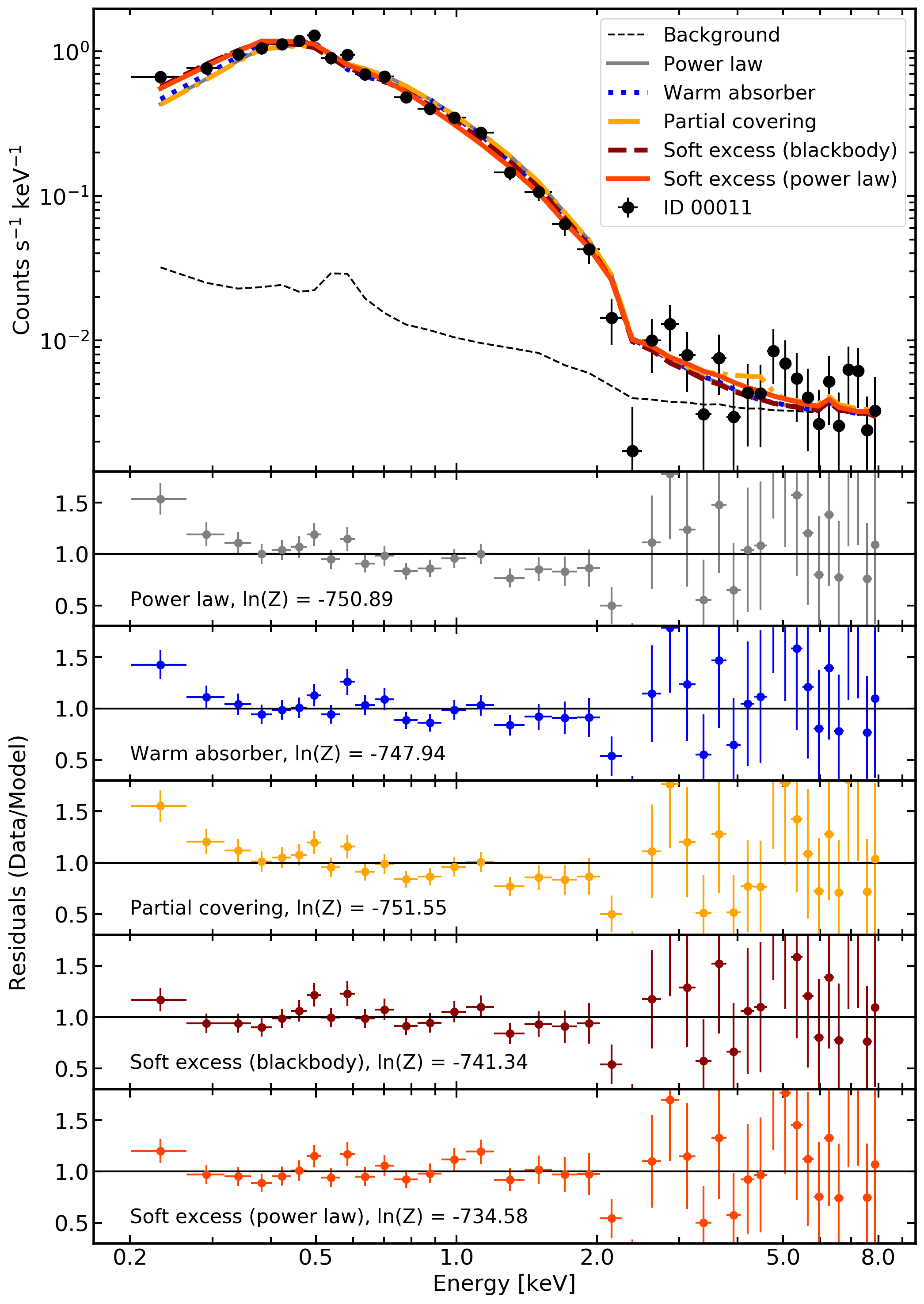}
    \caption{Comparison of different models and residuals for ID 00011 (z = 0.5121), a source best fit with a double power law soft excess model. The top panel shows the folded spectrum along with each of the best fit models, and the second, third, fourth, fifth and sixth panels show the residuals for the power law (grey), warm absorber (blue), partial covering (orange), blackbody (dark red) and double power law (red) models, respectively. The Bayesian evidence is also given for each model, to ease comparison. The spectrum and residuals are re-binned for clarity. The best fit is a power law soft excess, with a Bayes factor of K$_{pl} \sim 1.22\times10^7$ and a significance of >99\%. Data have been re-binned for display purposes.} 
    \label{fig:allmodex}

\end{figure}

\subsection{Baseline model: Absorbed power law}
\label{sect:base}

\begin{table}
\centering
\caption{Abbreviated model names as referenced in this work as well as their implementation in {\sc xspec}. }
\label{tab:models}
\resizebox{\columnwidth}{!}{%
\begin{tabular}{ll}
\hline
Name & {\sc xspec} implementation \\
 \hline
PL & \texttt{tbabs} $\times$ \texttt{ztbabs} $\times$ \texttt{powerlaw} \\
PL+PCF & \texttt{tbabs} $\times$ \texttt{ztbabs} $\times$ \texttt{zpcfabs} $\times$ \texttt{powerlaw}\\
PL+WA & \texttt{tbabs} $\times$ \texttt{ztbabs} $\times$ cwa18 $\times$ \texttt{powerlaw}\\
PL+BB & \texttt{tbabs} $\times$ \texttt{ztbabs} $\times$ (\texttt{powerlaw} + \texttt{constant} $\times$ \texttt{blackbody}) \\
PL+PL & \texttt{tbabs} $\times$ \texttt{ztbabs} $\times$ (\texttt{powerlaw} + \texttt{constant} $\times$ \texttt{powerlaw}) \\
\hline
\hline
\end{tabular}
}
\end{table}

Each source is first fit with an absorbed power law model (PL). Two absorption components are added; one component has a redshift of zero and the column density fixed to the value of the Milky Way (taken from \citet{2013Willingale} for each source), while the other has a redshift matching the host galaxy and column density, log(NH$_z$), left free to vary in order to account for absorption in the host galaxy (e.g. originating in galaxy-scale gas, torus, other absorbing material). The column density of the host galaxy absorber is allowed to vary between $\simeq10^{20}\pscm$ and $\simeq10^{25}\pscm$, where the lower limit is much smaller than the absorption in the Milky Way and is thus difficult or impossible to measure, and the upper limit represents an entirely obscured spectrum. The index of the power law component is constrained to be between one and three. This will allow for the identification of very hard and soft sources while ensuring that most sources have reasonable values of $\Gamma$ $\sim1.8-2.0$ \citep[e.g.][]{1994Nandra, 2000Reeves, 2007Nandra, 2020Waddell, 2022Waddell, 2022Liu}. For the normalisation, a broad, log-uniform prior ensures that all of the broad range of AGN fluxes found in the eFEDS field can be adequately characterised. The full list of priors are summarised in Table~\ref{tab:priors}. 

\begin{figure}
   \centering
    \includegraphics[width=44mm]{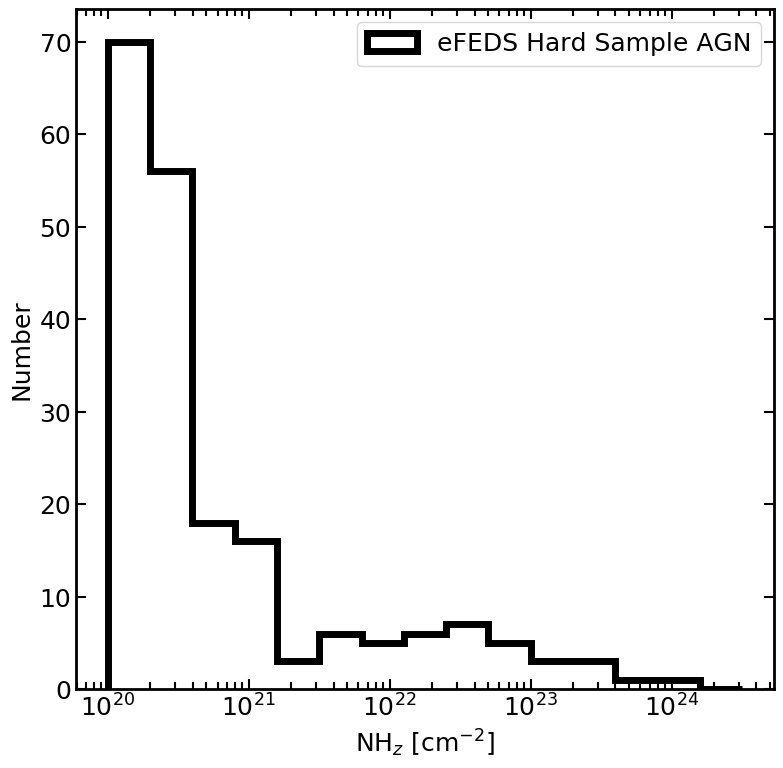}
    \includegraphics[width=44mm]{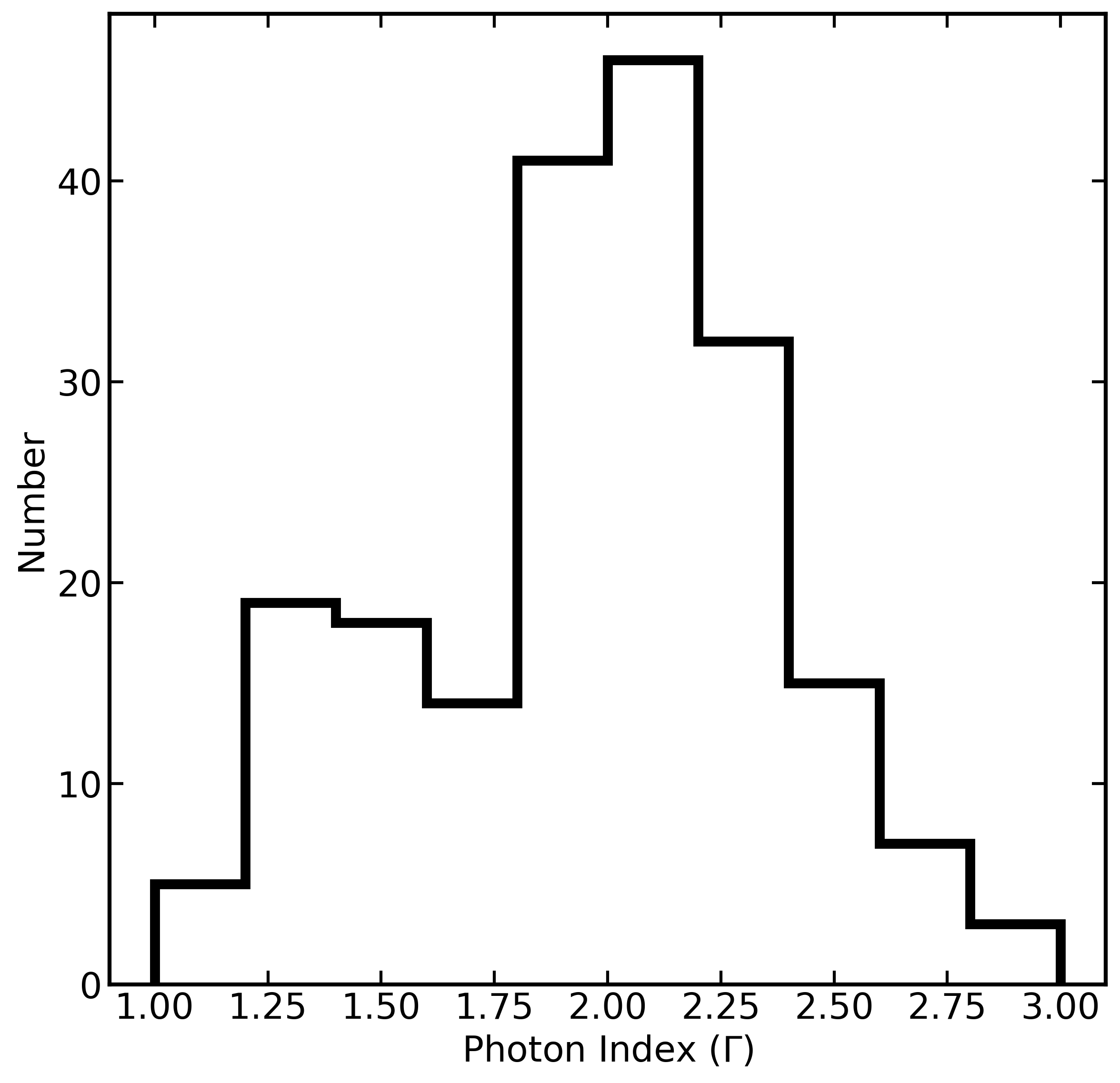}
    \caption{Distributions of absorbed power law (PL) model parameters. Left: Distribution of host galaxy column density measured using the baseline, absorbed power law model. There is a clear primary peak at NH$_z$ $\simeq10^{20}\pscm$, with a secondary peak at NH$_z$ $\simeq10^{22.5}\pscm$. Right: as top left, but shown for the photon index. There is a clear peak at $\Gamma \simeq2.0$, and a second cluster of objects with an unusually flat $\Gamma$ of $\sim1.4$. }
    \label{fig:toyhist}

\end{figure}

\begin{table*}
\centering
\caption{Priors used for simple spectral models. Column (1) gives the shortened name of the model. Columns (2) shows the lower and upper limit of the host galaxy absorption. Columns (3) shows the lower and upper limit of the hard (coronal) power law component. Columns (4) and (5) show the upper and lower limits of the warm or partial covering absorber column density and ionisation component, respectively. Column (6) shows the upper and lower limit placed on the neutral absorption covering fraction. Column (7) shows the upper and lower limit placed on the soft X-ray photon index. Column (8) gives the constraints placed on the cross-normalisation placed on the soft excess component, and column (9) shows the constraints placed on the blackbody temperature. Column (10) gives the constraint placed on the hard power law normalisation. }
\label{tab:priors}
\resizebox{0.85\textwidth}{!}{%
\begin{tabular}{llllllllll}
\hline
(1) & (2) & (3) & (4) & (5) & (6) & (7) & (8) & (9) & (10) \\
Model name & log(NH$_z$)   & $\Gamma_h$ & log(nH$_{abs}$) & log($\xi_{wa}$) & Covering & $\Gamma_s$ & log(c) & $kT$ & norm \\
 & [cm$^{-2}$] & & [cm$^{-2}$] & & fraction & & & [keV] & \\
 \hline
Prior & log-uniform & uniform & uniform & uniform & uniform & uniform & log-uniform & log-uniform & log-uniform \\
\hline
PL    & 20 -- 25  & 1 -- 3  & - & - & - & - & - & - & -10 -- 1 \\
PL+PCF   & 20 -- 25  & 1 -- 3  & 20 -- 25  & - & 0 -- 1 & - & - & - & -10 -- 1 \\  
PL+WA    & 20 -- 25  & 1 -- 3  & 20 -- 24  & -4 -- 4  & - & - & - & - & -10 -- 1 \\
PL+BB    & 20 -- 25  & 1 -- 3  & - & - & - & - & -3 -- 1 & 0.04 -- 0.4 & -10 -- 1 \\
PL+PL    & 20 -- 25  & 1 -- 3  & - & - & - & 2.25 -- 10 & -3 -- 1 & - & -10 -- 1 \\
\hline
\hline
\end{tabular}
}
\end{table*}

Some results from this preliminary fit are shown in Fig.~\ref{fig:toyhist}. The left-hand panel shows the distribution of median values of the host-galaxy column density, NH$_z$. There is clearly a large peak at column densities of NH$_z$ $\simeq10^{20}\pscm$ (at the limit of the prior so consistent with no additional absorption component beyond Galactic absorption), with a smaller, secondary peak at $\simeq10^{22.5}\pscm$. Since most sources have low column densities, this suggests that the soft excess component or warm absorption, if present, should be easily detectable in most sources. The right-hand panel shows the distribution of median values of the photon indices ($\Gamma$), with most having values of $\Gamma \simeq2$. This is likely a selection effect, as the value is slightly steeper than found by some samples \citep[e.g.][]{1994Nandra, 2000Reeves, 2020Waddell, 2022Waddell}, but in agreement with previous eROSITA modelling presented in \citet{2022Liu} and Nandra \et (in prep.). There are also a number of sources with very steep ($\Gamma >2.3$) or very flat ($\Gamma <1.4$) values. With typical error bars of the order of $\pm0.2$, these values are not in agreement with the expected median of $\Gamma \simeq2$. These likely indicate the presence of more complex spectral features, such as soft excess emission or complex absorption, motivating further investigation.

\subsection{Model comparison summary}
In the rest of this work, the best model for each spectrum is identified with Bayesian model comparison. This relies on the computation of Bayes factors, which examine the Bayesian evidence for two models to determine which is preferred.

Simulations are used to assess the significance of selecting one model over another. These are described in detail in Appendix A. One-thousand simulated spectra are generated using an absorbed power law (PL) model using the average sample properties, and these spectra are then fit with each of the more complex models subsequently defined in this work. Purity thresholds can then be defined based on the Bayes factor values which yield a given number of instances where modelling falsely selects the more complex model as the correct one; in this work, thresholds of 95\%, 97.5\% and 99\% are considered, and more detail is given on these selections in Appendix A. For these simulated spectra, false detections are defined when a model both has the lowest Bayesian evidence of all models, and the Bayes factor exceeds the threshold. In this way, for a real source to be classified as having a warm absorber, soft excess, or partial covering absorber, this model must have the lowest Bayesian evidence, and the Bayes factor must exceed the threshold. For this reason, there is no overlap between the true soft excess, partial covering, or warm absorber samples. 

\renewcommand{\arraystretch}{1.3}
\begin{table}
\centering
\caption{ Definitions of Bayes factors used throughout this work for model comparison.} 
\label{tab:bayes}
\resizebox{\columnwidth}{!}{%
\begin{tabular}{lllll}
\hline
(1) & (2) & (3) & (4) & (5) \\
K$_{wa}$ & K$_{pcf}$ & K$_{pl}$ & K$_{nth}$ & K$_{rel}$ \\
\hline
$Z_{WA}$/$Z_{PL}$ & $Z_{PCF}$/$Z_{PL}$ & $Z_{PL+PL}$/$Z_{PL}$ & $Z_{NTH}$/$Z_{PL}$ & $Z_{REL}$/$Z_{PL}$ \\
\hline
\hline
\end{tabular}
}
\end{table}

\subsection{Absorption modelling}
\label{sect:wafit}
The first model used to model a complex absorption component in this work is a neutral partial covering scenario (PL+PCF), where emission from the corona passes through an absorber before reaching the observer \citep[e.g.][]{2004Tanaka}. This absorbs hard X-ray photons while allowing leakage in the soft X-ray, which flattens the observed power law slope, gives the appearance of a soft excess, and can produce a deep edge at $7\kev$ depending on the column density. However, often multiple absorbing zones with a variety of ionisation states, column densities and covering fractions are required to fit the observed spectral shape.

In this work, one neutral partial covering absorber is applied to each source. The full XSPEC implementation is given in Table~\ref{tab:models}. Two more free parameters are present; the absorption column density is allowed to vary between $10^{20}$ and $10^{25}\pscm$, and the fraction of emission which passes through the absorber (the covering fraction) is allowed to vary between zero (no absorption) and one (full covering). The redshift of the absorber is set to that of the host galaxy such that absorption in the vicinity of the corona is modelled. All other parameters and priors are the same as the baseline PL model. While the use of a single, neutral absorber to explain the observed curvature in the spectral shape may be an over-simplification of a true physical absorber, this simple implementation still allows for a preliminary model check, and the Bayesian evidence can be compared with other physical scenarios. 

In the warm absorber model (PL+WA), rather than passing through a neutral absorber, the X-ray emission passes through an ionised medium, which produces absorption features in the soft X-ray spectrum due to partially ionised materials including Neon, Oxygen and Iron \citep[e.g.][]{1998George, 2000Kaastra, 2000Kaspi, 2004Blustin, 2005Blustin, 2004Gierlinski, 2007Mckernan, 2014Laha, 2019Mizumoto}. These warm absorber features have been physically linked to low velocity (e.g. 100s-1000s km~s$^{-1}$) outflows or disc winds which intercept the line of sight \citep[e.g.][]{2019Kallman}. To model the warm absorber, an XSPEC-compatible table model (cwa18.fits; cwa18) was generated using XSTAR. The construction of this model is described in \citet{2007Nandra}. This model also has two more free parameters than the baseline PL model; the column density and the ionisation of the absorber. The ionisation of the absorber is allowed to vary broadly between 10$^{-4}$ and 10$^4$\xii to account for a broad range of wind ionisation states, and the column density is between $10^{20}$ and $10^{24}\pscm$. The full list of priors for the partial covering absorber and warm absorber models is summarised in table~\ref{tab:priors}. 

The significance of each of the absorption components is assessed using simulations, described in detail in Appendix A, and all individual fit parameters are given in Appendix B. Using these simulations and the Bayes factor as given in table~\ref{tab:bayes}, it is found that 29/200 sources ($14.5\%$) have evidence for a warm absorber and 25 sources ($12.5\%$) have evidence for partial covering absorbers, both at the 97.5\% confidence level. For completeness, most figures will show all three determined purity levels (95\%, 97.5\%, and 99\% significance) for comparison.

\begin{figure}
   \centering
    \includegraphics[width=0.95\columnwidth]{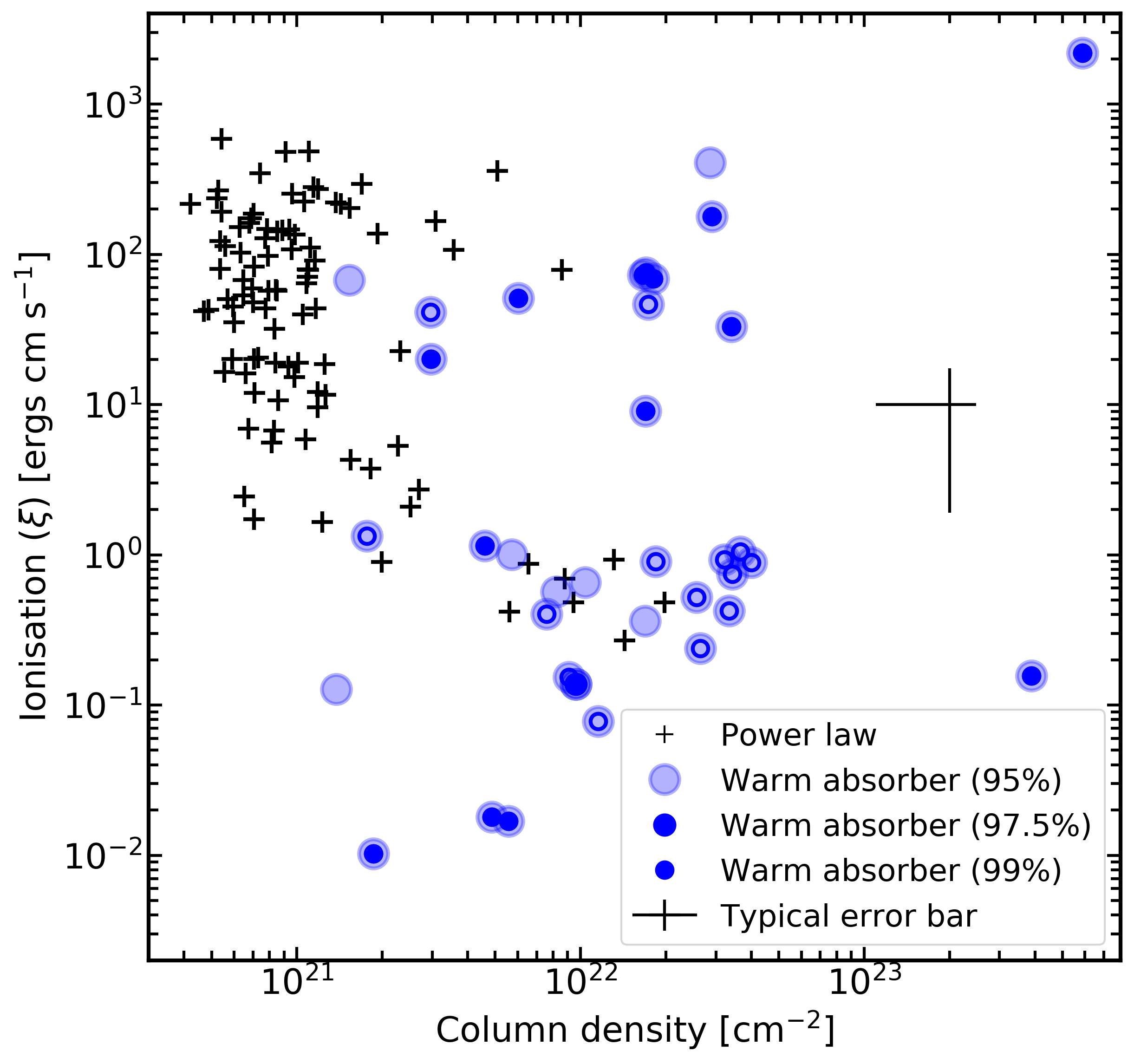}
    \caption{Warm absorption parameters for all AGN in the sample. Sources with warm absorption components of various purity levels (95\%, 97.5\% and 99\%) are indicated with blue circles (translucent, unfilled and opaque, respectively). Typical error bars are indicated with a black cross. }
    \label{fig:warmabs}

\end{figure}

Fig.~\ref{fig:warmabs} shows the distribution of the warm absorber parameters; column density and ionisation ($\xi$). All three purity levels are shown; sources which have purity at the 95\% level (K$_{wa}$ > 0.815) are shown as translucent blue circles, sources at the 97.5\% level (K$_{wa}$ > 1.126) are shown as blue rings, and sources with the 99\% purity (K$_{wa}$ > 2.040) are shown as dark blue circles. Sources which do not show significant evidence for warm absorbers, and are indicated with black crosses. Sources lacking evidence for warm absorption have lower column densities, while sources with warm absorbers have higher column densities of $\geq10^{21}\pscm$, with typical values around $10^{22}-10^{23}\pscm$. In general, the column densities are not well constrained and can extend to lower values, as there is significant degeneracy between the column density and the ionisation of the absorber as well as between the host-galaxy absorbing column density and the warm absorber column density. 

Interestingly, the significant warm absorbers show a wide range of column densities and ionisations, suggesting some diversity in absorbers across different AGN. Typically, warm absorbers studied in the X-ray have been found to have ionisations of the order of $\xi$ $\sim 10-1000$\xii and column densities of the order of 10$^{20}$-10$^{23}\pscm$ \citep[e.g.][]{2004Blustin,2005Blustin,2007Mckernan,2010Tombesiufo,2019Mizumoto}. The results from this work are broadly in agreement with this; however, several low ionisation absorbers with $\xi$ $\sim 0.01-1$\xii are also found, including some with very high significance (e.g. a very large improvement of the Bayesian evidence compared to a power law). The error bars on the ionisation these sources are large, and the marginal posterior probability distributions can be complex (see sections 5.2 and 6.4 for more details). These results should be interpreted with caution due to the known degeneracy between the ionisation and the column density of the absorber, however they appear significantly different than sources best fit with neutral absorbers. It is also interesting to note that the sources with higher ionisations ($\sim10^2$\xii) have relatively low redshifts ($z<0.5)$, while the sources with lower ionisations occupy a much broader range of redshifts with many having $0.5 < z < 1$. These sources will be discussed in more detail in later sections. 

\begin{figure}
   \centering
    \includegraphics[width=0.95\columnwidth]{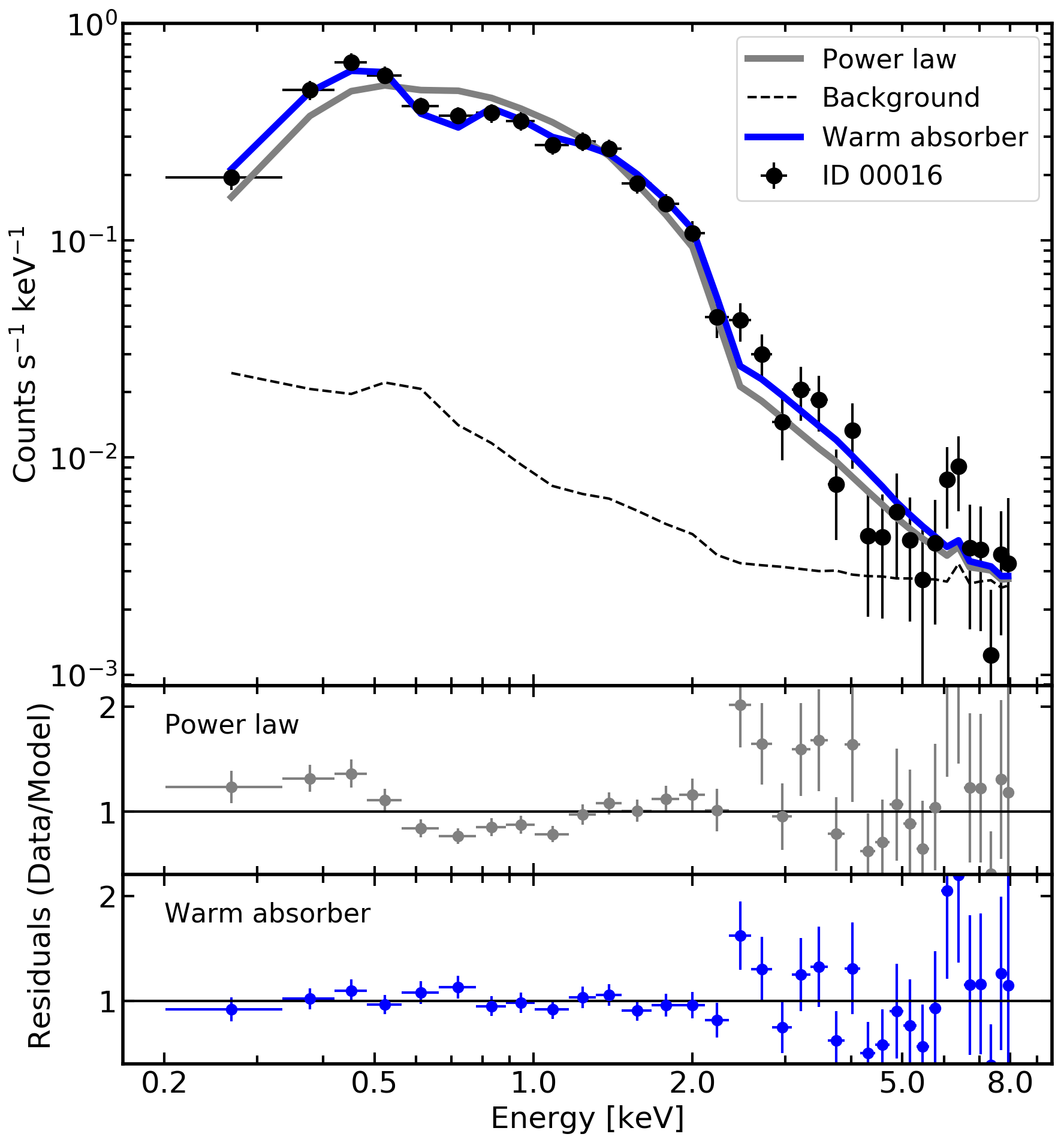}
    \caption{ID 00016 (z = 0.2907), a source best fit with a warm absorption model. The spectrum (re-binned for display) is shown in black, the background model is shown as a black dashed line, the power law model is shown as a grey line, and the warm absorber model is shown in blue. The bottom two panels show the residuals for the power law, and a warm absorber, respectively. The source has relatively high signal-to-noise, and has a warm absorber column density of $\sim10^{22}\pscm$ and an ionisation of $\xi\sim10^2$\xii. Data have been re-binned for display purposes.} 
    \label{fig:warmabsspec}

\end{figure}

An example of a source (ID 00016) best fit with a warm absorber model is shown in Fig.~\ref{fig:warmabsspec}. The background (black dashed line), the warm absorber model (blue) and an absorbed power law model (grey) are shown over-plotted with the folded spectrum. The warm absorber model clearly provides a better fit than the simple absorbed power law (PL) model, in particular to the softest X-ray energies, as well as in the $2-5\kev$ band. In this case, the partial covering absorber model fails to reproduce the absorption features. 

\begin{figure}
   \centering
    \includegraphics[width=0.95\columnwidth]{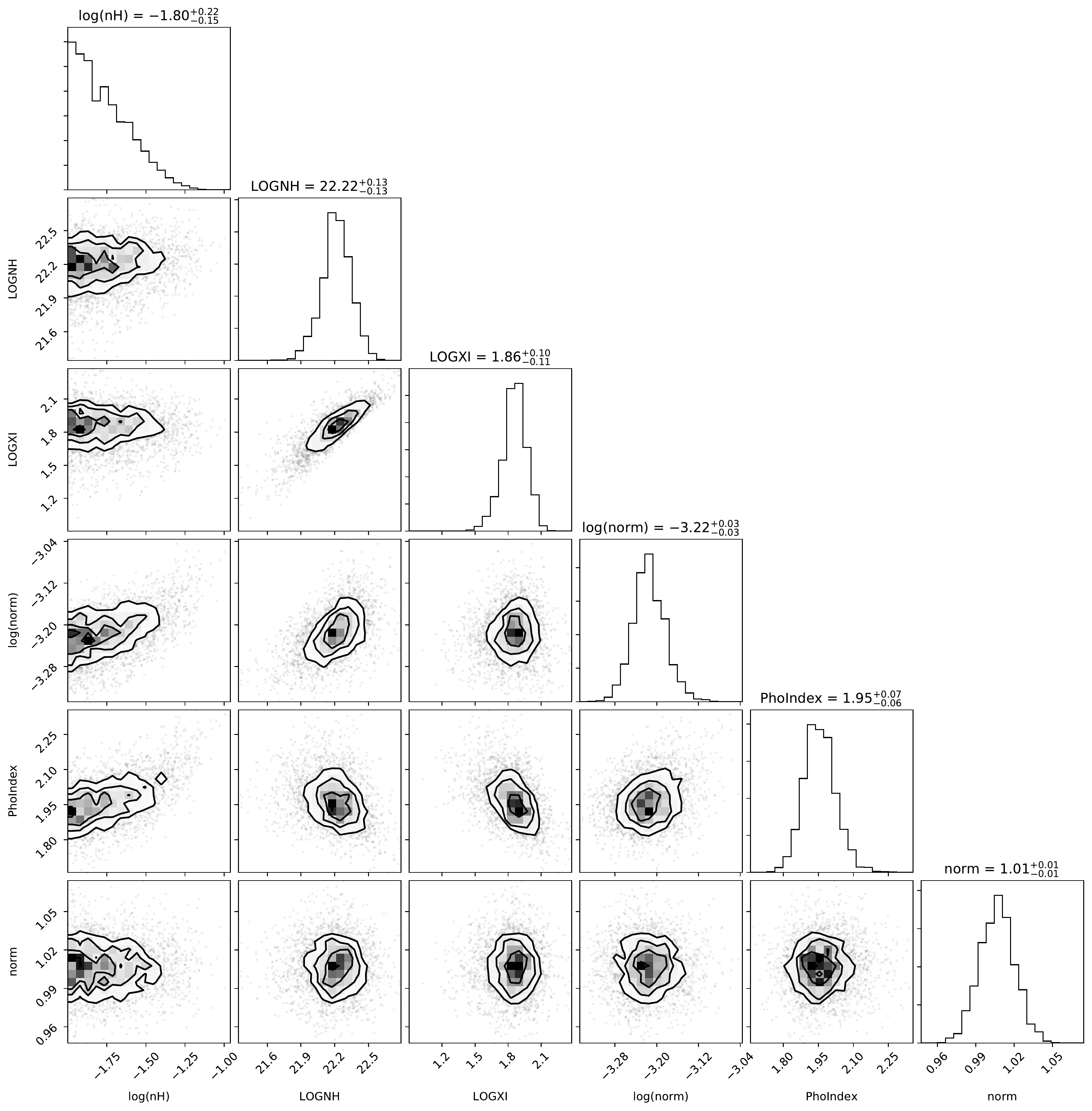}
    \caption{Corner plot for source ID 00016, best fit with a warm absorption model. The diagonal panels show the marginal posterior probability distribution for each parameter, while the other panels show the conditional probability distribution functions for each pair of parameters. Here, log(nH) is the host galaxy absorber column density (in units of $\times10^{22}\pscm$), LOGNH is the column density of the warm absorber ($\pscm$), LOGXI is the ionisation of the warm absorber (\xii), log(norm) is the power law normalisation, PhoIndex is the photon index of the power law, and norm is the relative renormalisation of the background model with respect to the source model, which is in agreement with 1. }
    \label{fig:wacorner}

\end{figure}

Fig.~\ref{fig:wacorner} is the corresponding corner plot for the warm absorber model fit to source ID 00016, with variable names provided in the figure caption. For this source, most parameters are very well independently constrained, though some degeneracy exists between the column density and the ionisation of the warm absorber. Here the column density of the warm absorber and host galaxy absorber are independently constrained, and the host galaxy column density is consistent with the minimum value. More discussion on parameter correlations and degeneracies for this model is found in sections 5.2 and 6.4.

\begin{figure}
   \centering
    \includegraphics[width=0.95\columnwidth]{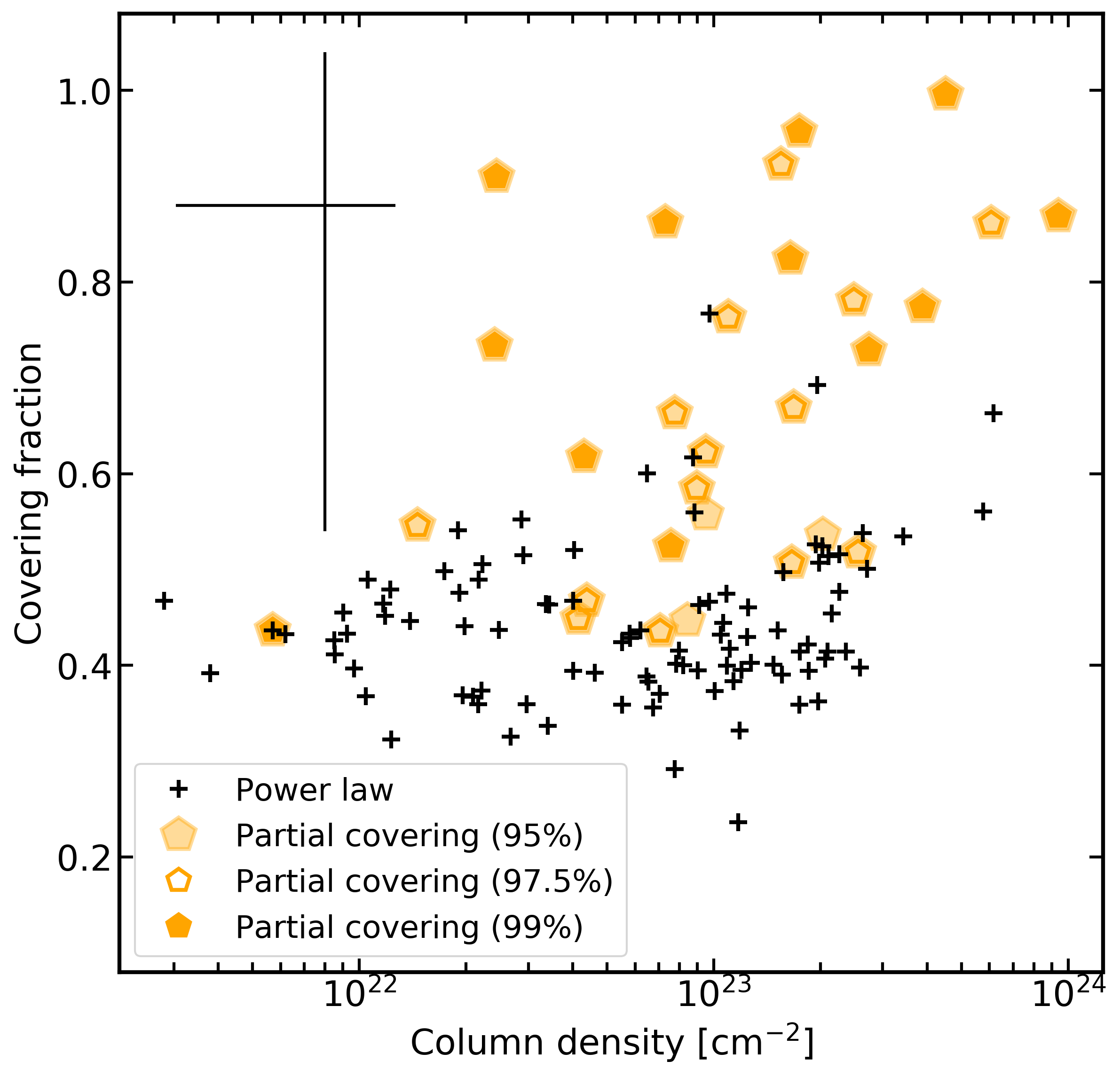}
    \caption{Partial covering parameters for all AGN in the sample. Sources with partial covering components of various purity levels (95\%, 97.5\% and 99\%) are indicated with orange pentagons (translucent, unfilled and opaque, respectively). Typical error bars are indicated with a black cross. }
    \label{fig:pcfabs}

\end{figure}

Finally, Fig.~\ref{fig:pcfabs} shows the distribution of partial covering column densities and covering fractions obtained from the PL+PCF model, with the Bayes factor as given in Table~\ref{tab:bayes}. As in Fig.~\ref{fig:warmabs}, sources which have purity at the 95\% level (K$_{pcf}$ > 1.392) are shown as translucent orange pentagons, sources at the 97.5\% level (K$_{pcf}$ > 1.555) are shown as orange unfilled pentagons, and sources with the 99\% purity (K$_{pcf}$ > 2.646) are shown as darker orange pentagons. It is clear that sources which are best fit with the partial covering model have higher covering fractions than those which do not; the median is $\sim0.4$ for sources with no evidence for partial covering absorption, and $\sim0.7$ for those with evidence for partial covering components. Typical column densities are $\sim10^{23}\pscm$, with a few sources having higher column densities near $10^{24}\pscm$. This is similar to the warm absorbers. Many of the highest significance sources (filled orange pentagons) also have very steep photon indices (e.g. those in the top right-hand corner of Fig.~\ref{fig:pcfabs}). The high covering fraction creates deep absorption edges around $7\kev$ for sources with sufficiently high column densities, which are difficult to see in the data due to high background at high energies. 

\begin{figure}
   \centering
    \includegraphics[width=0.95\columnwidth]{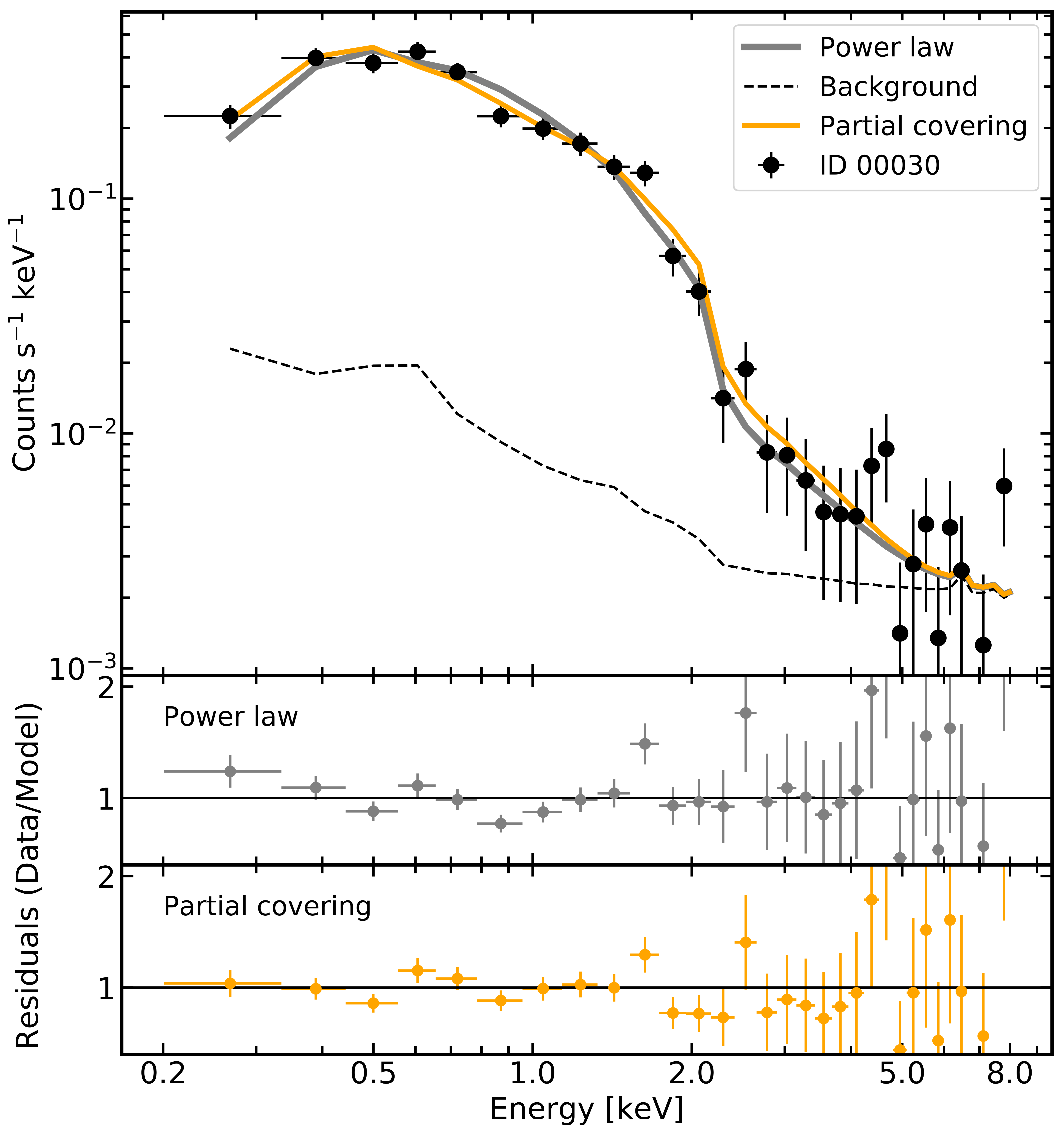}
    \caption{ID 00030 (z = 0.4263), a source best fit with a partial covering absorption model. The spectrum (re-binned for display) is shown in black, the background model is shown as a black dashed line, the power law model is shown as a grey line, and the neutral partial covering absorption model is shown in orange. The bottom two panels show the residuals for the power law, and a partial covering absorber, respectively. The source has a moderate covering fraction of $\sim0.6$ and a moderate column density of $4\times10^{22}\pscm$. Data have been re-binned for display purposes. } 
    \label{fig:pcfspec}

\end{figure}

\begin{figure}
   \centering
    \includegraphics[width=0.95\columnwidth]{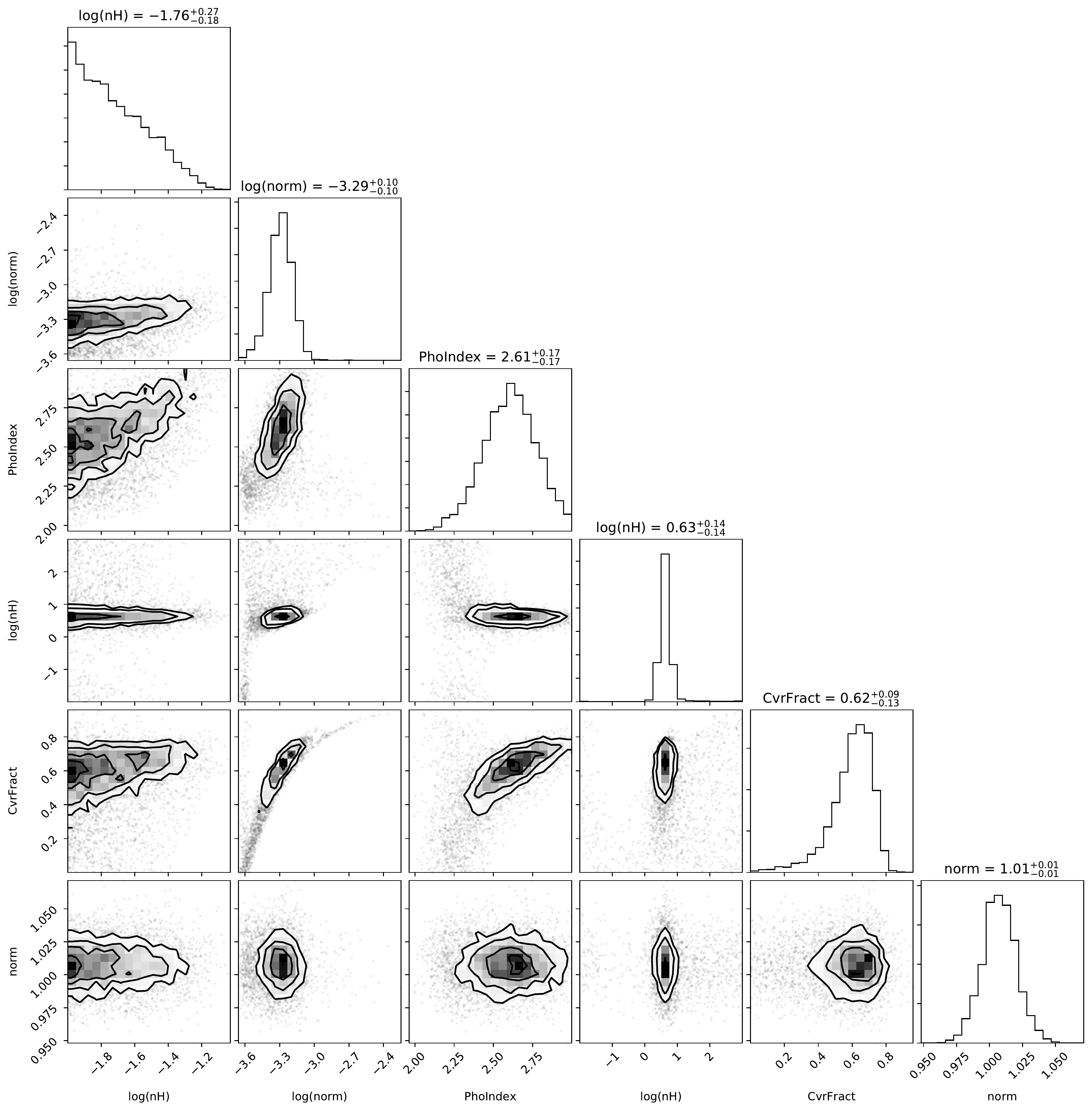}
    \caption{Corner plot for source ID 00030, best fit with a partial covering absorber. Here, the first instance of log(nH) is the host galaxy absorber column density (in units of $\times10^{22}\pscm$), log(norm) is the power law normalisation, PhoIndex is the photon index of the power law, the second instance of log(nH) is the partial covering absorber column density ($\times10^{22}\pscm$), CvrFrac is the covering fraction of the absorber, and norm is the relative renormalisation of the background model with respect to the source model, which is in agreement with 1. } 
    \label{fig:pcfcorner}

\end{figure}

An example of a source (ID 00030) best fit with the neutral partial covering model is shown in Fig.~\ref{fig:pcfspec}. As for the warm absorber spectrum, the background (black dashed line), the partial covering model (orange) and an absorbed power law model (grey) are shown over-plotted with the folded spectrum. This source is also well fit with a power law soft excess model, but evidence comparison reveals that the partial covering absorption model provides the best fit, highlighting the importance of considering a variety of models to explain the soft spectrum. 

The corner plot for the partial covering model of ID 00030 is shown in Fig.~\ref{fig:pcfcorner}. Some parameters are less well constrained for this model, and the photon index is found to be extrmely high compared to expected values of $\sim1.9-2.0$. In this source, there are degeneracies between the partial covering fraction and photon index, as well as between the partial covering fraction and the normalisation on the coronal power law component. In some sources, there are also degeneracies between the partial covering fraction and column density. These column density degeneracies will be discussed in Sect. 6.4.

\subsection{Soft excess modelling}
\label{sect:sefit}
In order to account for an intrinsic soft excess component, two separate spectral models are used. First, a \texttt{blackbody} component is used (PL+BB). The normalisation component of the blackbody is linked to that or the coronal power law, with a constant factor applied to set the relative spectral flux density of the power law and blackbody components at $1\kev$. For the second model, the blackbody component is replaced by a soft power law component (PL+PL), where the relative normalisations of the soft and hard power law is again fit using a constant factor. Motivated by the results of \citet{2022Liu} from fitting the eFEDS main sample, a wide range of values are adopted for the priors of the blackbody temperature $kT$ and the soft power law index $\Gamma_s$ in order to characterise a wide variety of soft excess shapes. Priors are listed in Table~\ref{tab:priors}. 

\begin{figure}
   \centering
    \includegraphics[width=0.95\columnwidth]{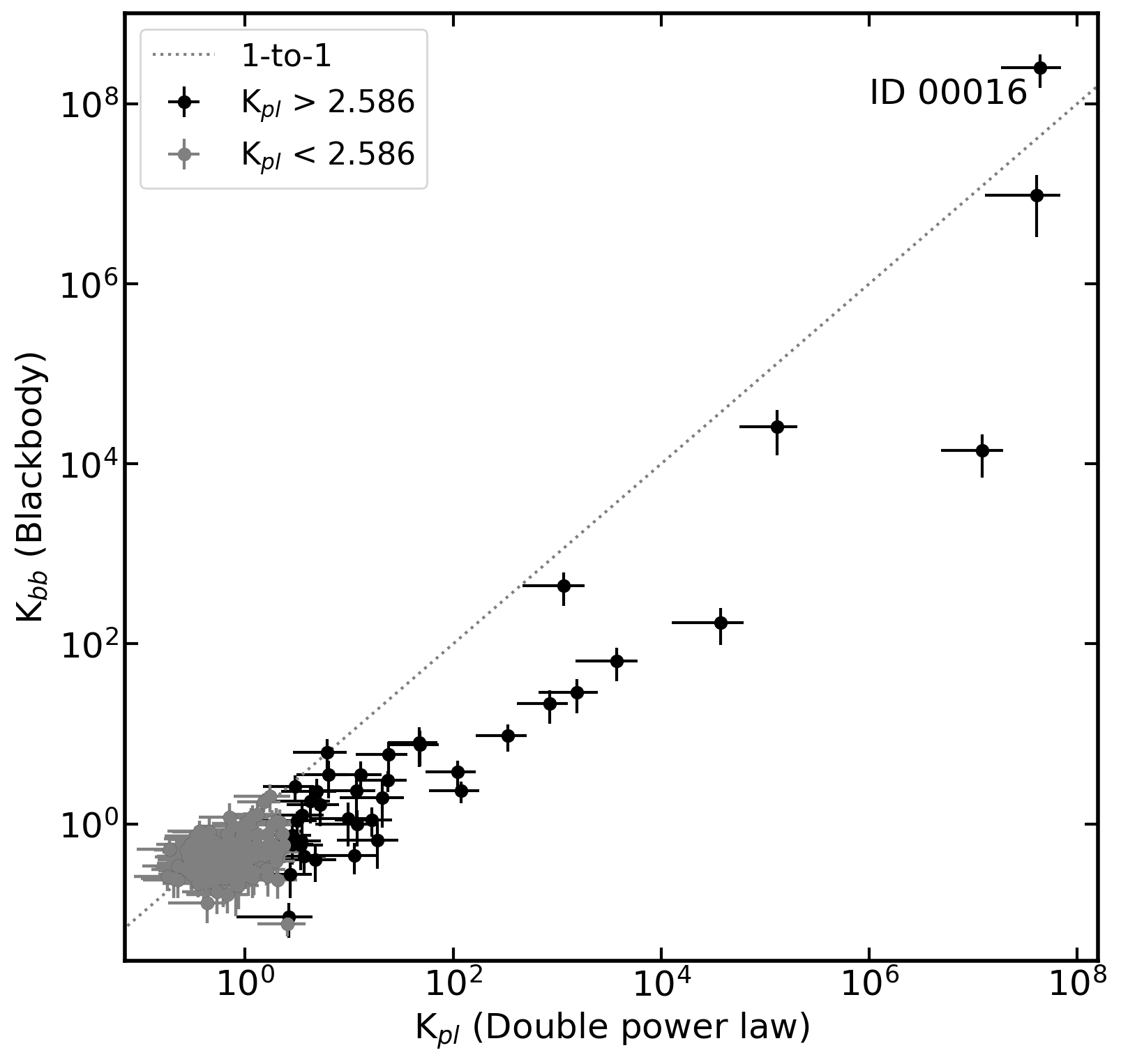}
    \caption{Comparison of Bayes factors between soft excess models. The Bayes factor for the double power law is shown on the horizontal axis, while the Bayes factor for the blackbody model is shown on the vertical axis. The dashed grey line shows the one-to-one relation, with sources lying below the line being better fit with the PL+PL model, and sources above the line being better fit with the PL+BB model. Sources which show evidence for a soft excess (K$_{pl}$>2.586, corresponding to a 97.5\% significance) are shown in black, and sources which do not show evidence for a soft excess are shown in grey.}
    \label{fig:plvsbb}

\end{figure}

The significance of the soft excess is evaluated for each source by computing the Bayes factor (equation~\ref{eqn:kbayes}), as given in Table~\ref{tab:bayes}. The resulting Bayes factors are then compared between models for each source to assess which model is better able to characterise the shape of the soft excess. The results are shown in Fig.~\ref{fig:plvsbb}, shown for all sources, including those which show evidence for obscuration. Sources which do not show evidence for a soft excess are shown in grey. Most sources are better fit with the PL+PL model (e.g. lie below the line), and all but one of the few sources which are better fit with the PL+BB model have very low K$_{pl}$ and K$_{bb}$ values and therefore likely do not have a soft excess. The sole exception of this is eFEDS ID 00016, which appears to have a strong soft excess with both models, but in fact is better fit with a warm absorber. Therefore, it can be concluded that the PL+PL model is a better representation of the soft excess, therefore the PL+BB model will be discarded for the remainder of this work. The double power law model will be used to select sources with significant soft excesses. 

\begin{figure}
   \centering
    \includegraphics[width=0.95\columnwidth]{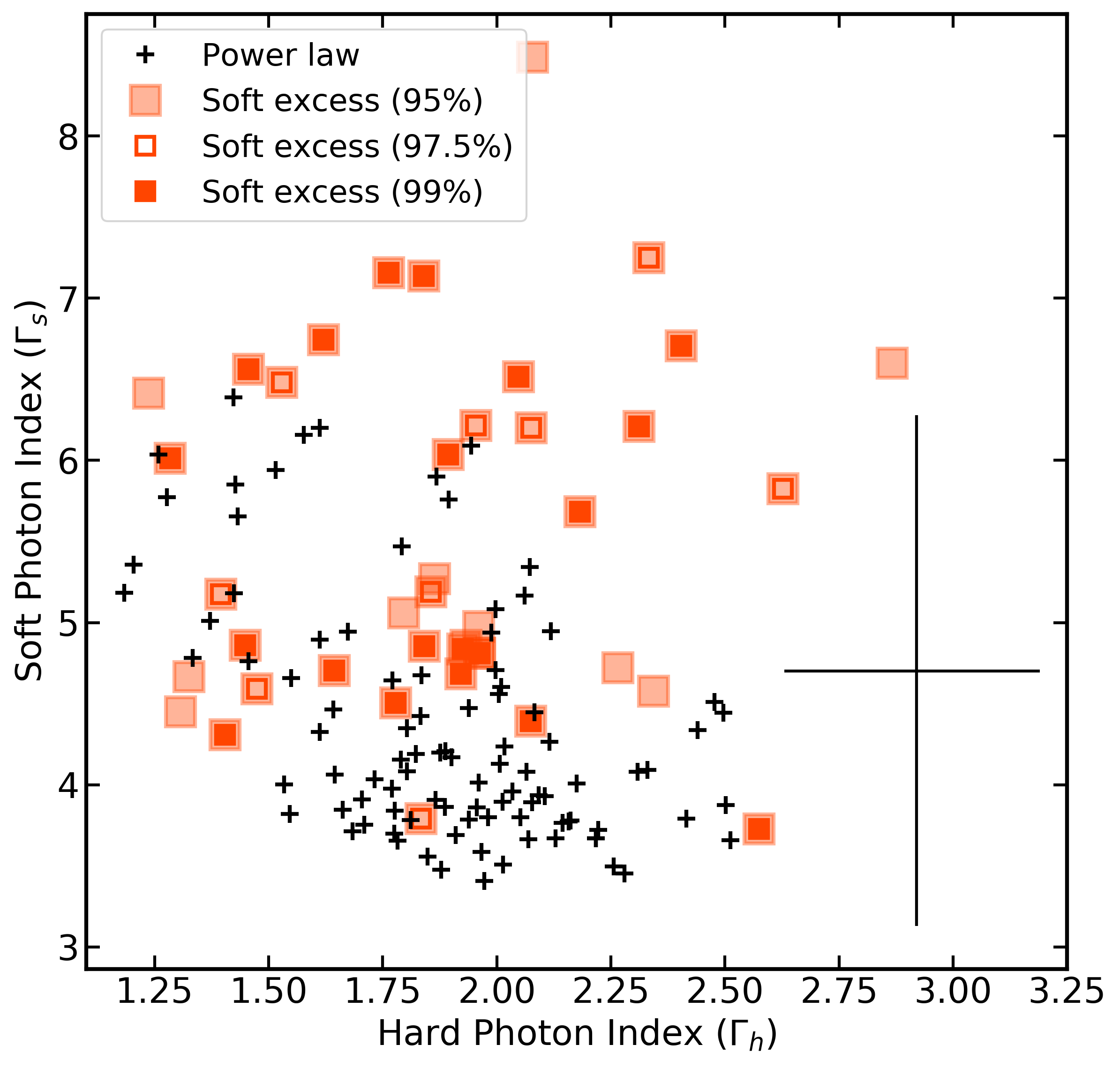}
    \caption{Soft excess (double power law) parameters for all AGN in the sample. Sources with soft excess components of various purity levels (95\%, 97.5\% and 99\%) are indicated with red squares (translucent, unfilled and opaque, respectively). Typical error bars are indicated with a black cross. }
    \label{fig:plplmod}

\end{figure}

As with the complex absorber modelling, the significance of each of the soft excess components is assessed using simulations, described in detail in Appendix A. The distributions of parameters for the soft excess models are shown in Fig.~\ref{fig:plplmod}. All three purity levels are shown; sources which have purity at the 95\% level (K$_{pl}$ > 1.392) are shown as translucent red squares, sources at the 97.5\% level (K$_{pl}$ > 2.586) are shown as unfilled red squares, and sources with the 99\% purity (K$_{pl}$ > 8.613) are shown as dark red squares. Choosing the 97.5\% purity level, this leaves 29/200 sources with soft excesses, or $\sim14.5\%$ of the full sample, the same number as found for a warm absorption model and slightly more than found using a partial covering model. However, significantly more sources are found to have soft excesses at the 95\% and 99\% confidence levels. Sources with soft excess display a surprisingly large variation in primary photon index, which may suggest that some sources have additional absorption components not considered in this model, or that the two power law model is too simplistic to characterise the spectral shape and complexity for some sources. There also appears to be two clusters of soft photon indices for sources with soft excesses; a cluster around $\Gamma_s = 4.5$, and another around $\Gamma_s = 6.5$. Both of these values are too steep to be produced in a corona with reasonable opacity and temperature \citep[e.g.][]{2018Petrucci}, and depending on the assumed temperature, these may exceed the steepness of the exponential cut-off. Rather, they likely highlight some diversity in the shape of observed soft excesses, or may be competing with the host galaxy absorption to attempt to match the observed spectral shape. This will be explored further in Sect. 4.

\begin{figure}
   \centering
    \includegraphics[width=0.95\columnwidth]{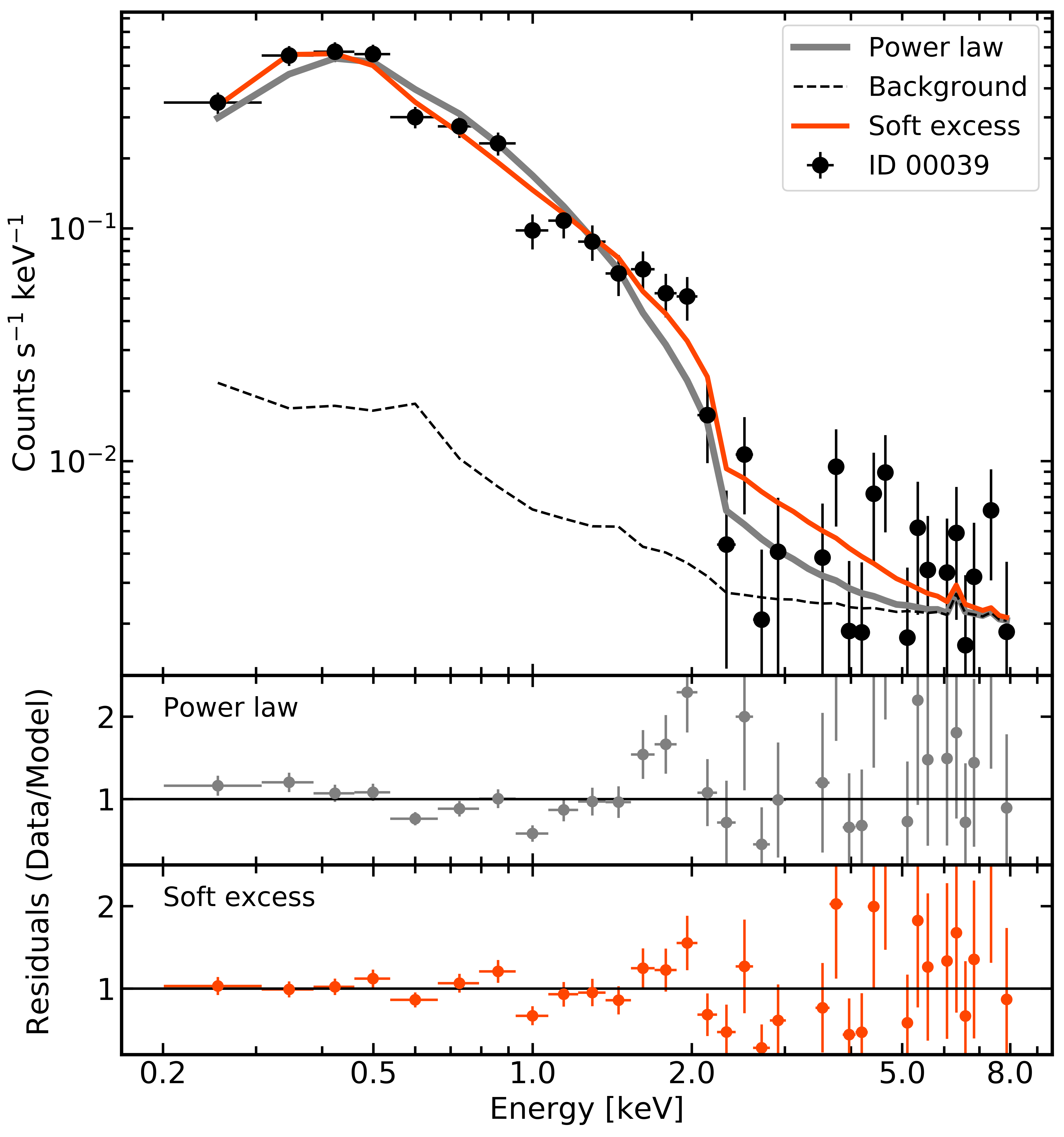}
    \caption{ID 00039 (z = 0.3893), a source best fit with a double power law soft excess model. The spectrum (re-binned for display) is shown in black, the background model is shown as a black dashed line, the power law model is shown as a grey line, and the double power law model is shown in red. The bottom two panels show the residuals for the power law, and a soft excess, respectively. Data have been re-binned for display purposes. }
    \label{fig:plplspec}

\end{figure}

\begin{figure}
   \centering
    \includegraphics[width=0.95\columnwidth]{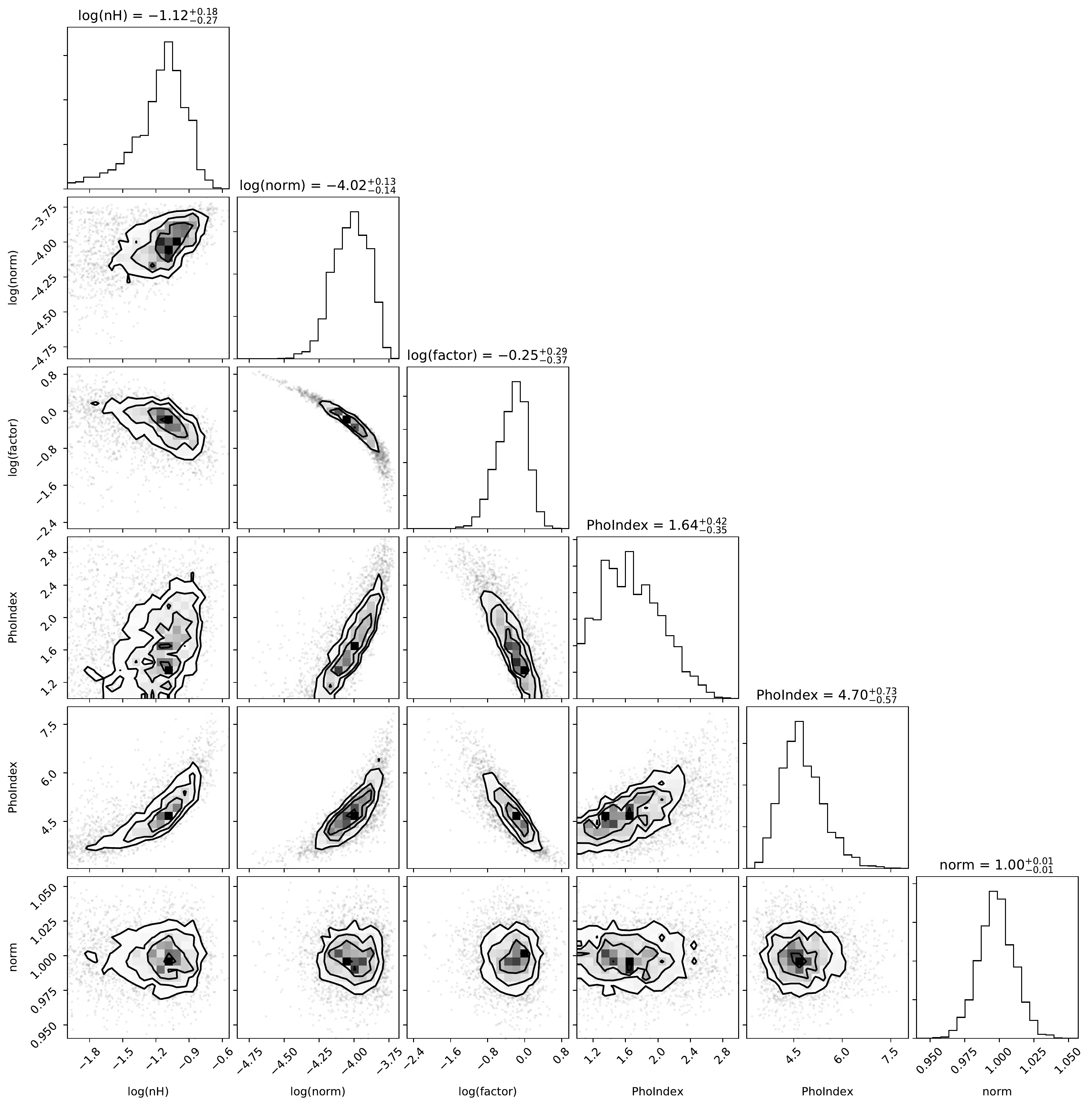}
    \caption{Corner plot for source ID 00039, best fit with a double power law soft excess. Here, the first instance of log(nH) is the host galaxy absorber column density (in units of $\times10^{22}\pscm$), log(norm) is the power law normalisation, log(factor) is the relative normalisation of the soft power law with respect to the hard power law, the first instance of PhoIndex is the photon index of the hard power law, the second instance of PhoIndex is the photon index of the soft power law, and norm is the relative renormalisation of the background model with respect to the source model, which is in agreement with 1. }
    \label{fig:secorner}

\end{figure}

An example of a source (ID 00039) best fit with the double power law model is shown in Fig.~\ref{fig:plplspec}. As for the previous models, the background (black dashed line), the double power law soft excess model (red) and an absorbed power law model (grey) are shown over-plotted with the folded spectrum. The fit improvement by adding the second power law is visually apparent throughout the spectrum, and in particular in the hard band - the single power law model tends to try to approximate the softer end of the spectrum where the instrument is more sensitive, while the second power law provides a better fit across all energies. The corresponding corner plot for this source is shown in Fig.~\ref{fig:secorner}, and the component labels are explained in the caption. Here there are some additional degeneracies between parameters, including between the photon indices and normalisations, as well as between the photon indices and the covering fraction. While some parameters are less well independently constrained, the double power law model is still highly informative in characterising the shape of the soft excess.

\subsection{The soft excess, warm absorber and partial covering samples}
\label{sect:sesamp}
\begin{figure}
   \centering
    \includegraphics[width=0.95\columnwidth]{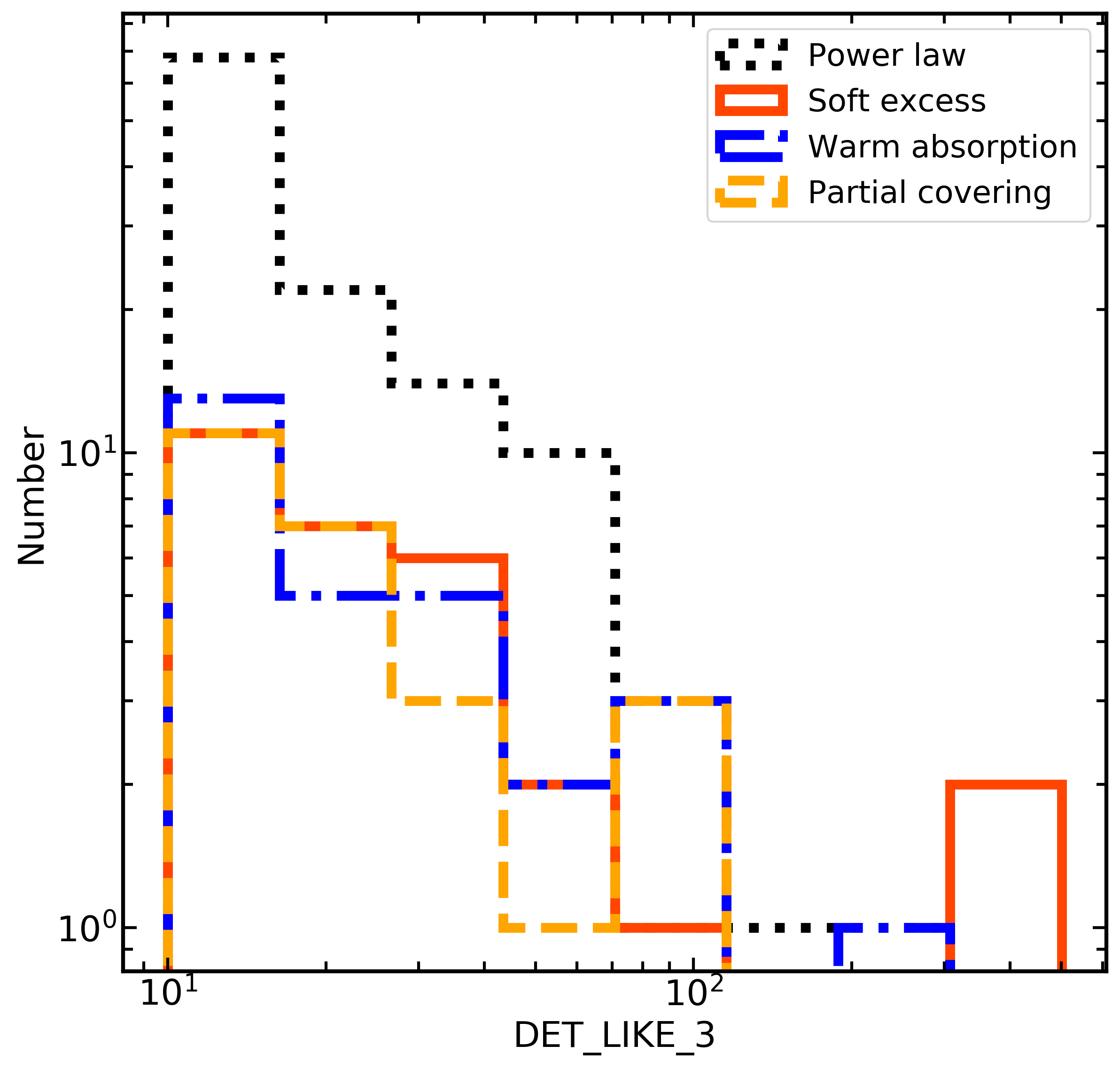}
    \caption{Distributions of detection likelihood in the $2.3-5\kev$ band (DET\_LIKE\_3) shown for sources best fit by each model. The vertical axis is given in log-space to highlight the true distribution of the sample. Sources best fit with a power law are shown with a black dotted line, sources best fit with a soft excess are shown as a red solid line, sources best fit with a warm absorber are shown as a blue dash-dot line, and sources best fit with partial covering absorption are shown with an orange dotted line. }
    \label{fig:bias}

\end{figure}

After analysing these phenomenological models, based on selecting samples with 97.5\% purity, this work finds 29 sources with true soft excesses, 29 with warm absorbers, and 25 sources with partial covering absorbers, where by definition there is no overlap between each of these samples. This is because we require the model to provide the best fit to the data as well as satisfying the Bayes factor criteria. With these defined samples, it is possible to search for possible sources of bias in the relatively limited parent sample used in this work. Figure~\ref{fig:bias} shows the detection likelihood in the $2.3-5\kev$ (DET\_LIKE\_3) band for sources best fit with each model. To assess any potential differences in the distributions, a Kolmogorov-Smirnov test (KS-test) is used, which compares two samples to compute the likelihood that they are drawn from the same parent sample. No differences between distributions of hard band detection likelihoods are found, with KS-test p-values $>0.1$ when comparing each sub-sample, and soft excesses and warm absorbers are found in sources with the lowest and highest computed DET\_LIKE\_3 values alike. This shows that the selection of the hard X-ray--selected sample does not heavily bias the detection of soft excesses or complex absorbers, and indeed, likely facilitates these measurements as the hard power law can better be constrained. This also motivates using hard X-ray--selected samples of AGN for further investigations of future eROSITA samples.

\begin{figure}
   \centering
    \includegraphics[width=0.95\columnwidth]{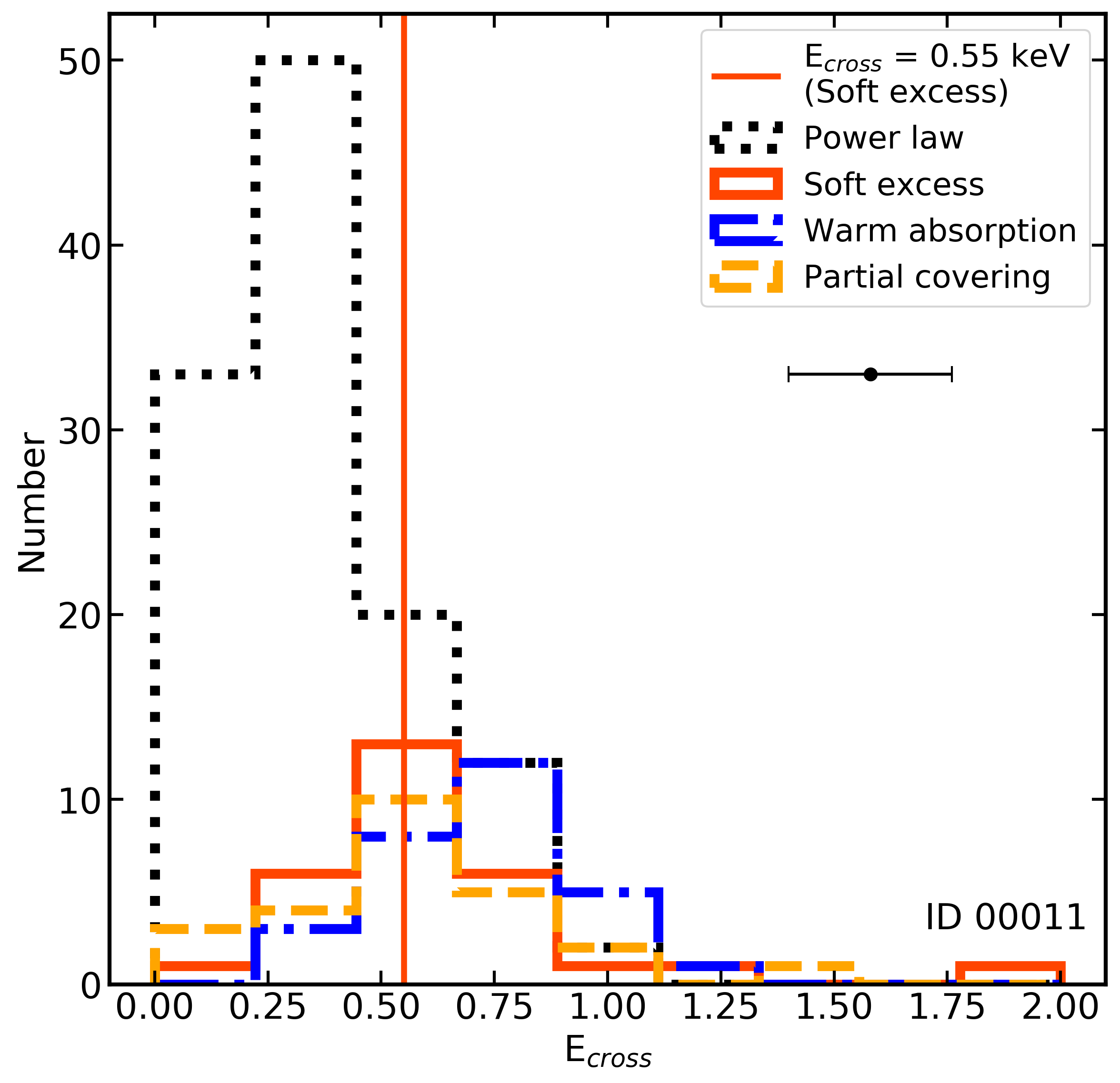}
    \caption{Distribution of computed rest-frame E$_{\rm cross}$ values. Sources best fit with a power law are shown with a black dotted line, sources best fit with a soft excess are shown as a red solid line, sources best fit with a warm absorber are shown as a blue dash-dot line, and sources best fit with partial covering absorption are shown with an orange dotted line. The median of E$_{\rm cross} = 0.55\kev$ for the soft excess sample is indicated with a solid red line. The typical error bar is shown in black. }
    \label{fig:ecross}

\end{figure}

Considering the true soft excess, it is also of interest to examine the energy at which the two power laws have the same flux. This point marks where the soft excess begins to dominate over the hard corona power law. This value (E$_{\rm cross}$) has been computed, and the resulting histogram is shown in Fig.~\ref{fig:ecross}, using the 97.5\% purity samples. Regardless of which model is the best fit, the PL+PL model is necessarily used to compute the E$_{\rm cross}$ value. All results are shown in the rest-frame, and the typical error bar is also shown in black in the top right-hand corner. Most sources have E$_{\rm cross}$ values of less than $1\kev$, and all are below $2\kev$. The median value for the soft excess sample of E$_{\rm cross} = 0.55\kev$ is indicated with a solid red vertical line. This demonstrates the importance of having a good fit in the softest energy X-rays in order to properly characterise the soft excess; fits performed only above $0.5-1\kev$ will likely not be able to fit the true soft X-ray shape.

In addition to examining the Bayes factors and crossing energies, the soft excess can also be defined in terms of the soft excess strength (SE). In this work, this is defined as; 

\begin{equation}
    \rm{SE} = \frac{F_{SE}}{F_{PL}}
\end{equation}

\noindent where F$_{SE}$ is the unabsorbed, rest-frame flux of the soft power law in the $0.2-1\kev$ band, and F$_{PL}$ is the unabsorbed, rest-frame flux of the hard power law in the $0.2-1\kev$ band. The soft excess strength therefore indicates how much excess flux is provided by the soft excess component in the $0.2-1\kev$ band. Separately, the soft flux fraction (SFF) can also be defined; 

\begin{equation}
    \rm{SFF} = \frac{F_{0.2-1}}{F_{0.2-10}}
\end{equation}

\noindent where F$_{0.2-1}$ is the unabsorbed, rest-frame flux in the $0.2-1\kev$ band, and F$_{0.2-10}$ is the unabsorbed, rest-frame flux in the $0.2-10\kev$ band. Fluxes are measured using the PL+PL model. Therefore, this flux ratio indicates the fraction of the broad-band flux which is emitted in the soft X-ray band. 

\begin{figure}
   \centering
    \includegraphics[width=85mm]{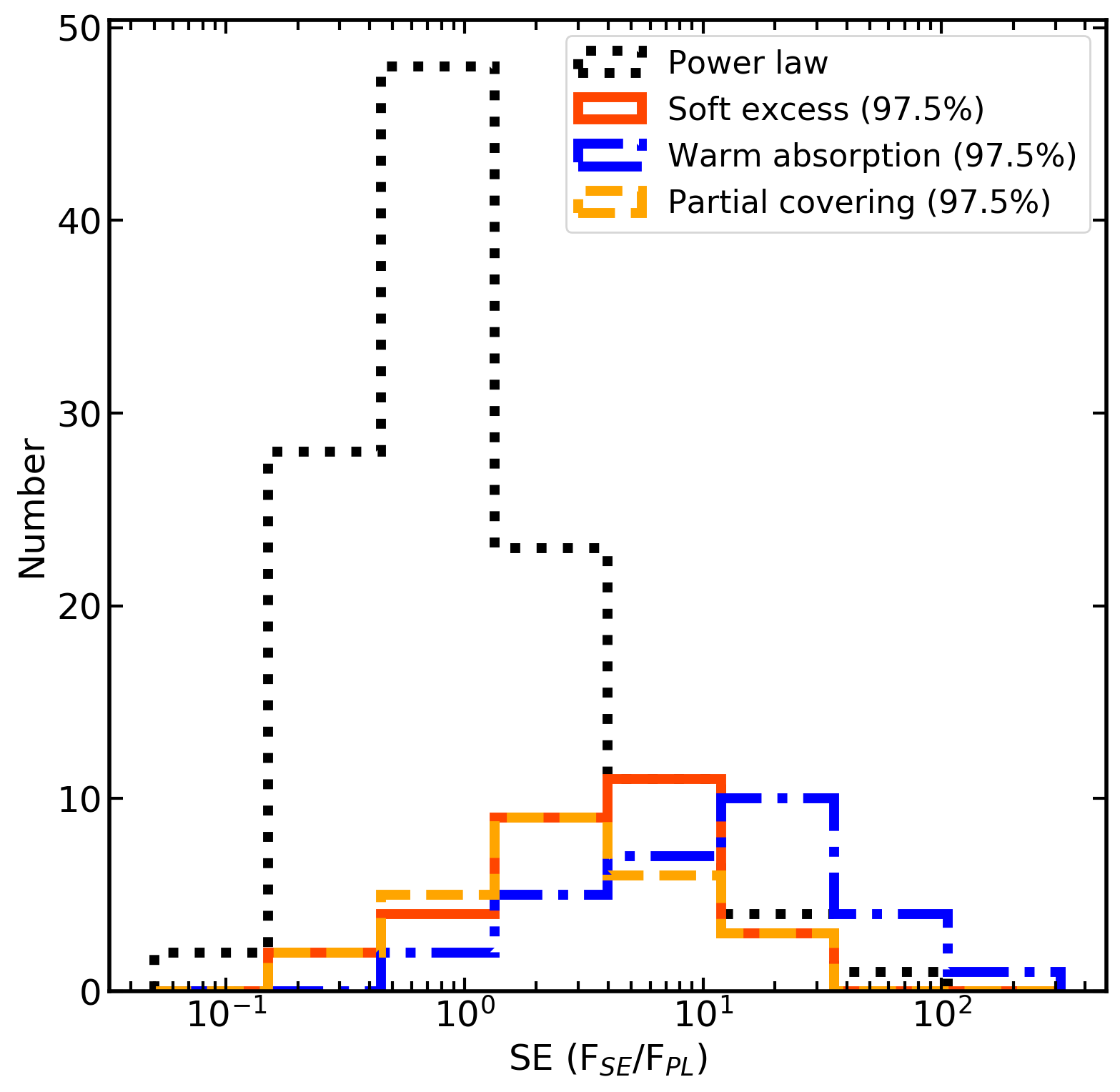}
    \includegraphics[width=85mm]{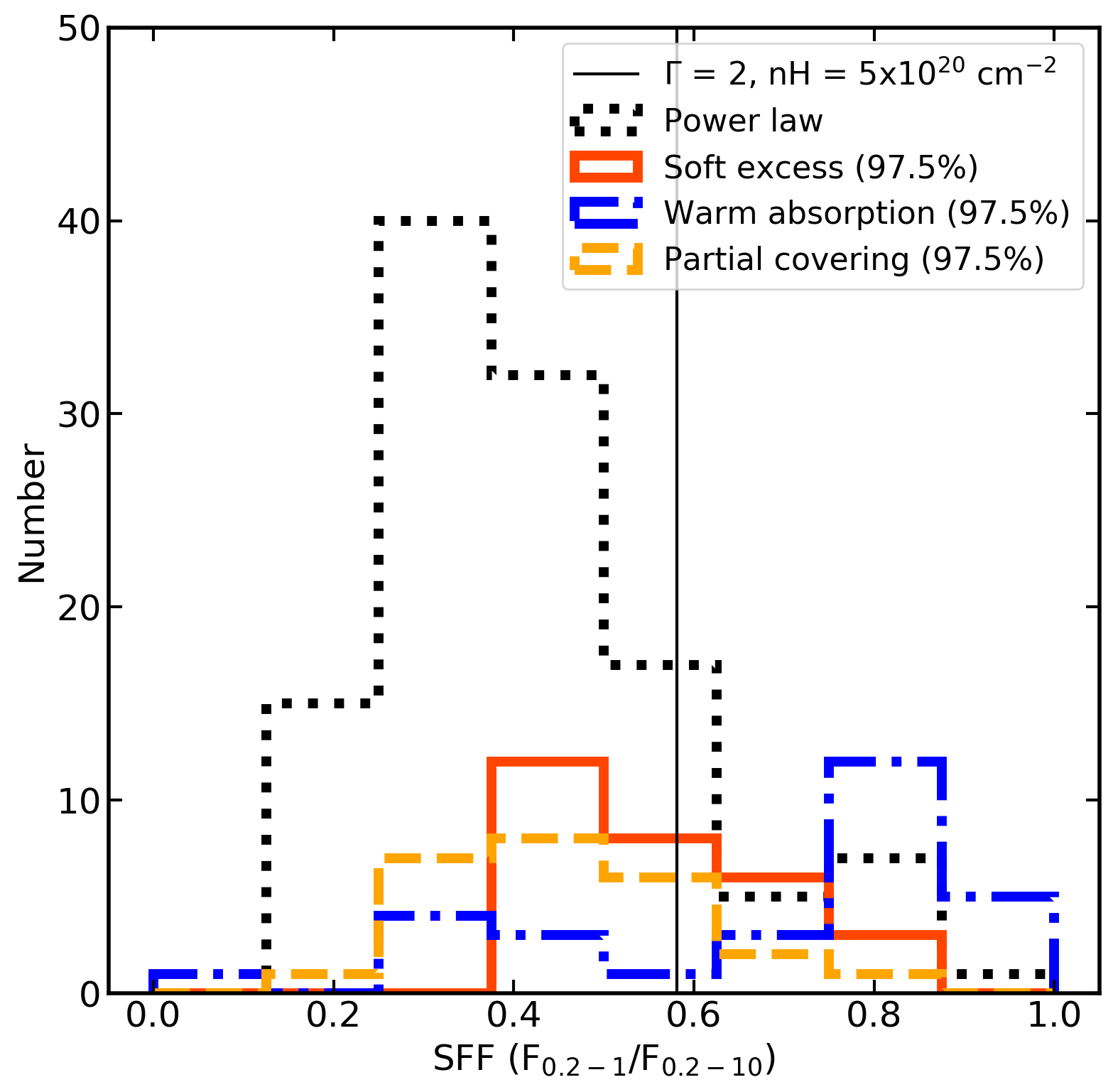}
    \caption{Distributions of soft excess strength and soft flux fraction. Top: Distribution of soft excess strengths ($F_{SE}/F_{PL}$) shown for sources best fit by each model. Sources best fit with only with a single absorbed power law are shown with a black dotted line, sources best fit with a soft excess (PL+PL) are shown as a red solid line, sources best fit with a warm absorber are shown as a blue dash-dot line, and sources best fit with partial covering absorption are shown with an orange dotted line. Bottom: same as top, but shown for the soft flux fraction, $F_{0.2-1}/F_{0.2-10}$. The vertical black line shows the expected SFF for a source with $\Gamma$ = 2 and nH = $5\times10^{22}\pscm$.  }
    \label{fig:softex}

\end{figure}

Histograms for both of these are shown in Fig.~\ref{fig:softex}, with the soft excess strength shown in the top panel and the soft flux fraction in the bottom panel. In both panels, it is apparent that the SE and SFF values for sources with soft excesses are higher on average than those which do not. This seems reasonable, as it would be expected that sources with statistically significant soft excesses would have stronger soft emission and would therefore emit a higher fraction of their total flux in the soft X-ray. Furthermore, it is shown that sources with warm absorbers also appear to have very strong soft excesses and soft flux fractions with this model, likely due to the fact that very steep soft photon indices are preferred for these sources in order to approximate the shape of the absorption features, which highlights the important of using the correct model for characterising the soft excess.

To better quantify these differences, a KS-test is performed comparing the distribution of soft excess sources to those best fit with a power law. For the soft excess strength and for the soft flux fraction, the KS-test returns a significance level of $4\times10^{-6}$ and $6\times10^{-9}$, respectively. This implies that for both qualifications of the soft excess, the null hypothesis that both distributions are drawn from the same parent sample can be rejected with $>99.999\%$ confidence. As would be expected, the soft excess strength and soft flux fractions are significantly higher for sources with soft excesses are higher on average than for those which do not. This may suggest that sources with and without soft excesses are two distinct populations of AGN. This conclusion, however, still cannot confirm the physical origin of the soft excess; to do this, physically motivated models must be fit to each spectrum, and the evidence compared.

\section{Physical interpretation for the soft excess}

\begin{table}
\caption{Physical soft excess model abbreviation, and corresponding {\sc xspec} implementations.  }
\label{tab:pmodels}
\resizebox{\columnwidth}{!}{%
\begin{tabular}{ll}
\hline
Name & {\sc xspec} implementation \\
 \hline
NTH & \texttt{tbabs} $\times$ \texttt{ztbabs} $\times$ (\texttt{nthComp} + \texttt{constant} $\times$ \texttt{nthComp}) \\
REL & \texttt{tbabs} $\times$ \texttt{ztbabs} $\times$ \texttt{relxill} \\
\hline
\hline
\end{tabular}
}
\end{table}

\begin{table*}
\centering
\caption{ Priors used for the physical true soft excess models. Column (1) gives the shortened name of the model. Column (2) shows the lower and upper limit of the host galaxy absorption. Column (3) shows the lower and upper limit of the hard (coronal) power law component. Columns (4) through (7) show the upper and lower limits of the warm Comptonisation parameters, and columns (8) through (10) shows the constraints placed on the blurred reflection parameters. Column (11) gives the constraint placed on the power law normalisation. }
\label{tab:sepriors}
\resizebox{0.95\textwidth}{!}{%
\begin{tabular}{lllllllllll}
\hline
(1) & (2) & (3) & (4) & (5) & (6) & (7) & (8) & (9) & (10) & (11) \\
Model name & log(NH$_z$)   & $\Gamma_h$ & $\Gamma_s$ & log(c) & $kT_e$ & q$_1$ & Inclination & log($\xi$) & Reflection & norm \\
 & [cm$^{-2}$] & & & & [keV] & & (degrees) & & fraction & \\
 \hline
Prior & log-uniform & uniform & uniform & log-uniform & log-uniform & uniform & uniform & log-uniform & log-uniform & log-uniform \\
\hline
NTH   & 20 -- 25  & 1 -- 3  & 2 -- 3.5 & -3 -- 1 & 0.1 -- 1 & - & - & - & - & -10 -- 1 \\
REL   & 20 -- 25  & 1 -- 3  & - & - & - & 3 -- 10 & 10 -- 80 & 0 -- 4 & 0.1 -- 10 & -10 -- 1 \\  
\hline
\hline
\end{tabular}
}
\end{table*}

\subsection{Soft Comptonisation}
One physical interpretation for the soft excess is that it is produced via Comptonisation of disc blackbody photons in a secondary warm corona, which is cooler than the hot corona responsible for the primary hard power law $\Gamma_{h}$. The warm corona is hypothesised to have a higher optical depth than the hot corona \citep[e.g.][]{2012Done,2018Petrucci,2020Petrucci}, but as the temperature is lower, the resulting X-ray emission will be a steeper power law which dominates at low energies. This interpretation has been used successfully to model steep but very smooth soft excesses in type-1 AGN, as the soft Comptonisation will not produce emission or absorption features. To model this in XSPEC, \texttt{nthComp} \citep{1996Zdziarski,1999Zycki} is used, with one \texttt{nthComp} component modelling the optically thin hot corona, and another modelling the optically thick, warm corona. The blackbody seed temperature, $kT_{\rm bb}$, is fixed at $1$ eV and is linked between the two coronae \citep[e.g.][]{2018Petrucci,2020Petrucci}. The X-ray spectral shape is not dependant on this parameter so long as it remains at a reasonable disc temperature of a few eV \citep{2018Petrucci}. 

For the hot corona, the photon index is allowed to vary uniformly between one and three in order to capture the likely parameter space. The electron temperature is frozen at $100\kev$, well outside the eROSITA bandpass and in agreement with the assumptions from other works \citep[e.g.][]{2015Fabian}. For the warm corona, the photon index is allowed to vary uniformly between two and 3.5, and the electron temperature is allowed to vary uniformly between $0.1-1\kev$. These parameter ranges are based on fits and simulations performed by, for example, \citet{2018Petrucci,2020Petrucci}, where it is demonstrated that these parameters are reasonable when assuming an optical depth of $\sim10-20$. Finally, as in the PL+PL model, the normalisations of the two \texttt{nthComp} components are linked, and the flux of the soft corona relative to the hard is set using a cross-normalisation constant. The normalisation component is given a log-uniform prior between -10 and 1, and the cross-normalisation is given a log-uniform prior between -3 and 1. All priors of free parameters are listed in Table~\ref{tab:sepriors}.

\subsection{Blurred reflection}
In a blurred reflection scenario, some of the X-ray photons emitted from the corona are reflected from the innermost regions of an accretion disc, producing a reflection spectrum \citep[e.g.][]{1999Ross,2005Ross,2012Dauser,2013Garcia}. As photons strike the disc, they are absorbed, producing deep absorption features and edges. As the atoms de-excite, they produce emission features mostly. For an ionised disc, the reflected emission is concentrated in the soft X-ray, but also includes a prominent \feka\ emission line and hard reflection continuum. Due to the fast rotation of the disc and gravitational redshift due to the central black hole, the features in the reflection spectrum are relativistically blurred, producing a soft excess \citep[e.g.][]{2006Crummy,2019Jiang}. Understanding the reflection spectrum can reveal many properties of the innermost regions of the AGN, including the height and structure of the corona, the ionisation and abundances in the accretion disc, and changes in these parameters over time. This model has been successfully used to probe the geometry of the corona as well as successfully explain the variability and spectral shape of many type-1 AGN \citep[e.g.][]{2008Zoghbi, 2012Dauser, 2019Gallo, 2019Waddell, 2021Boller}. 

Here, the reflection spectrum and power law are both modelled using the \texttt{relxill} model \citep{2012Dauser,2014Dauser,2013Garcia}. There are many free parameters in this model, including the inner emissivity index $q_1$, which describes the illumination pattern of the corona onto the accretion disc. This parameter is allowed to vary uniformly between three and ten, while the outer emissivity index ($q_2$) is fixed to 3. The inclination, or viewing angle, is allowed to vary uniformly between ten to 80 degrees. This parameter should actually be evenly distributed in cosine space, however, the inclination is typically poorly constrained and difficult to measure correctly, so the uniform prior is acceptable. The black hole is assumed to have maximum spin, in part due to selection effects which make maximum spin AGN brighter and thus easier to detect in flux limited samples \citep[e.g.][]{2016Vasudevan,2018Baronchelli,2019Arcodia}, and also due to the fact that the spin is difficult to constrain without high signal-to-noise data in the $4-10\kev$ band, where the iron K$\alpha$ line can be modelled \citep[e.g.][]{2016Bonson}. The inner radius of the disc is fixed at the innermost stable circular orbit (ISCO; $1.235r_g$ for a maximum spin black hole with $a = 0.998$), and the outer radius is fixed somewhat arbitrarily to $400r_g$, beyond where significant reflection of X-ray photons is possible for moderate coronal heights. The iron abundance in the disc is fixed to solar, and the disc ionisation ($\xi = 4\pi F/n$, where F is the illuminating flux and n is the hydrogen number density of the disc) is allowed to vary between log($\xi$) of zero and four.

Since the coronal power law component is also included in \texttt{relxill}, no separate power-law model is included. The photon index, $\Gamma$, is allowed to vary between one and three, to account for a very broad range of possible indices. The reflection fraction, which describes the fraction of flux from the corona which is reflected from the accretion disc, is allowed to vary uniformly between 0.1 and 10. Here, a reflection fraction (R) of 0.1 would indicate strong beaming (e.g. the corona is outflowing or forms the base of the jet), a reflection fraction of one suggests that half the flux from the corona is reflected off the disc, while the other half is observed directly, and a reflection fraction of 10 is a strong indicator for light bending (e.g. the corona is close to the disc such that the gravitational pull of the black hole bends the path of the light towards the disc). Finally, the normalisation component is given a log-uniform prior between -10 and one so that AGN of an extreme range of fluxes can be modelled. All priors of free parameters are listed in Table~\ref{tab:sepriors}.

There are several different flavours of \texttt{relxill}, all intended to model different physical properties of the innermost regions of the AGN \citep[e.g.][]{2016Dauser,2019Jiang}. Users can choose to assume a lamp-post geometry (\texttt{relxilllp}), a varying disc density (\texttt{relxillD}), among other changes. We also freeze many parameters in our analysis (e.g. the iron abundance, outer emissivity index, and black hole spin, although these have been shown to vary, in some cases dramatically, between AGN \citep{2008Zoghbi, 2009Fabian, 2014Daly, 2019Reynolds}. In particular, many parameters are best constrained using the iron K$\alpha$ line profile, as the iron line is broadened due to the strong relativistic effects in the central region. Given the limited eROSITA sensitivity and high background levels above $\sim5\kev$, this is very difficult for most sources, even those in the hard sample presented in this work. Nevertheless, this simplified treatment of relativistic reflection still has the potential to capture sources which display the typical characteristics of a blurred reflection spectrum.

\subsection{Model selection}

\begin{table}
\centering
\caption{Evidence comparison for the 29 sources in the soft excess sample. The source ID is listed in column (1). Columns (2) and (3) show the Bayesian evidence, where the values are all normalised by subtracting the highest fit value. A negative number in the column therefore indicates the worse fitting model, while a value of zero shows the preferred model. In the final row, the same exercise is performed for the full sample. Column (4) lists the name of the best fitting model. }
\label{tab:phys}
\resizebox{1.0\columnwidth}{!}{%
\begin{tabular}{cccc}
\hline
(1) & (2) & (3) & (4) \\
eROID & ln(Z$_{nth}$) - ln(Z$_{BEST}$) & ln(Z$_{rel}$) - ln(Z$_{BEST}$) & Best model \\
\hline
00001  &  0.0  &  -0.69  &  NTHCOMP \\ 
00004  &  0.0  &  -3.28  &  NTHCOMP \\ 
00007  &  0.0  &  -1.26  &  NTHCOMP \\ 
00011  &  0.0  &  -2.83  &  NTHCOMP \\ 
00029  &  0.0  &  -0.60  &  NTHCOMP \\ 
00034  &  0.0  &  -2.77  &  NTHCOMP \\ 
00035  &  -0.25  &  0.0  &  RELXILL \\ 
00038  &  0.0  &  -0.85  &  NTHCOMP \\ 
00039  &  0.0  &  -2.80  &  NTHCOMP \\
00045  &  0.0  &  -2.30  &  NTHCOMP \\
00054  &  0.0  &  -1.88  &  NTHCOMP \\
00057  &  -0.16  &  0.0  &  RELXILL \\
00076  &  0.0  &  -0.32  &  NTHCOMP \\
00121  &  0.0  &  -0.92  &  NTHCOMP \\
00122  &  -0.15  &  0.0  &  RELXILL \\
00153  &  0.0  &  -0.98  &  NTHCOMP \\
00176  &  -0.05  &  0.0  &  RELXILL \\
00200  &  0.0  &  -0.75  &  NTHCOMP \\
00204  &  -0.39  &  0.0  &  RELXILL \\
00216  &  0.0  &  -2.01  &  NTHCOMP \\
00237  &  0.0  &  -0.74  &  NTHCOMP \\ 
00288  &  0.0  &  -1.15  &  NTHCOMP \\
00340  &  0.0  &  -1.18  &  NTHCOMP \\
00358  &  0.0  &  -1.36  &  NTHCOMP \\ 
00426  &  0.0  &  -1.64  &  NTHCOMP \\
00760  &  -0.26  &  0.0  &  RELXILL \\ 
00784  &  0.0  &  -0.27  &  NTHCOMP \\ 
01136  &  0.0  &  -0.44  &  NTHCOMP \\ 
01736  &  0.0  &  -1.16  &  NTHCOMP \\
\hline
ALL    & 0.0  & -30.93    & NTHCOMP \\
\hline
\hline
\end{tabular}
}
\end{table}

Having fit both of the models described above to the 29 sources in the soft excess sample, the evidence for each model can be computed and compared in order to determine which model is preferred, with Bayes factors computed as given in Table~\ref{tab:bayes}.  The results of this comparison are presented in Table~\ref{tab:phys}. The final column in the table indicates which is the preferred model for each source. Out of the 29 sources, six are better fit with blurred reflection and the remaining 23 are best fit with soft Comptonisation. 

\begin{figure}
   \centering
    \includegraphics[width=0.95\columnwidth]{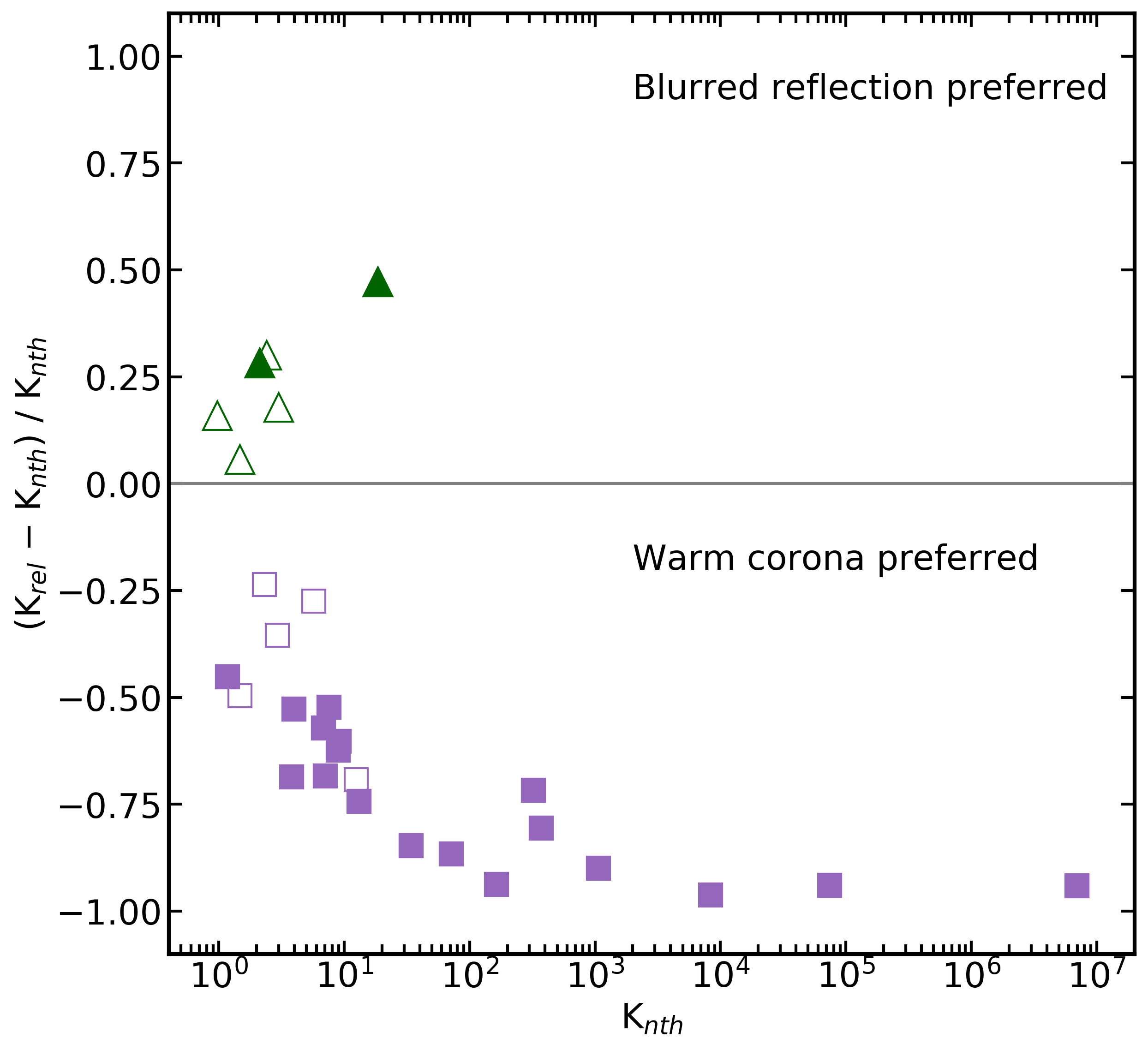}
    \caption{Comparison of Bayes factors for the warm corona and relativistic blurred reflection models. Bayes factors for the warm corona model are shown on the horizontal axis, and the difference between Bayes factors normalised by the warm corona Bayes factor is shown on the vertical axis. Open shapes indicate soft excesses with 97.5\% confidence, and filled shapes indicate 99\% confidence. Sources best fit with a warm corona are shown as purple squares, and sources best fit with blurred reflection are shown as green triangles. }
    \label{fig:krelknth}

\end{figure}

More closely examining the sample, it is apparent that many of the sources which are best fit with blurred reflection have very small differences in Bayes factors between models. This is not the case for sources best fit with soft Comptonisation, where some sources have much larger differences in Bayesian evidence than with blurred reflection. This indicates that while not all sources are fit well with all models, all sources are relatively well fit with soft Comptonisation. This effect can also be seen in Fig.~\ref{fig:krelknth}, which shows the Bayes factor (K$_{nth}$) for the warm corona for each source plotted with the normalised difference in Bayes factor values, (K$_{rel}$ - K$_{nth}$)/K$_{nth}$. Sources which lie above zero on the y-axis (shown with green triangles) are best fit with blurred reflection, and sources which lie below zero (shown in purple squares) are best with with the warm corona model. Many sources best fit with the warm corona model are much better fit with this model. Furthermore, sources which have more statistically significant soft excesses (shown with filled in symbols) are far more likely to be best fit with a warm corona, with only 2/20 preferring a blurred reflection model.

The last line of Table~\ref{tab:phys} shows the evidence comparison for the full sample, following, e.g. \citet{2018Baronchelli}. Unsurprisingly, the best fitting model for the full sample is the soft Comptonisation model. Individual model parameters and their errors for each source are given in Appendix B, and we note that many parameter values are poorly constrained and have large errors for these complex models.

\subsection{Model parameters}
\begin{figure}
   \centering
    \includegraphics[width=0.95\columnwidth]{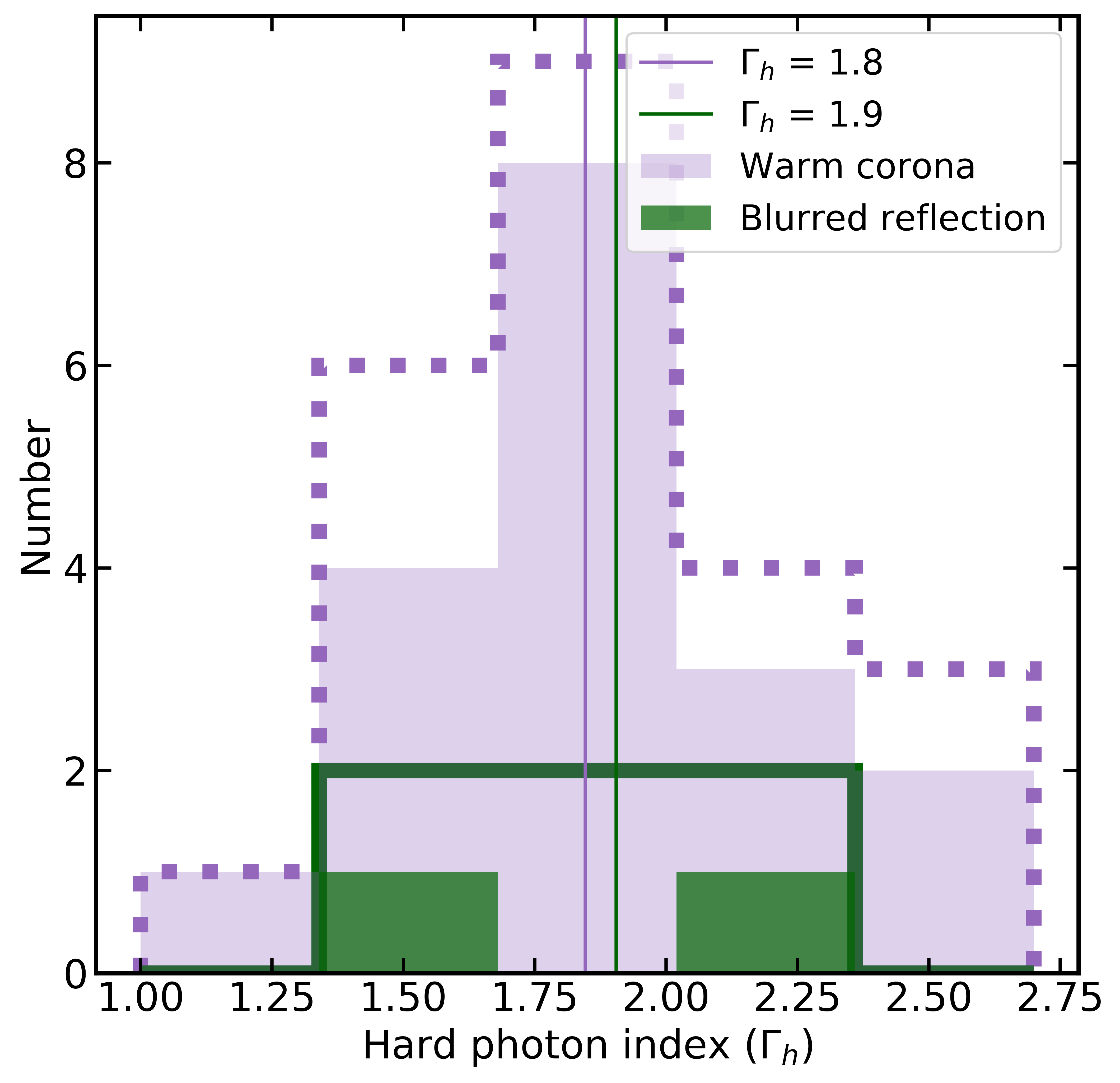}
    \includegraphics[width=0.95\columnwidth]{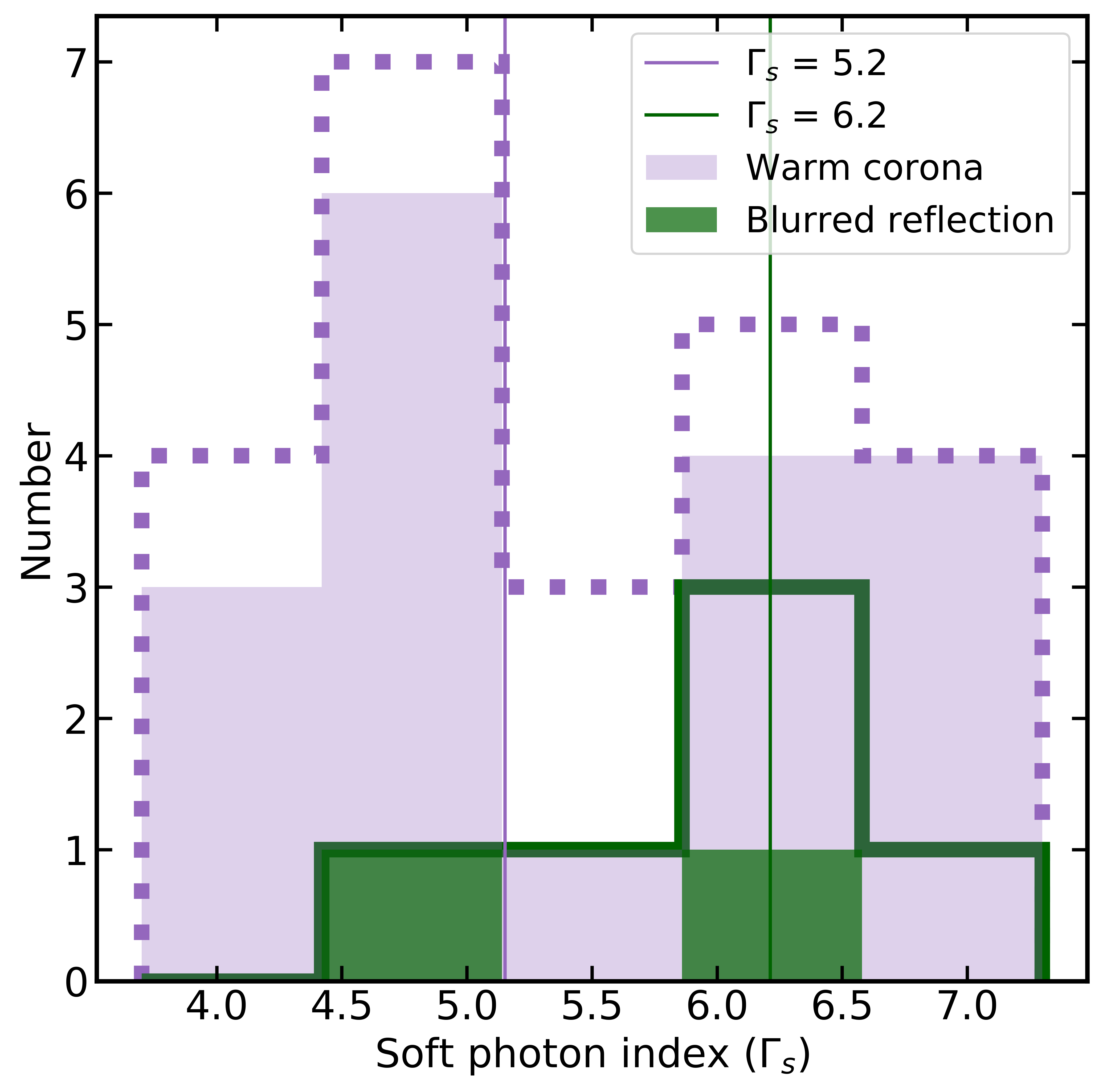}
    \caption{Distributions of soft and hard photon indices separated by best soft excess model. Top: Distributions of hard photon index obtained from the PL+PL modelling for sources in the soft excess sample. Sources which are best fit with soft Comptonisation are shown with a purple dotted line and sources best fit with blurred reflection are shown with a solid dark green line. Median values for each sample are shown with vertical lines in the corresponding colours. The shaded histograms indicate the sources with 99\% significance on the soft excess. Bottom: As top, but showing the soft X-ray photon index. }
    \label{fig:splhpl}

\end{figure}

It is also of interest to compare the properties derived from the phenomenological double power law model of the soft excess sources, shown in Fig.~\ref{fig:splhpl}. Sources best fit with blurred reflection are again shown in dark green, and sources best fit with a warm corona model are shown in purple. Median values for each sub-sample are shown with vertical lines. While the distributions of hard X-ray photon indices are very similar, the distributions of soft photon indices differ, with the median value being much higher for sources best fit with blurred reflection. While this result is not significant when using e.g. the Anderson-Darling test, there are very few sources best fit with blurred reflection so it is difficult to make firm conclusions. Nevertheless, the result is intriguing, as it may present a diagnostic tool to differentiate between a warm corona and a blurred reflection soft excess, and will be discussed further in Sect. 6.2. 

\begin{figure}
   \centering
    \includegraphics[width=0.95\columnwidth]{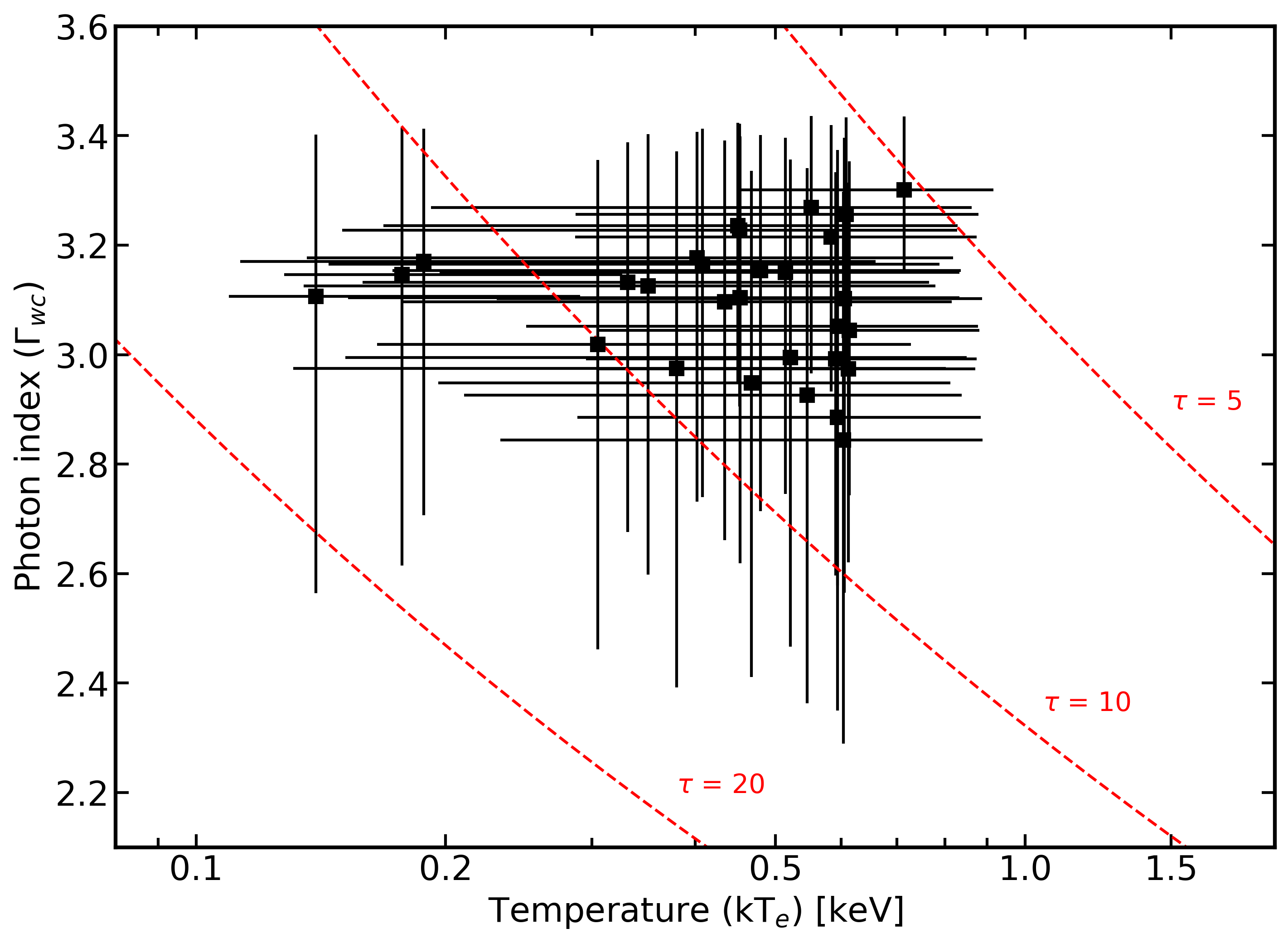}
    \caption{Warm corona photon indices and temperatures derived from the soft Comptonisation modelling. Lines of constant optical depth are shown in red. }
    \label{fig:warmcorona}

\end{figure}

Regarding the parameters of the warm Comptonisation models, the median hot corona photon index is $\Gamma = 1.61$, and the median warm corona photon index is $\Gamma = 3.15$. The median warm corona temperature is kT$ = 0.45\kev$, which is consistent with other studies \citep{2018Petrucci,2020Petrucci}. The distributions of the best-fit parameters for all soft excess sources are shown in Fig.~\ref{fig:warmcorona}, where the warm corona values and errors are indicated with black squares. The red lines show different values of the optical depth, where the warm corona optical depths are between $\tau=5$ and $\tau=20$ and the optical depths for the hot corona are $\simeq1$. While the warm corona photon index is also consistent with previous works, the hot corona has a much flatter spectral index than expected (e.g. $\Gamma \sim 1.8-1.9$ in previous studies, and $\Gamma \sim 2.0$ in typical eROSITA sources). This result is unexpected and is further discussed in Sect. 6.3. 

Moving to the parameters of the blurred reflection modelling, it is apparent that many parameters are not well constrained, likely due to the lower quality of many eFEDS spectra as well as the absence of a high signal-to-noise iron line. Examining the best-fitting parameters for the two blurred reflection sources with soft excesses at >99\% significance (corresponding to the filled green triangles in Fig.~\ref{fig:krelknth}), both have intermediate disc ionisations of $\xi \sim 100$, intermediate inclinations of $\sim 40$ degrees (which are expected for type-1 AGN), and of particular note, high reflection fractions $R >> 1$. In fact, examining all sources best fit with blurred reflection, it is found that all have best fit $R>1$, although not all are constrained to be $>1$. This suggests that the spectral fitting method presented in this work is preferentially identifying sources with very strong reflection components. Indeed, in some cases of sources best fit with blurred reflection, it seems there is more excess emission around $0.7-0.9\kev$, which may correspond to iron-L which is present in the reflection spectrum but not the Comptonisation spectrum, which may explain why the blurred reflection model is preferred. Spectral modelling with brighter sources with more counts in the $4-7\kev$ band, or including data above $8\kev$, would help to identify the iron line and Compton hump if present and thus better constrain blurred reflection parameters.

\renewcommand{\arraystretch}{1.3}
\begin{table}
\caption{Summary of blurred reflection parameters for the two sources with highly significant soft excess ($>99\%$) which are best fit with blurred reflection. Column (1) gives the eROSITA ID, column (2) gives the emissivity index, column (3) gives the disc inclination, column (4) gives the photon index, column (5) gives the disc ionisation and column (6) gives the reflection fraction. }
\label{tab:ref}
\resizebox{\columnwidth}{!}{%
\begin{tabular}{lccccc}
\hline
(1) & (2) & (3) & (4) & (5) & (6) \\
eROID & q$_{1}$ & Inclination & $\Gamma$ & log($\xi$) & log($R$) \\
 \hline
00035  &  7.1 $_{- 2.3 } ^{+ 2.0 }$ &  36 $_{- 18 } ^{+ 21 }$ & 2.45 $_{- 0.11 } ^{+ 0.10 }$ & 1.8 $_{- 1.2 } ^{+ 1.4 }$ & 0.4 $_{- 0.8 } ^{+ 0.5 }$ \\
00204  &  4.3 $_{- 1.0 } ^{+ 2.6 }$ &  43 $_{- 21 } ^{+ 19 }$ & 1.86 $_{- 0.25 } ^{+ 0.22 }$ & 2.0 $_{- 0.9 } ^{+ 0.8 }$ & 0.8 $_{- 0.3 } ^{+ 0.1 }$ \\
\hline
\hline
\end{tabular}
}
\end{table}

\begin{figure}
   \centering
    \includegraphics[width=0.99\columnwidth]{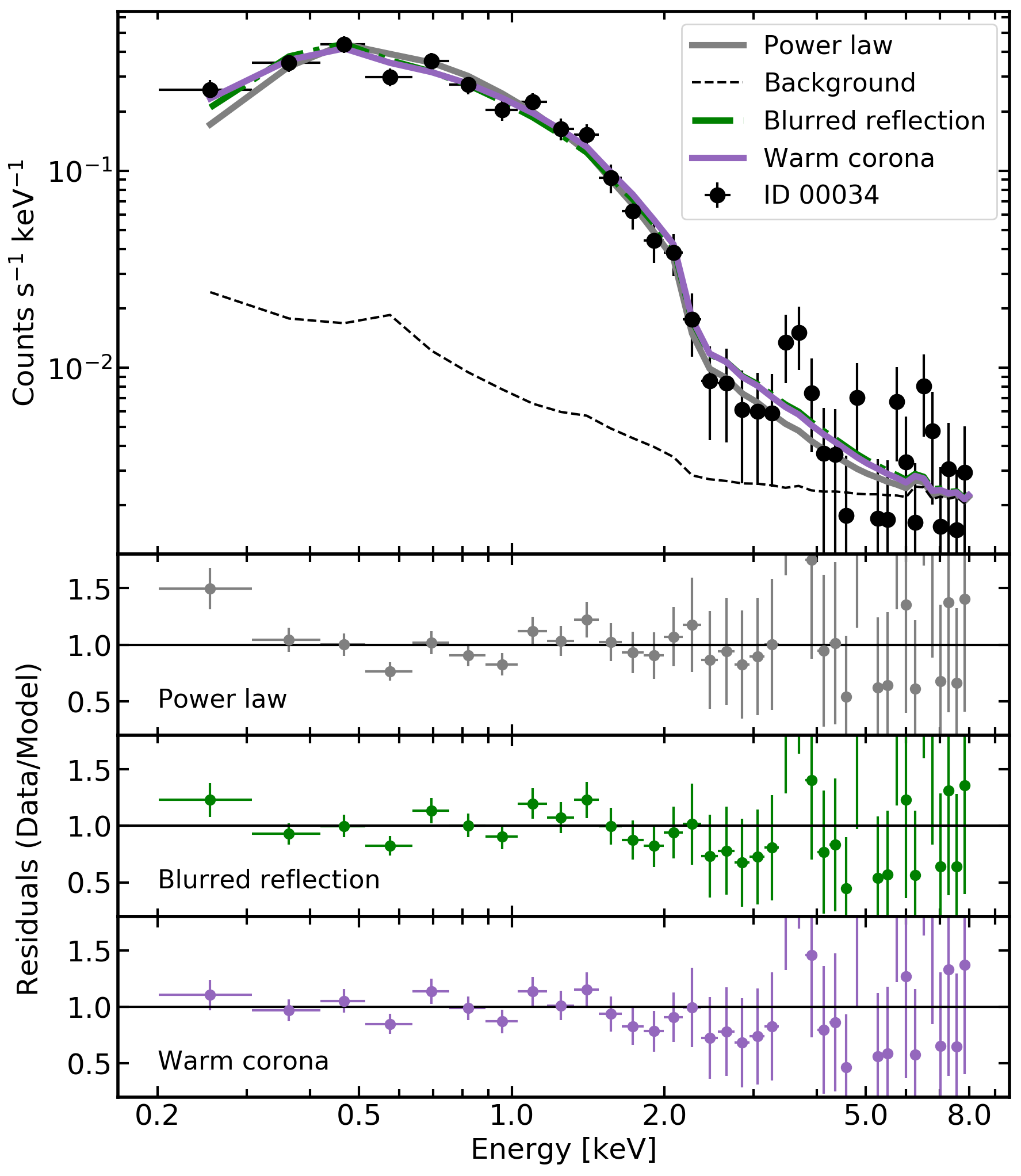}
    \caption{ID 00034 (Z=0.1027), a source best fit with a warm corona model. The spectrum (re-binned for display) is shown in black, the background model is shown as a black dashed line, the power law model is shown as a grey dotted line, a blurred reflection model is shown in green dash-dot line, and the best fitting warm corona model is shown in purple. The bottom three panels show the residuals for the power law, blurred reflection, and a warm corona, respectively. Data have been re-binned for display purposes. }
    \label{fig:nthrelspec}

\end{figure}

To visually demonstrate differences between the spectral models, Fig.~\ref{fig:nthrelspec} shows an example source, ID 00034, which is best fit with a warm corona model. The data are shown in black along with the background model in a black dashed line, an absorbed power law in a grey dotted line, the blurred reflection model in a green dash-dot line, and the best fit warm corona model in purple. From the spectrum, it can be seen that the blurred reflection model under-estimates the flux in the $0.2-0.3\kev$ energy band, and over-estimates the flux around $0.5\kev$ where the blurred reflection model features strong iron emission. Both models however clearly provide a much better fit than the absorbed power law model, which fails to reproduce the spectral shape at most energies. 

\section{Spectral properties}

\subsection{Luminosity-redshift plane}
\begin{figure}
   \centering
    \includegraphics[width=0.95\columnwidth]{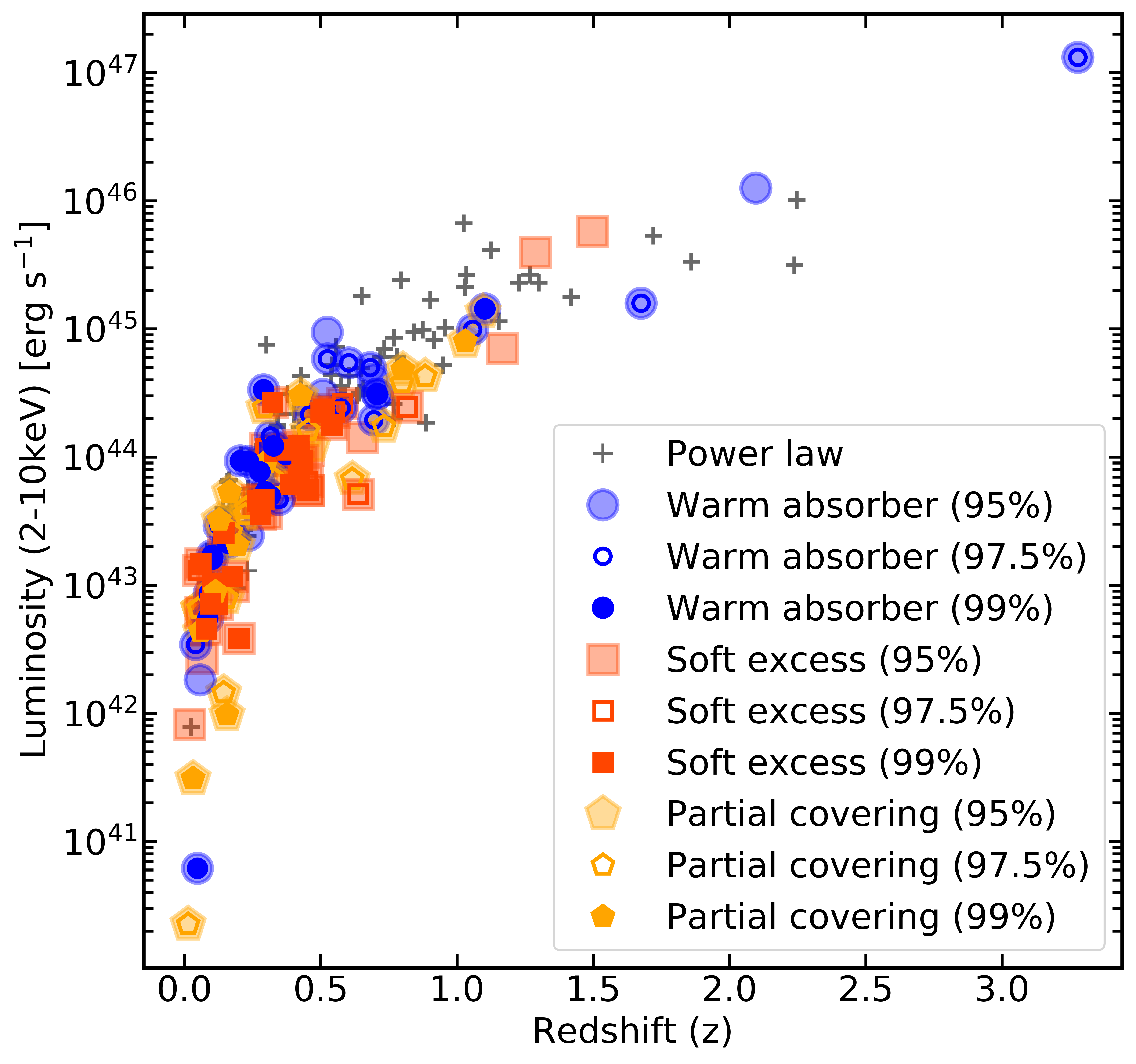}
    \caption{Distributions of redshifts and $2-10\kev$ un-absorbed X-ray luminosity for each source. Sources best fit with an absorbed power law are shown as grey crosses, sources with soft excesses are shown as red squares, and sources with warm absorbers are shown with blue circles.  }
    \label{fig:lz}

\end{figure}

To study the distribution of soft excesses and warm absorbers in a parameter space which can easily be compared to other surveys, sources are plotted in the L-z plane in Fig.~\ref{fig:lz}, where most redshifts are spectroscopic. The $2-10\kev$ luminosity is estimated from the baseline absorbed power law model. Sources best fit with an absorbed power law are shown as grey crosses, sources best fit with warm absorbers are shown as blue circles, sources best fit with partial covering are shown as orange pentagons, and sources best fit with soft excesses are shown as red squares. In this way, we differentiate from the luminosity redshift plane already presented in Nandra et al. (in prep). Different opacities and fill-styles indicate different purity thresholds for the best fit model, as previously defined. The rest-frame, absorption corrected X-ray luminosities of sources with soft excesses and warm absorbers follow those of sources best fit with an absorbed power law. Most sources with soft excesses can be detected up to about $z = 0.5-0.6$, above which the soft excess is presumably shifted out of the observed band, while complex absorption can be detected up to about $z = 1$, with a few sources at higher redshifts. Interestingly, the highest redshift source in the sample, with a spectroscopic redshift of $z = 3.277$ (see Nandra \et in prep.), shows complex absorption best fit by a warm absorber. However, because this source has very few counts below $\sim1\kev$, it is hard to determine definitively the nature of the soft X-ray complexity without data with a higher signal-to-noise ratio. 

\subsection{Characterising the soft excess and complex absorption}
\begin{figure}
   \centering
    \includegraphics[width=0.95\columnwidth]{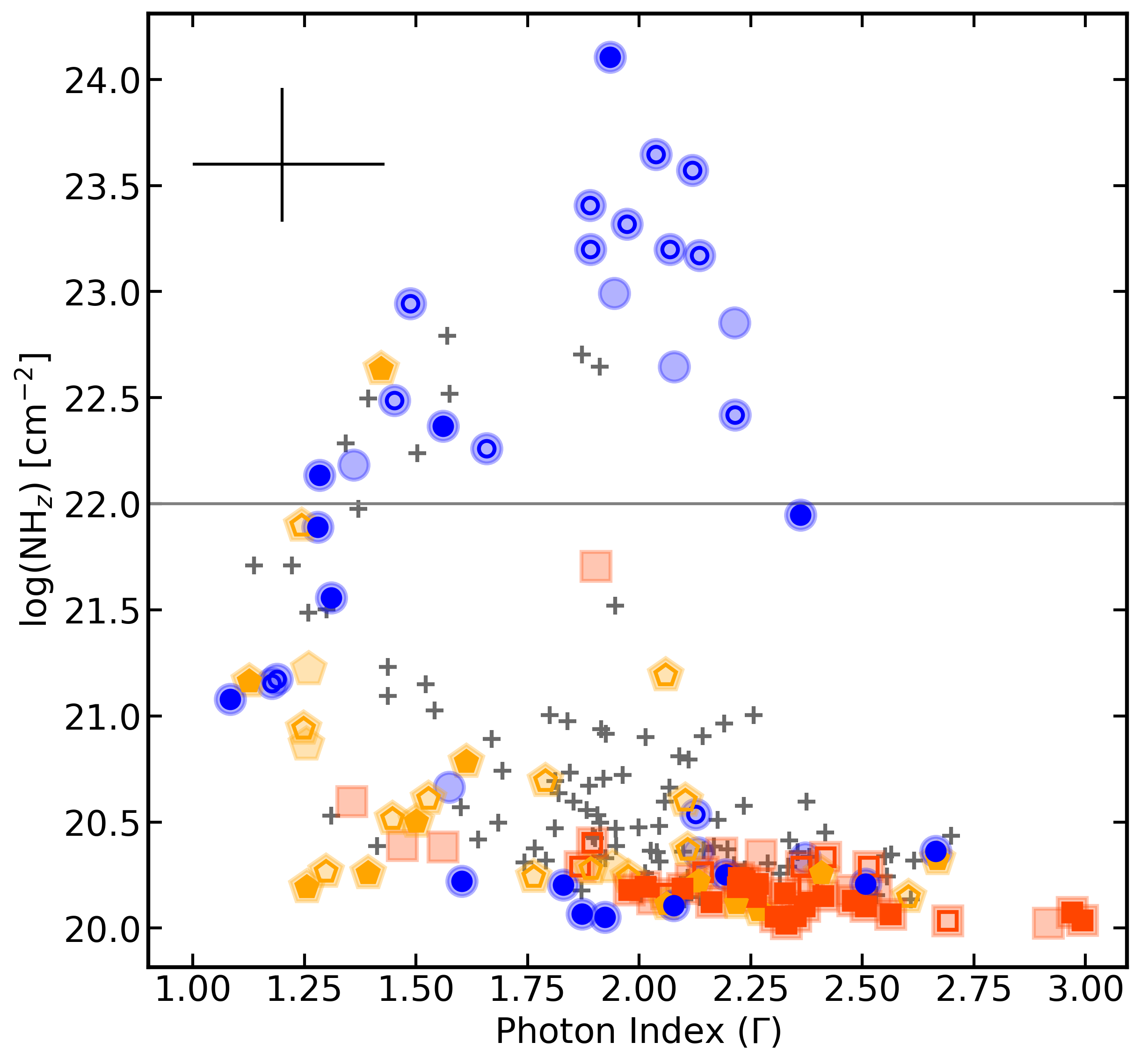}
    \caption{Distributions of photon indices and the host-galaxy absorption column densities, These values are always measured using the baseline PL model, irrespective of the true best-fit model for each source. Sources with soft excesses are shown as red squares, sources with warm absorbers are shown with blue circles, sources best fit with partial covering are shown with orange pentagons, and sources best fit with the baseline power law model are shown as black crosses. Marker styles represent samples of different purities, as described in Sect. 3. The typical error bars are shown in the top right corner, and the horizontal grey line indicates a column density of $10^{22}\pscm$.  }
    \label{fig:sespace}

\end{figure}

Having completed the modelling for all 200 sources in the hard X-ray--selected sample of AGN in eFEDS, the soft excess, warm absorption and partial covering sub-samples can be examined in more detail to search for distinctive characteristics. When considering the full sample of hard X-ray--selected AGN, only $\sim15$\% of sources show strong statistical evidence for a soft excess. However, this fraction can increase significantly when only considering a small parameter space. Figure~\ref{fig:sespace} shows the distribution of photon index and host-galaxy column density. These values are obtained from the \texttt{ztbabs} component of the baseline PL model (and thus not including any additional spectral components), regardless as to the best-fit model for each source. In this way, the properties of sources can be compared for a naive approach wherein it is assumed that all spectra can only be fit with a power law.

Here, the sources with soft excesses are heavily clustered at large photon indices and small host galaxy column densities. This is sensible, as the single power law attempts to explain both the high energy component and the steep soft excess with a single power law, increasing the slope. Selecting only sources with photon indices larger than two and host galaxy column densities less than $2\times10^{20}\pscm$, 42\% of the sources have soft excesses. These are not intrinsic properties of these sources; indeed, it is found that when the soft excess is modelled correctly, the measured mean photon index decreases by 0.35, from a mean of 2.15 when using only one power law to a mean of 1.8 for the hard photon index when the second power law is added. These values are similar to the mean value of the sample of sources best fit with only a power law.

Examining now the parameter space region populated by the complex absorption sources, many of the sources with extremely low photon indices ($\Gamma < 1.4$) and a range of column densities are best fit with warm absorption or partial covering, which likely explains why these photon indices appear so flat compared to the more typical values of $\Gamma \sim 1.8$. When the correct absorption model is applied, the photon index increases to more reasonable values for many of these sources. However, there also appears to be a large cluster of sources with column densities of $>10^{22}$ and photon indices around $\Gamma \sim 2$ (see upper middle of Fig.~\ref{fig:sespace}). These sources may be of particular interest, as they suggest the presence of Compton-thin AGN in eFEDS, which may show absorption and scattered emission from the torus.

Almost all sources with column densities $>10^{22}\pscm$ (18/26), and all eight sources with column densities $>10^{23}\pscm$ have evidence for a warm absorber. This raises the very interesting possibility that many of the AGN that might have been classified as Compton-thin obscured AGN are actually better described with a warm absorber model, and suggests that eROSITA is more likely to probe these complex absorption sources as opposed to AGN obscured by neutral, distant gas. Such column densities are likely too large to be associated with absorption on host-galaxy scales, and must instead originate in the torus. However, these large column densities would not be expected in type-1 AGN, which are typically unobscured. Searching for obscured ($>10^{22}\pscm$) sources which also have SDSS spectra, 17 sources were observed. Of these, six are type-1 AGN with constrained black hole masses and accretion rates (with broad H$\beta$, Mg II or CIV lines), and four of these six have evidence for a warm absorber. The other optical spectra do not have sufficient data quality to confirm whether they are type-1 or type-2 AGN (see Sects. 5.3 and 6.4 for more details).

\begin{figure}
   \centering
    \includegraphics[width=0.95\columnwidth]{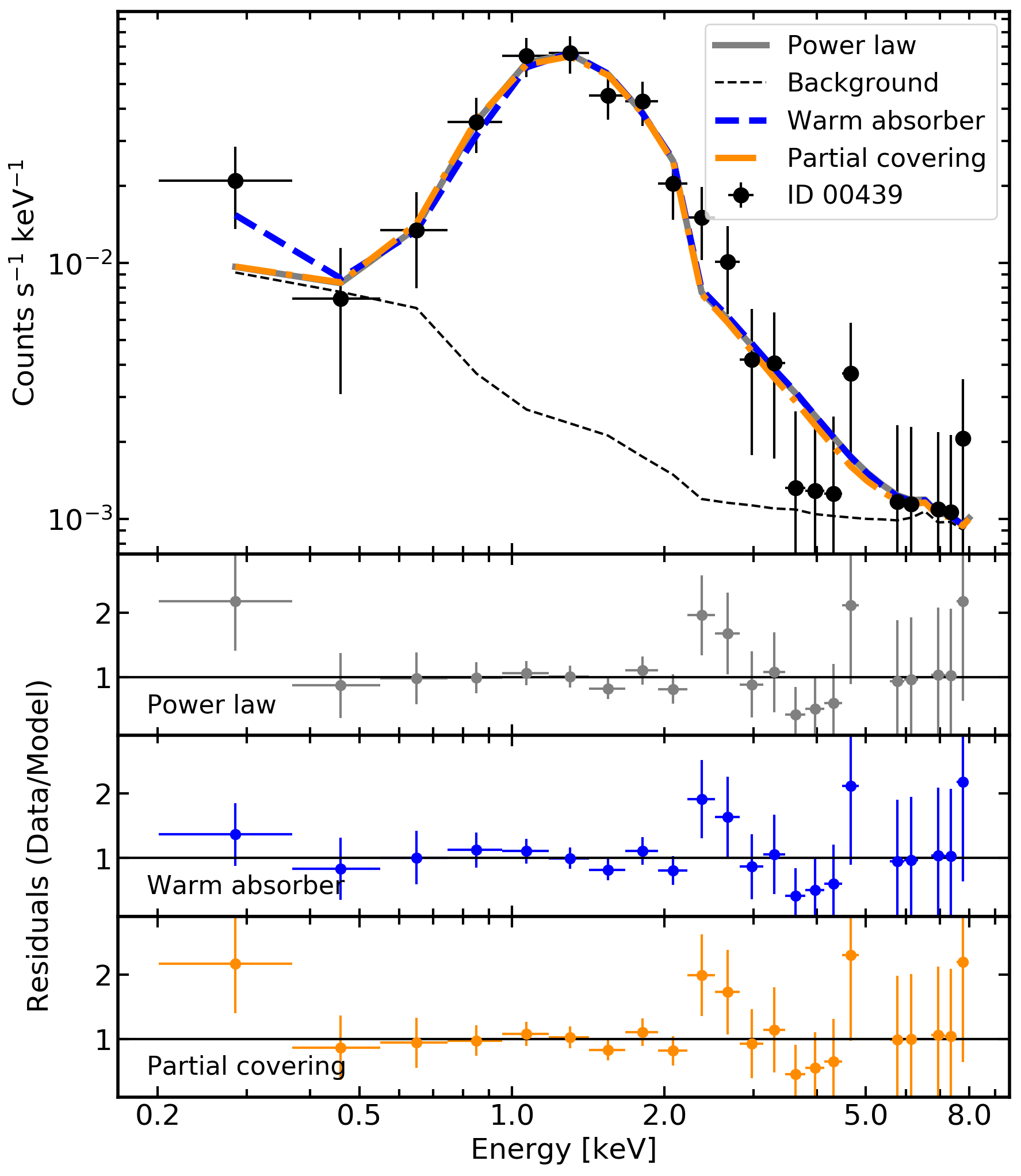}
    \caption{Example spectrum of ID 00439 (z = 0.6027), which is best fit fit a warm absorber model. The spectrum is shown along with the background model (black dashed line), the absorbed power law model (grey), a partial covering absorption model (orange), and the best-fit warm absorber model (blue). The bottom three panels show the residuals for the power law, warm absorber and partial covering models, respectively. Data have been re-binned for display purposes.  }
    \label{fig:warmex}

\end{figure}

\begin{figure}
   \centering
    \includegraphics[width=0.95\columnwidth]{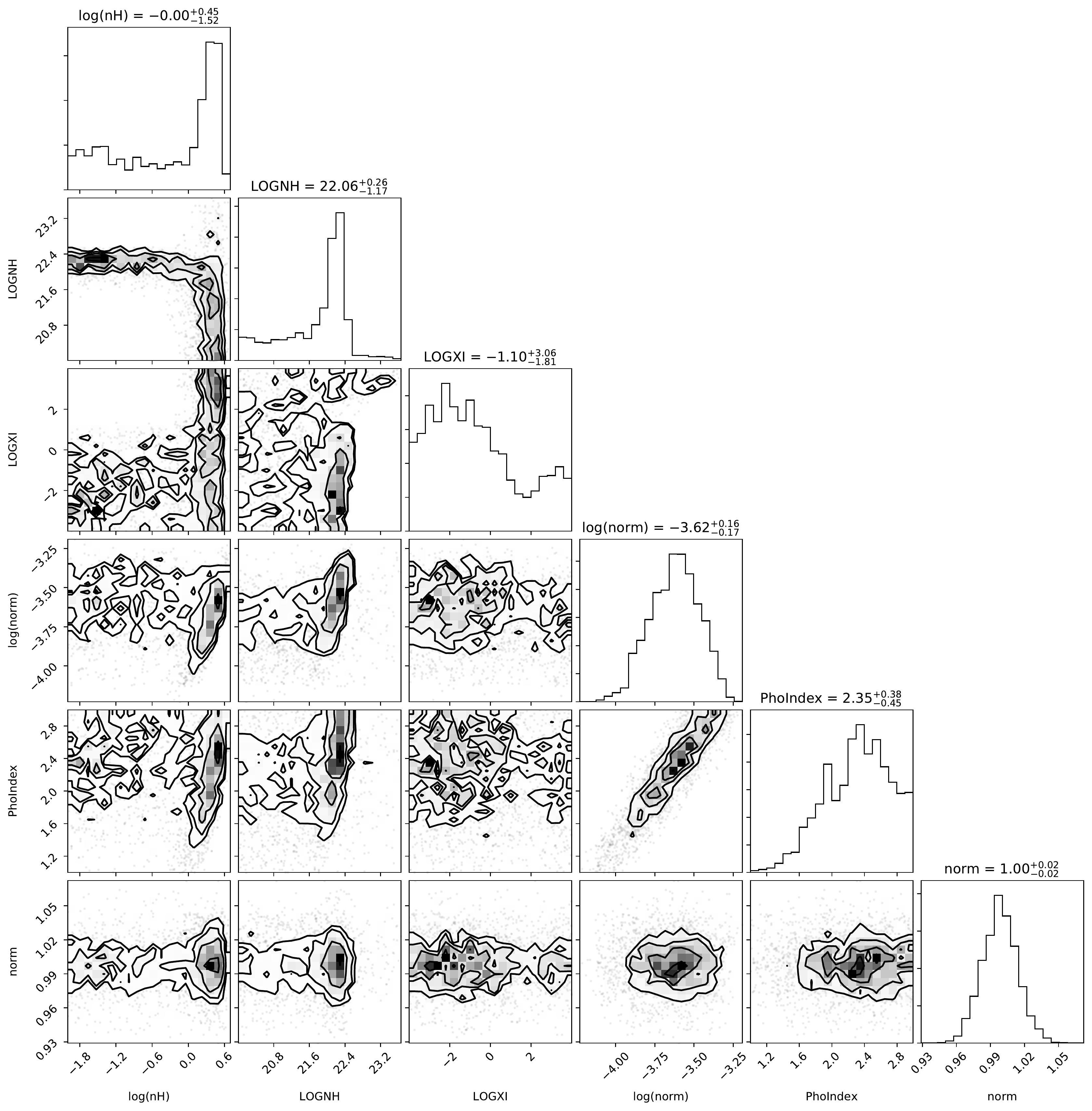}
    \caption{Corner plot for source ID 00439, best fit with a warm absorption model but which shows evidence for Compton-thin absorption when modelled with an absorber power law. The diagonal panels show the marginal posterior probability distribution for each parameter, while the other panels show the conditional probability distribution functions for each pair of parameters. Here, log(nH) is the host galaxy absorber column density (in units of $\times10^{22}\pscm$), LOGNH is the column density of the warm absorber ($\pscm$), LOGXI is the ionisation of the warm absorber (\xii), log(norm) is the power law normalisation, PhoIndex is the photon index of the power law, and norm is the relative renormalisation of the background model with respect to the source model, which is in agreement with 1. }
    \label{fig:abscorner}

\end{figure}

Examining each of these sources individually, many show deep absorption edges in the $\sim0.5-1\kev$ band. When modelled using a single absorbed power law, the absorption edge is well fit with the single absorber, but the emission is significantly under-fit at low energies (e.g. $\sim0.2-0.5\kev$). An example of this is shown in Fig.~\ref{fig:warmex}, which shows a spectrum \citep[ID 00439, also presented in][]{2022Brusa} folded with the background model, an absorbed power law model, a partial covering absorber, and a warm absorber. Without a warm absorber, the flux is significantly under-estimated in the softest X-ray energies (below $0.4-0.5\kev$). When the warm absorber is added, the host galaxy column density modelled using \texttt{ztbabs} is consistent with $10^{20}\pscm$, and the absorption edge better describes the emission below $\sim1\kev$. The partial covering absorber also fails to produce the correct spectral shape. Furthermore, when the warm absorber is added to these sources, the host-galaxy column densities are all consistent with $10^{20}\pscm$, which is consistent with values found in the Milky Way and is consistent with other host galaxies. This can also be seen in the corner plot for the warm absorption model of source ID 00439, shown in Fig.~\ref{fig:abscorner}. Interestingly, the optical spectrum for source was found to show evidence for an ionised outflow \citet{2022Brusa}, further supporting the warm absorber X-ray model.  This source has relatively fewer X-ray counts as it is heavily absorbed in the soft band, and key parameters are less well constrained. There is some degeneracy between the power law index and normalisation, and the ionisation appears to be low, but the marginal posterior probability distribution has a small secondary peak at a higher ionisation. Most strikingly, the column densities of the host galaxy and warm absorber cannot be independently constrained, though the host galaxy absorption is consistent with $10^{20}\pscm$, and the warm absorber component significantly improves the fit. The constraining of the absorbing column densities will be discussed in further detail in Sect. 6.4.

\subsection{Relationship with optically derived properties}
A total of 172 of the AGN in our eFEDS hard X-ray--selected sample have optical spectra available from SDSS I-V \citep{2006Gunnsdss, 2013Smeeboss, 2016Dawsoneboss, 2017Blantonsdssiv, 2020sdssiv,1973Bowensdss, 2006Gunnsdss, 2013Smeeboss, 2017sdssv, 2019Wilsonsdss}. By fitting these spectra, and using certain assumptions, the optical luminosity obtained from spectral fitting can be used to estimate the bolometric luminosity, L$_{\rm bol}$ \citep[see e.g.][]{2011Shen}, and the width of the broad optical lines can be used to estimate the black hole mass. The spectral fitting procedure applied to AGN in the hard X-ray--selected sample of eFEDS AGN is described in more detail in Nandra \et (in prep.). In short, the optical spectral fitting programme PyQSOFit \citep{2018Guo} is used to measure the continuum luminosity and optical/UV broad lines (H$\beta$, Mg II or C IV) width, such that L$_{\rm bol}$ and M$_{\rm BH}$ can be estimated.

\begin{figure}
   \centering
    \includegraphics[width=0.95\columnwidth]{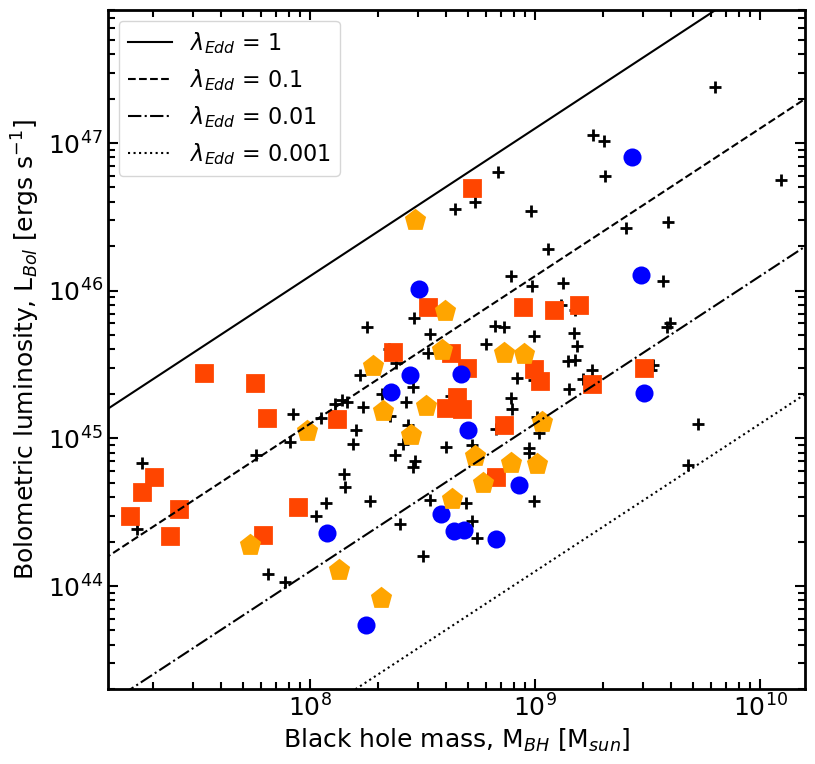}
    \caption{Distribution of black hole masses and bolometric luminosities calculated from the optical spectra. Sources best fit with an absorbed power law are shown as grey crosses, sources with soft excesses (97.5\%) are shown as red squares, sources with warm absorbers (97.5\%) are shown with blue circles, and sources with partial covering (97.5\%) absorbers are shown with orange pentagons. }
    \label{fig:mbhlbol}

\end{figure}

Using the black hole masses and bolometric luminosities (derived from the optical luminosity), the Eddington luminosity (L$_{\rm Edd}$) can be defined as

\begin{equation}
    L_{\rm Edd} = 1.26 \times 10^{38} \bigg( \frac{M_{\rm BH}}{M_{\rm sun}} \bigg) ~~ \rm{[erg s^{-1}]},
\end{equation}

\noindent and the Eddington ratio, L$_{\rm Edd}$, can then be defined as 

\begin{equation}
    \lambda_{\rm Edd} = \frac{L_{\rm bol}}{L_{\rm Edd}}.
\end{equation}

We then apply the selection criteria presented in \cite{2022Wu} to select a clean sample of line measurements, namely that the flux of the line divided by the error on the flux must be > 2, which excludes three sources from the sample. The analysis is restricted to sources where the accretion rate can be constrained, that is, where the error on the accretion rate
does not exceed the value of the accretion rate, which removes an additional 13 sources. Using this approach a total of 154 AGN have constrained accretion rates, measured from the H$\beta$, Mg II or C IV optical lines. 

Fig.~\ref{fig:mbhlbol} shows the distribution of black hole masses and bolometric luminosities for the sources included in the sample. There are several sources with relatively low black hole masses which appear to host highly significant soft excesses. Furthermore, there is a large cluster of sources with black hole masses of the order of $5 \times 10^8 M_{\rm sun}$. At these masses, there appears to be some evidence that the sources with soft excesses have higher bolometric luminosities than those with partial covering absorbers or warm absorbers.

\begin{figure*}
   \centering
    \includegraphics[width=0.97\columnwidth]{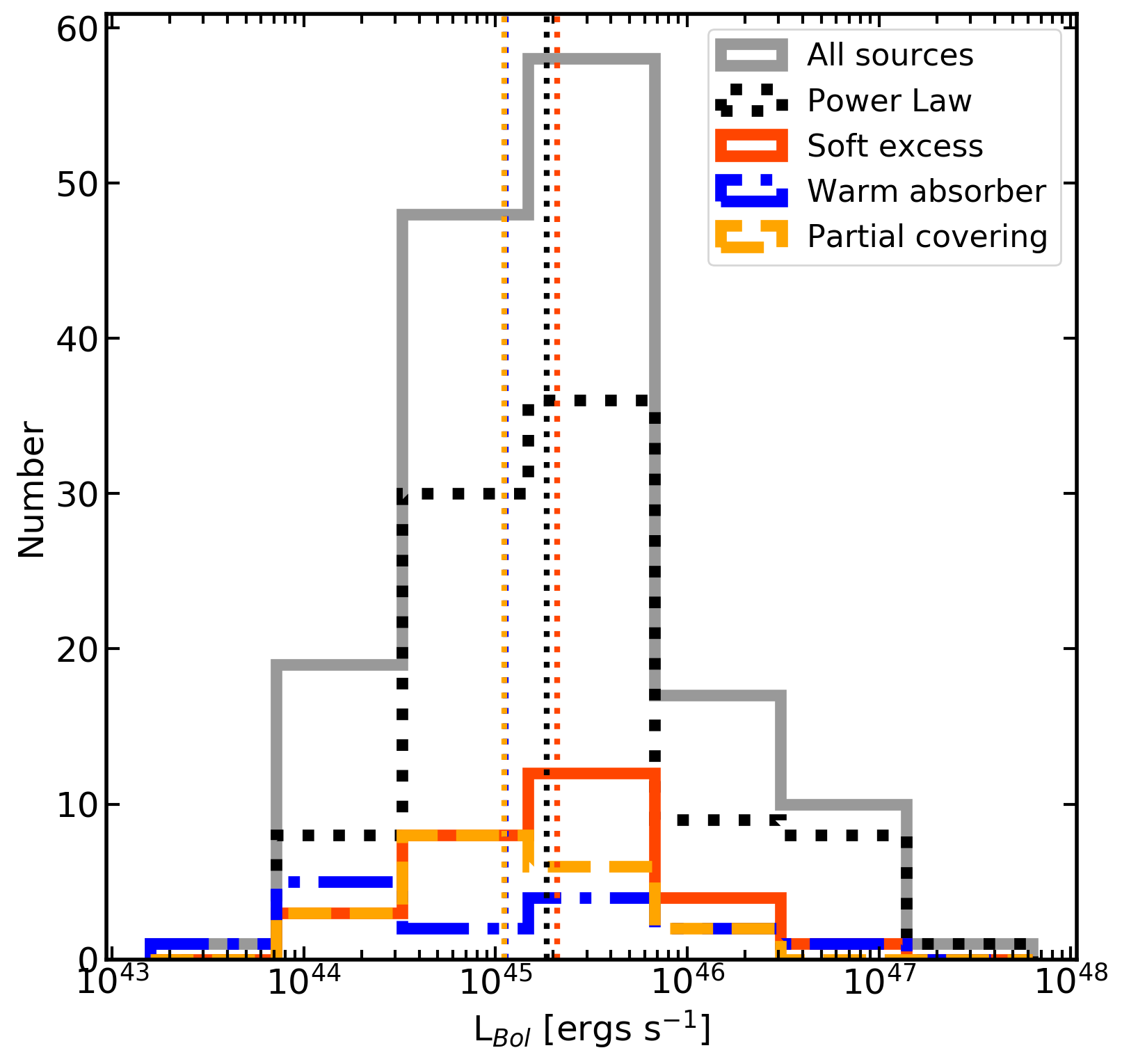}
    \includegraphics[width=0.95\columnwidth]{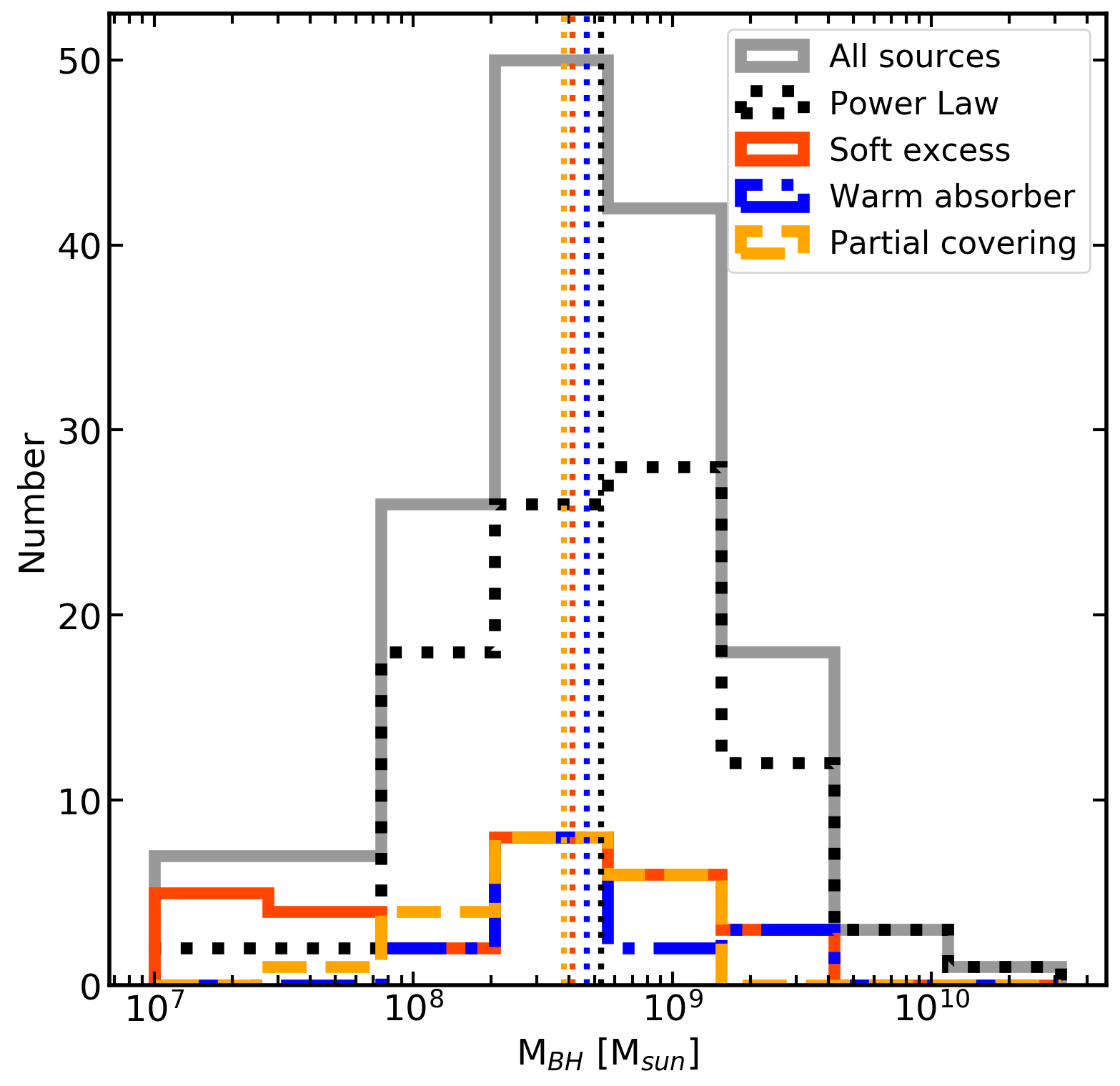}
    \includegraphics[width=0.95\columnwidth]{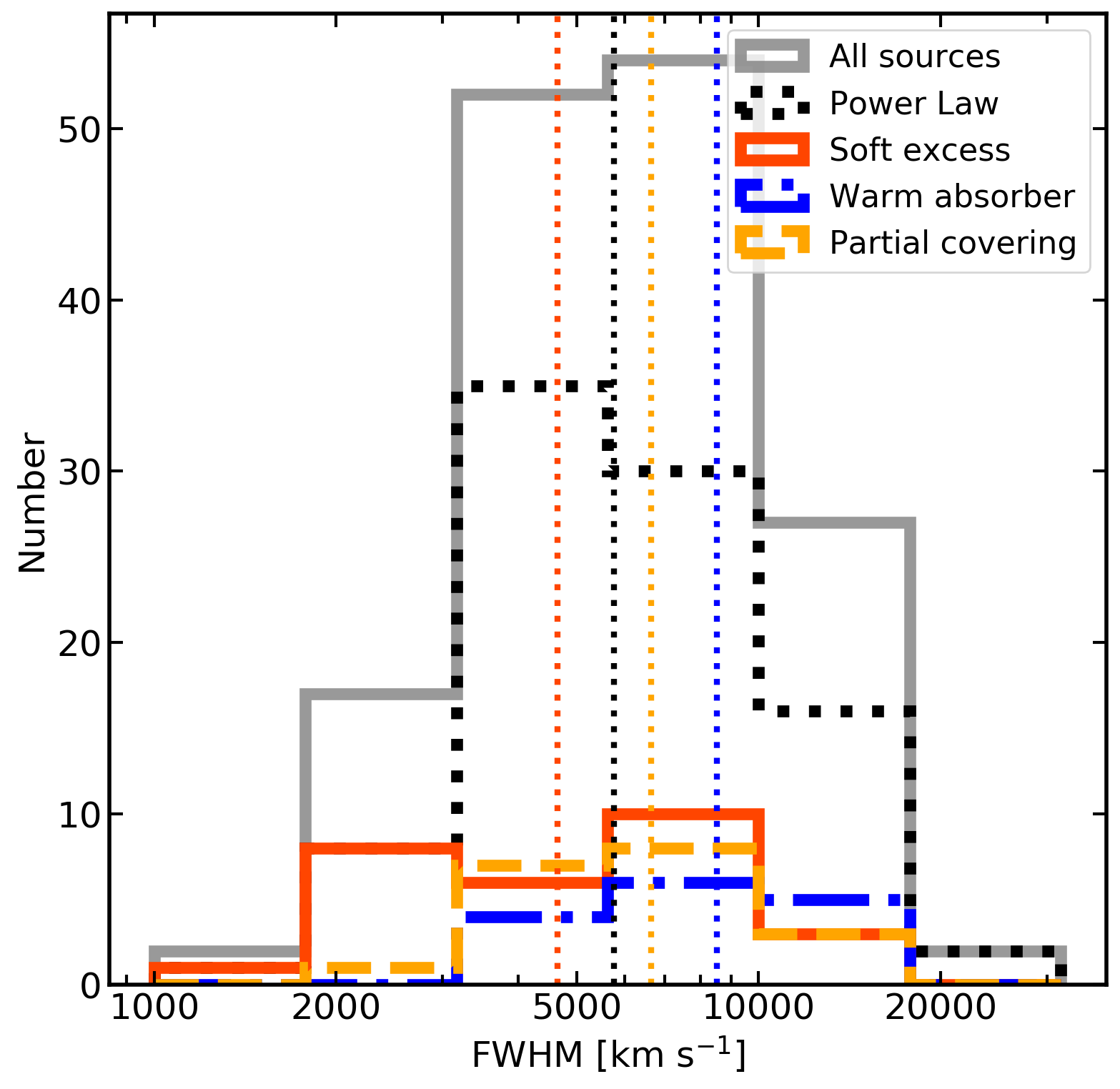}
    \includegraphics[width=0.95\columnwidth]{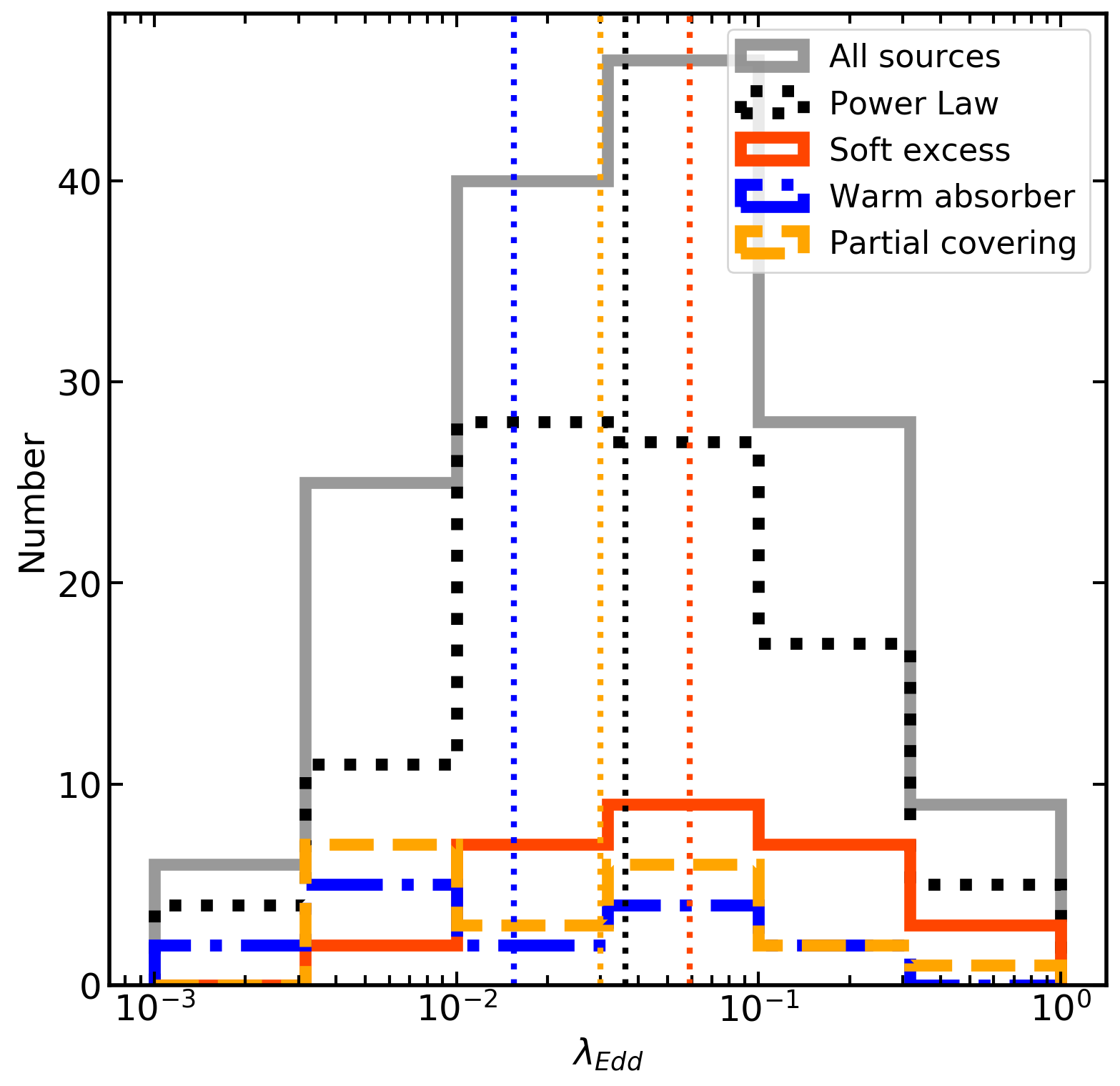}
    \caption{Histograms of key optical-derived properties, shown as: all sources (grey), sources best fit with a power law (black dotted line), sources best fit with a soft excess (red solid line), sources best fit with a warm absorber (blue dot-dash line) and sources best with with partial covering (orange dashed line). Median values for each sub-sample are indicated with vertical lines in corresponding colours. Top left: distributions of bolometric luminosities. Top right: distributions of black hole masses. Bottom left: distributions of FWHM of the H$\beta$, Mg II, or C IV optical lines. Bottom right: distributions of the Eddington ratios ($\lambda_{\rm Edd}$).  }
    \label{fig:opthist}

\end{figure*}

In Nandra \et (in prep.), the distributions of redshift, black hole mass, bolometric luminosity and Eddington ratio are discussed in detail. Here, these distributions are re-examined, but separating the sources based on the best fitting model from this work. These distributions are shown in Fig.~\ref{fig:opthist}, with median values for each sub-sample indicated with vertical dashed lines in the corresponding colours. No significant differences were found in the bolometric luminosities (top left) or black hole masses (top right) between models. However, the distributions of FWHM of the optical broad lines (bottom left), as well as the distributions of Eddington ratios (bottom right), differ significantly. To quantify this, we use the Anderson-Darling (AD test), which is more sensitive to small changes in the wings of the distributions than the KS-test, and is ideal for treatment of smaller samples. Using this, it is found that the measured FWHM values for sources with warm absorbers are significantly higher (AD test p-value 0.012) than those with soft excesses. In general, it is also seen that sources with soft excesses tend to have lower than average FWHM values (as compared to the full sample), while those with warm absorbers tend to have larger FWHM. Also of interest is to examine the sources with FWHM values consistent with those found in narrow-line Seyfert 1 (NLS1) galaxies \citep[e.g.][]{1996Boller}, which are classified based on their H$\beta$ line widths and other optical line properties \citep{1985Osterbrock,1989Goodrich} and are believed to host younger, lower mass black holes accreting at a high fraction of the Eddington limit \citep[e.g.][]{1995Pounds, 2004Grupe, 2018Gallo, 2020Waddell}. For sources with FWHM < 2000\kms, four out of five sources show evidence for a soft excess, which is consistent with previous findings that NLS1s typically have strong and steep soft excesses \citep[e.g.][]{1996Boller, 2020Waddell}.

Given that the two parameters that are used to construct the Eddington ratio are the FWHM and luminosity, and that there are no differences in luminosities between samples, it should be assumed that differences in FWHM are likely driven by differences in the Eddington ratios of the respective samples. Indeed this is seen, and the bottom-right panel of Fig.~\ref{fig:opthist} shows the distributions of Eddington ratios for sources best fit with each model, as well as the median values for each sample, which are indicated with vertical lines in the corresponding colours. While the median Eddington ratio for the partial covering sample agrees with the power law sample, the median values for the soft excess and warm absorber samples differ significantly, with the soft excess sources having a much higher median Eddington ratio (AD test p-value 0.0059). The median value for the warm absorbers is also significantly lower than for the power law sample (AD test p-value 0.029). This may indicate some intrinsic differences between the sources which show evidence for soft excesses compared to those which show evidence for warm absorbers. 

\begin{figure}
   \centering
    \includegraphics[width=0.95\columnwidth]{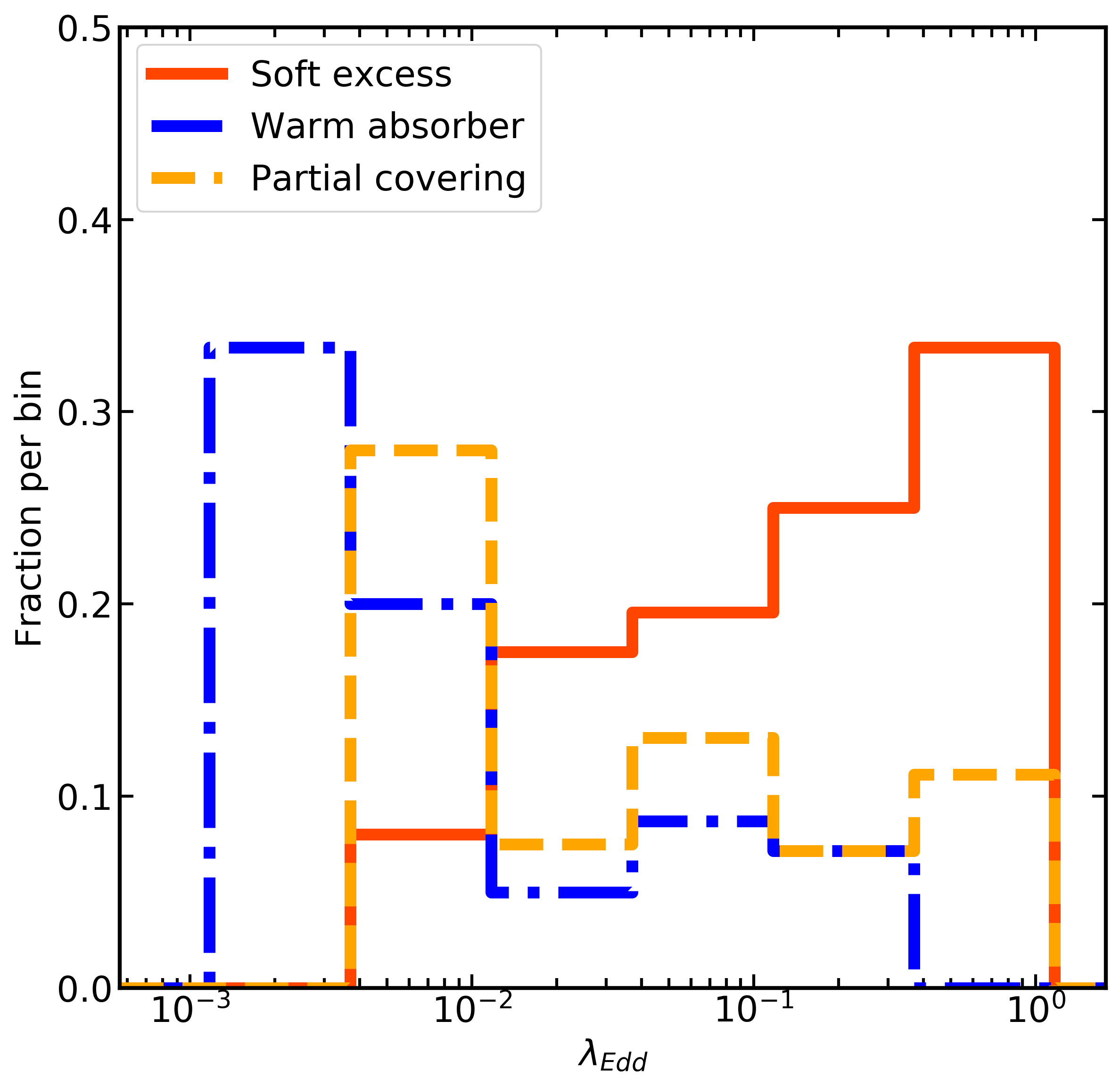}
    \caption{Fraction of the total sources best fit with each model (soft excess, warm absorber, partial covering), shown per bin using the same binning as the bottom-right panel of Fig.~\ref{fig:opthist}. Sources best fit with soft excesses are shown in red, sources best fit with warm absorbers are shown as blue dashed lines, and sources best fit with partial covering are shown as orange dash-dot lines. }
    \label{fig:eddfrac}

\end{figure}

To look at this in more detail, it is useful to examine the fraction of sources best fit by each model as a function of accretion rates. This is shown in Fig.~\ref{fig:eddfrac}, where the fraction of sources best fit with a soft excess, warm absorber or partial covering component are shown, using the same binning as in the bottom-right panel of Fig.~\ref{fig:opthist}. Sources best fit with soft excess tend to have higher Eddington ratios, and the fraction of sources with soft excess increases with increasing Eddington ratio. By contrast, the sources with warm absorbers typically have lower Eddington ratios, with a small secondary peak of $\lambda_{\rm Edd} \sim 0.1$. The sources with partial covering appear unremarkable, with most sources having median Eddington ratios of $\lambda_{\rm Edd} \sim 0.005-0.1$.

To confirm these results, several potential sources of bias were examined. First, the lowest and highest Eddington ratio bins had very few sources, which may lead to erroneously large fractions of sources best fit with a given model in those bins. The distribution was then recomputed using just four bins, where the fractions from the lowest two and highest two bins were summed. Again, the fraction of sources with warm absorbers appeared to decrease with increasing Eddington ratio with a small secondary increase around $\lambda_{\rm Edd} \sim 0.1$, and the fraction of sources best fit with a soft excess increases with increasing Eddington ratio, suggesting that the binning is not significant influencing this result. Next, since soft excesses were mostly detected up to redshifts of $z \sim 0.5-0.6$ (see Fig.~\ref{fig:lz}), it was found that warm absorbers could be found up to much higher redshifts, with some even found above $z=1$ (Fig.~\ref{fig:lz}). The analysis was therefore repeated using only sources with $z \leq 0.6$. The median Eddington ratio was again found to be significantly higher for sources with a soft excess than for those with a warm absorber (AD test p-value 0.0031), and the fractions of sources per bin followed the same trends as for the full sample. 

Sources with soft excesses typically have higher X-ray spectral counts than sources best fit with a power law, as there are additional counts in the soft band. The analysis is therefore repeated by selecting a narrow range of counts (in the $0.2-5\kev$ band), here between 25 and 1000 counts. This removes the very high signal to noise spectra, but also removes the very low count spectra where additional spectral components may be difficult to identify. The differences remain significant, with an AD test p-value of 0.015. If the minimum counts is increased to 50 counts, the AD test p-value increases slightly to 0.035. Many of the higher accretion rate sources have soft excesses, but also have high counts. By contrast, many of the warm absorber sources have lower counts as they are heavily absorbed, and also have lower accretion rates. However, when only using sources with between 50 and 1000 counts, a large fraction of sources have been removed from the sample and it is difficult to make statistical evaluations. We note that no correlations exist between the number of counts and the statistical significance of a warm absorber, partial covering, or soft excess components, and that all these components are found in sources with a broad range of counts. 

Finally, to maximise the number of sources with measurable accretion rates, there is no signal to noise cut placed on the optical spectra. However, this can lead to erroneous measurements of the black hole mass (e.g. \citet{2019Coffey} and references therein). Therefore, the analysis is again repeated using a minimum median SDSS spectrum signal-to-noise ratio of 5, as proposed in these works. Again, the results are confirmed - the Eddington ratios were found to be significantly higher for sources with a soft excess than for those with a warm absorber (AD test p-value 0.0042), and the trends observed in the fractions per bin were unchanged. These suggest that the differences observed are intrinsic to the sources and not simply related to biases in the analysis. For the subsequent analysis and discussions, no cuts are made on the redshift, X-ray counts, or SDSS signal-to-noise.

\subsection{X-ray and optical-derived parameter correlations}

\begin{figure*}
   \centering
    \includegraphics[width=1.5\columnwidth]{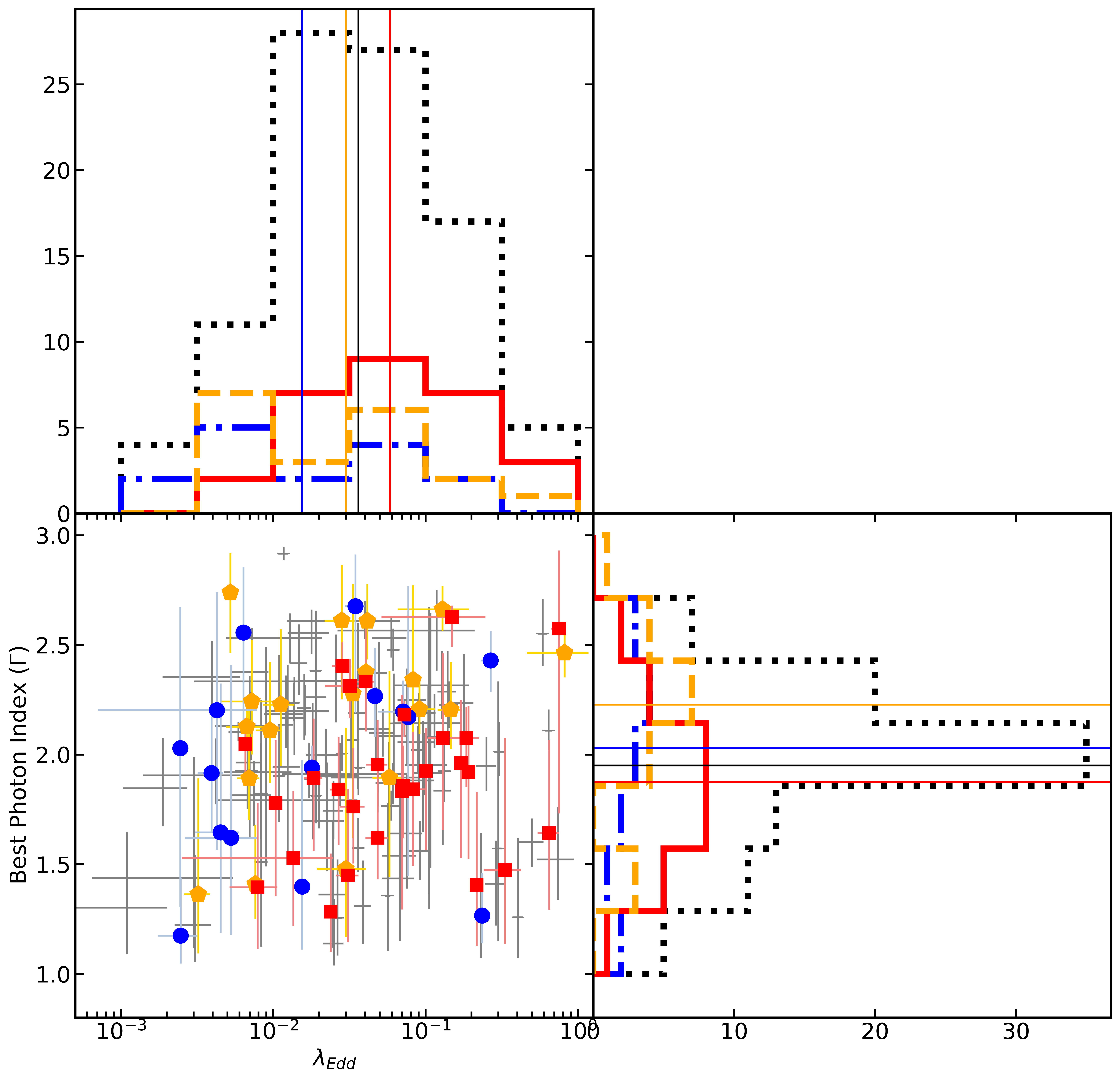}
    \caption{Distribution of Eddington ratios and photon indices measured from the best fitting spectral model. Sources shown in grey are best fit with a power law, sources shown with red squares are best fit with a soft excess, sources shown in blue circles are best fit with warm absorbers, and sources shown as orange pentagons are best fit with partial covering. Histograms for each parameter are also shown, and median values are indicated with solid lines in the corresponding colours. }
    \label{fig:eddpl}

\end{figure*}

Previous works \citep[e.g.][]{2008Shemmer, 2009Risaliti, 2013Brightman, 2017Trakhtenbrot} have shown a shallow correlation between the X-ray photon index and the Eddington ratio, although this has been debated or refuted in other works \citep[e.g.][]{2022Laurenti}. This correlation is often explained in the context of the eigenvector 1 (EV1) space, wherein the main variance in samples of quasars is shown to take the form of an anti-correlation between the strength of the FeII and [OIII] optical emission, which has been confirmed to be driven by the Eddington ratio \citep{1992Boroson,2014Shen,2020Wolf}. To examine this in the context of eFEDS, the (hard) photon index from the best fitting spectral model is shown in Fig.~\ref{fig:eddpl}, plotted with the Eddington ratios, with histograms also shown for each parameter. There is a general trend where sources with steeper photon indices also have higher Eddington ratios; the slope is constrained to be 0.07$\pm0.05$, which is significantly non-zero and in agreement with the slope found in \citet{2017Trakhtenbrot}. Performing a KS test on the photon indices for the sources best fit with warm absorbers and soft excesses, a p-value of 0.5 is obtained, suggesting that there is no difference in photon index. This also suggests that some additional intrinsic properties of these systems is responsible for the observed differences. Separately, the Eddington ratio was also checked along with the soft excess strength (F$_{SE}$/F$_{PL}$), however no significant correlation between these parameters was found. These findings will be discussed in more detail in Sect. 6. 

\begin{figure}
   \centering
    \includegraphics[width=0.95\columnwidth]{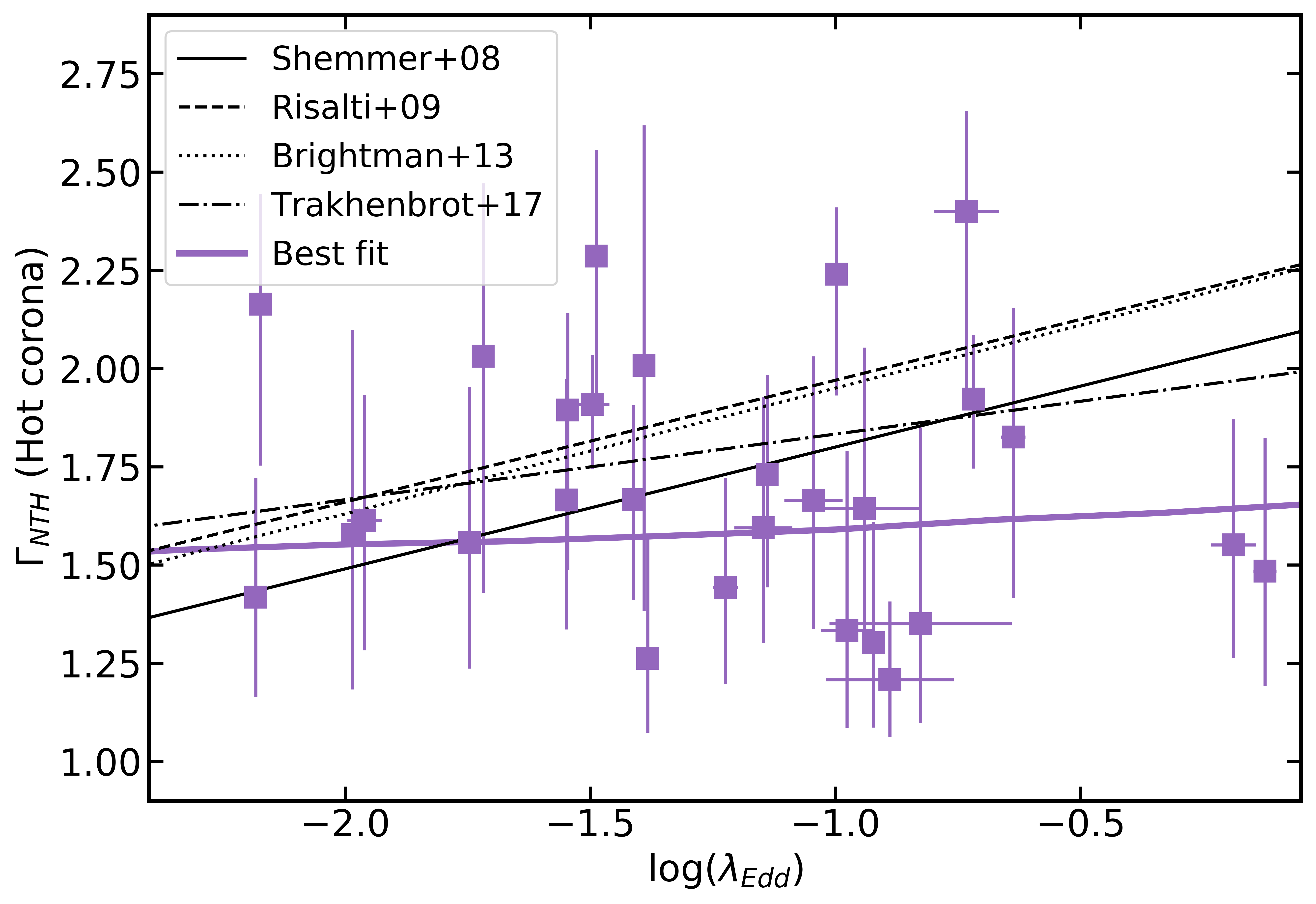}
    \includegraphics[width=0.95\columnwidth]{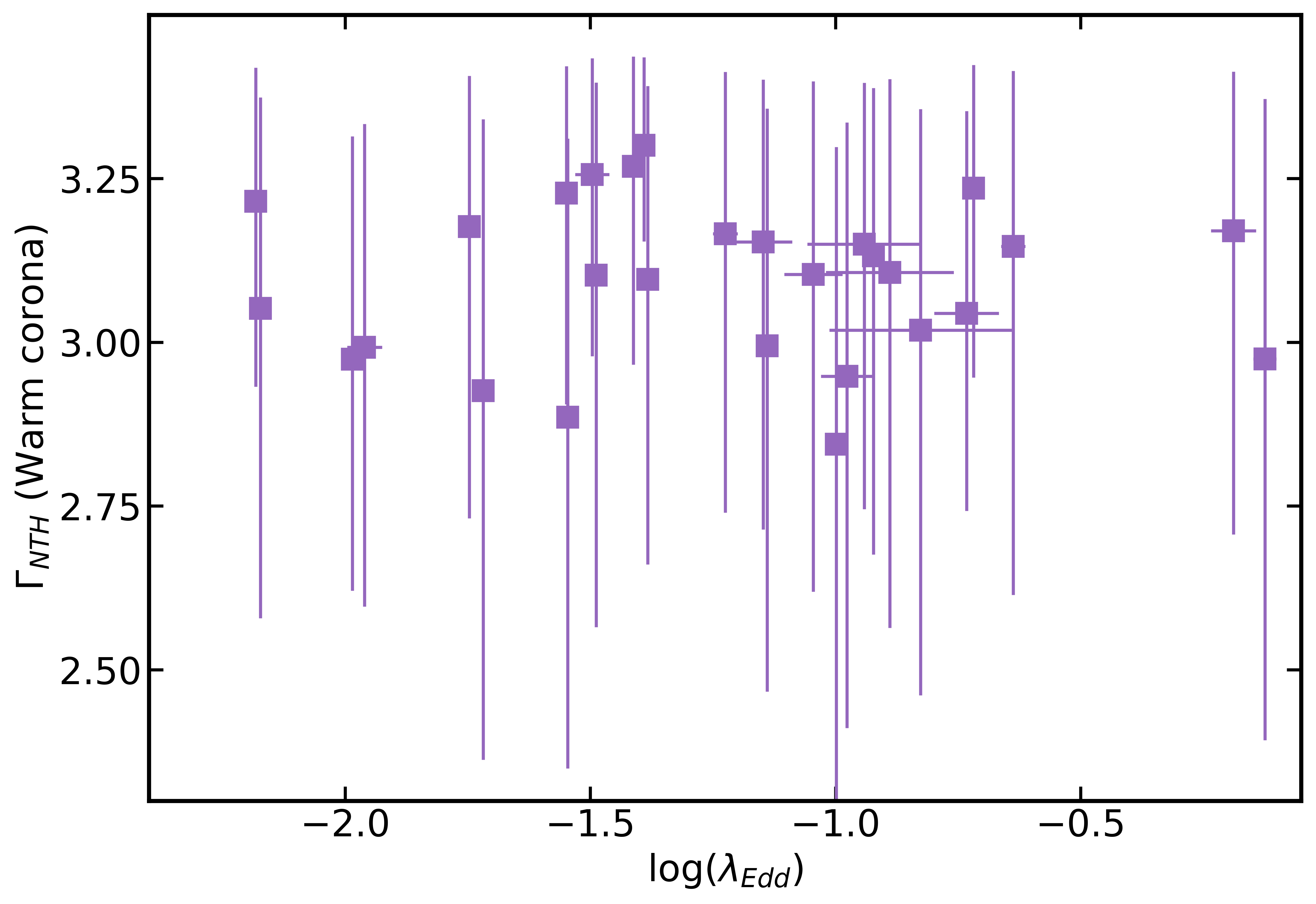}
    \includegraphics[width=0.95\columnwidth]{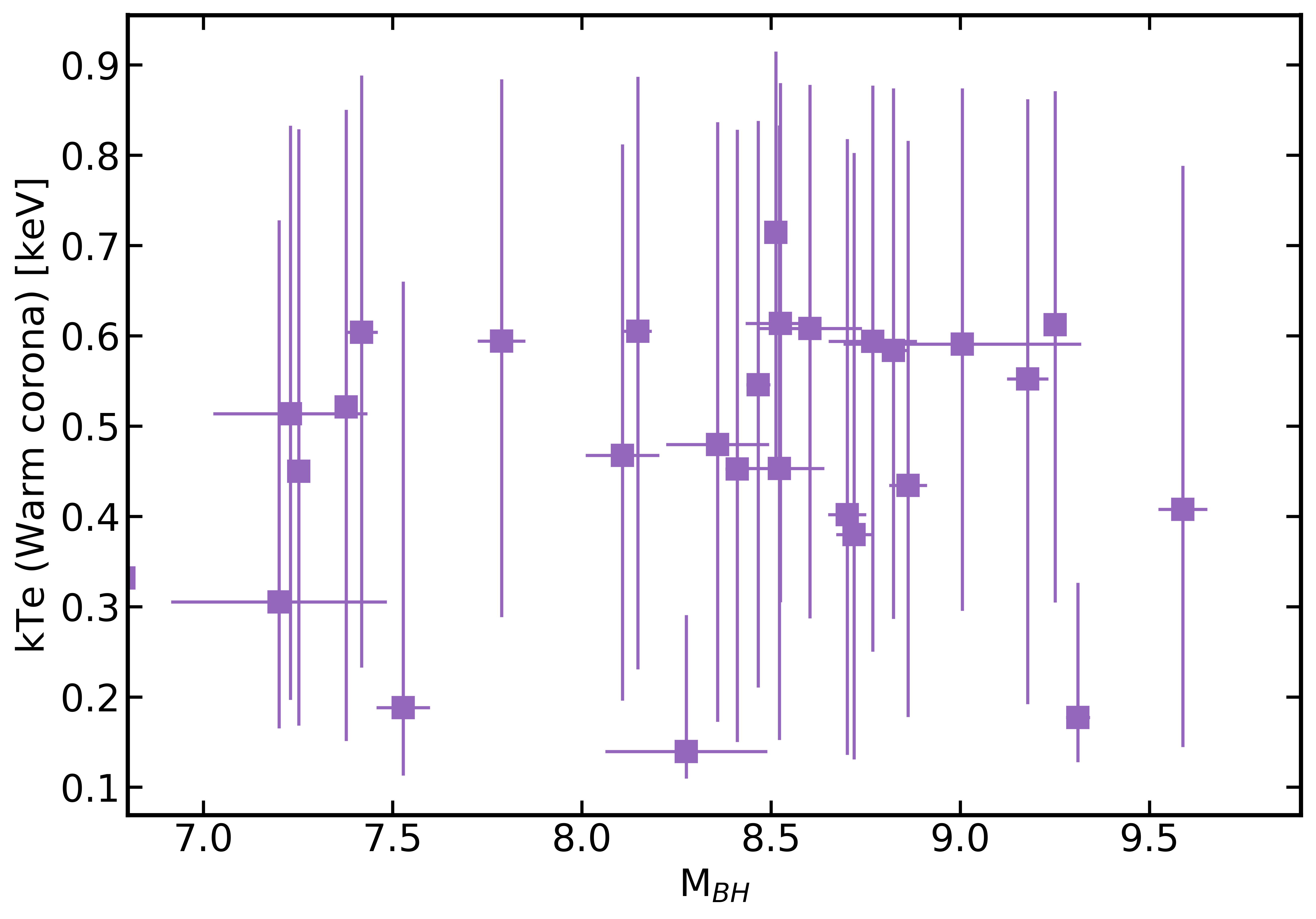}
    \caption{Relationship between warm corona parameters and optical derived parameter measurements. Top: Relationship between Eddington ratio and hot corona photon index, shown for all sources with soft excesses. The best fit line and errors are shown by the shaded purple region. Correlations between these parameters found in previous works are shown in black lines of various line styles. Middle: Same as top, but for the warm corona photon index. Bottom: Relationship between black hole mass and warm corona temperature, kT. }
    \label{fig:eddse}

\end{figure}

With the sources that show evidence for a soft excess and also have reliable SDSS spectra, we also attempted to identify a physical parameter that was strongly correlated with the Eddington ratio, black hole mass, or bolometric luminosity in an attempt to explain why sources with soft excesses are found in an increasing fraction at higher Eddington ratios. Examples of interest are shown in Fig.~\ref{fig:eddse}, with the hot corona photon index and Eddington ratio shown in the top panel, the warm corona photon index and the Eddington ratio shown in the middle panel, and the warm corona temperature and black hole mass shown in the bottom panel. The best fit lines obtained from a linear regression are shown as shaded purple regions in the top panel, and comparisons of best-fit lines found by \citet{2008Shemmer, 2009Risaliti, 2013Brightman} and \citet{2017Trakhtenbrot} are shown with black solid, dashed, dotted and dash-dot lines, respectively. Neither correlation is significant, due to large errors on the data points and shallow slopes. That being said, the photon indices obtained from the warm corona are lower than, but broadly consistent within errors with, those found by these previous studies. This is also discussed in Sects. 6.2 and 6.3. No correlation is apparent between the warm corona temperature and the black hole mass (or Eddington ratio), and errors for the warm corona temperature are large for most sources.

\section{Discussion}
\subsection{Soft X-ray spectral properties of eFEDS AGN}
The eFEDS survey covering $140$ deg$^2$ of the sky was designed to demonstrate the capabilities of eROSITA for extragalactic survey science, in anticipation of the all sky survey. Even with a very small fraction of the final survey area, eFEDS has demonstrated the power of eROSITA not only to detect new sources and source populations, but to characterise their properties and provide physical insights via their X-ray spectra. This work uses those capabilities to identify AGN with a soft excess, and to differentiate between various models for the excess, which may be an artefact of complex absorption or a true emission component. More specifically, this work uses Bayesian fitting methods and reliable model selection to characterise the X-ray spectral of a sample of 200 hard X-ray--selected AGN, finding the following results: 

\begin{itemize}
    \item In addition to the underlying continuum modelled with a power law, an additional power law describes the shape of the soft excess better than a blackbody component. This suggests a non-thermal origin for the soft excess is preferred over a blackbody originating in the inner disc.
    \item Making use of simulations and spectral fitting, 29 (14.5\%) sources show evidence for a soft excess, 29 (14.5\%) sources show evidence for a warm absorber, and 25 (12.5\%) sources show evidence for a partial covering absorber, all with 2.5\% ($\sim5$) spurious detections estimated from simulations. By design, there is no overlap between these samples.
    \item Examining these sources in colour-colour space reveals differences between sources best fit with soft excesses, complex absorbers, and only power laws, which can be used for comparison with soft X-ray--selected samples and the eROSITA all-sky survey.
    \item Of the 29 sources in the selected soft excess sample, 23 appear best fit with the soft Comptonisation interpretation, and six appear best fit with blurred reflection.
    \item Many sources which display evidence for Compton-thin absorbers ($>10^{22}\pscm$) from the baseline absorbed power law model are actually warm absorbers, including all sources with column densities $>10^{23}\pscm$, suggesting that a population of apparently absorbed type-1 AGN actually display evidence for warm absorption from e.g. a disc wind rather than absorption in the distant torus.
    \item Sources with lower Eddington ratios tend to more frequently host warm absorbers, and the fraction of sources with warm absorbers appears to decrease with increasing Eddington ratio.
    \item Sources with higher Eddington ratios more frequently have soft excesses, and the fraction of sources with soft excesses increases with increasing Eddington ratio.
\end{itemize}

These findings and the consequences for the all-sky survey are further discussed in the following sections.

\subsection{Soft excesses in eFEDS}
The sub-sample of 29/200 sources with soft excesses identified in this work prove a powerful tool for studying soft excesses in AGN. First, it should be noted that this fraction of sources with soft excesses is low, with less than half (83 sources, or 41.5\%) of sources showing evidence for some form of additional component (soft excess or complex absorption). Furthermore, only 29 sources (14.5\% of the full sample) show evidence for a soft excess which cannot be explained with complex absorption alone. Previous works studying samples of soft excesses typically assume that all sources have a soft excess and apply the models accordingly, and indeed some works have claimed ubiquitous or near-ubiquitous detections \citep{2005Piconcelli,2012Done,2012Scott},  while in this work it is demonstrated that little or no information can be gained from fitting a soft excess component to sources without a soft excess. Soft excess measurements are not biased by the selection of the hard X-ray sample (Fig.~\ref{fig:bias}), and that while sources with soft excess are typically higher fluxes, there are also very bright sources with no evidence for a soft component. It remains unclear if this work finds a lower number of AGN with soft excess due to low signal-to-noise for some data sets or overly stringent selection criteria, or if this is reflective of the true fraction of sources with soft excesses. Further investigation is needed to properly estimate this fraction.

While this work has found that most sources are best fit or at least well fit with soft Comptonisation, there are six sources which show some preference for blurred reflection. Sources best fit with blurred reflection have higher soft $\Gamma$ values when fitting with the PL+PL model (see Fig.~\ref{fig:splhpl}). Part of this may be due to modelling biases or uncertainties, however, this may imply that sources best fit with blurred reflection have steeper soft excesses. Furthermore, it is found that sources best fit with blurred reflection have a higher median soft excess strength, and high reflection fraction values. This suggests that the sources with extremely strong reflection spectra are still identified as best fit blurred reflection, while sources with weaker measured reflection spectra are in fact unlikely to be best fit with blurred reflection, and the soft excess is more likely produced with a warm corona. This may be a limitation of the eROSITA bandpass, where emission above $5\kev$ is very difficult to detect due to the combination of low effective area and high background. While the soft components can have a similar shape, reflection models predict a more prominent and relativistically broadened line at $\sim6\kev$ due to fluorescent emission from \feka, which may differ from the more limited reprocessing in the warm corona model. 

Strong and steep soft excesses are seen in a majority of X-ray observed NLS1 galaxies \citep[e.g.][]{1996Boller}. Individually, many NLS1s have been shown to be well fit with blurred reflection, absorption, or a combination of the two \citep[e.g.][]{2004Tanaka, 2009Fabian, 2015Gallo, 2019Gallo, 2019Jiang, 2019Waddell, 2021Boller}. On aggregate, NLS1s also appear to have stronger soft excess strengths, a strong correlation between the soft excess strength and the strength of the Compton hump, and a weak anti-correlation between the photon index and soft excess strength \citep{2020Waddell}. Together, this may suggest that some AGN display NLS1-type spectra, and that these sources are more likely fit with blurred reflection models compared to more typical broad-line Seyfert 1 (BLS1) sources with flatter, smoother and weaker soft excesses, which are well fit with the soft Comptonisation model \citep{2020Waddell}.

It is also noteworthy that while the warm corona model fits most sources in this sample, other works have clearly demonstrated the presence of broad lines and deep absorption edges associated with \feka\ emission and absorption. These features cannot easily be explained with the soft Comptonisation model; rather, this requires an absorption component, blurred reflection, or some combination of models. In this work, sources with a very strong reflection component (e.g. $R\gg1$) were typically best fit with a relativistic blurred reflection model, suggesting that sources with very strong reflection components are preferentially detected. Further investigation with a larger sample (e.g. eRASS:1) as well as follow-ups with high resolution spectroscopy \citep[e.g. XRISM;][]{2020Tashiro} will help to identify more sources with high energy spectral curvature and verify if there are sub-samples of AGN which require blurred reflection to explain the spectral shape.

\subsection{The warm corona model}
Despite the caveats discussed above, the evidence comparison for individual sources shows that most are well fit or best fit with a warm corona model (Table~\ref{tab:phys}). The combined evidence comparison  also shows that if we assume that the underlying model for all sources is the same, then the warm corona model is preferred over blurred reflection. 

From the top panel of Fig.~\ref{fig:eddse}, it is evident that the warm corona model tends to produce fairly flat hot corona photon indices, with a median value of $\sim1.6$. This is flatter than the typical population values of $\Gamma \sim1.8-1.9$. Interestingly, these flatter photon indices are not found when examining the PL+PL model (see the left-hand panel of Fig.~\ref{fig:splhpl}). As discussed in section 5.2, the hard X-ray photon index is correlated with the detection likelihood in the $2.3-5\kev$ band such that sources with very flat photon indices are also very close to the background in the hard X-ray, which may explain why flatter slopes are measured. However, the fact that the PL+PL modelling returns more reasonable slope values of $\Gamma \sim1.85$ may suggest that the stricter priors placed on the soft photon index may also be artificially flattening the hard X-ray photon index. This may also be a result of the combined shape of the two \texttt{nthComp} components, in particular when considering the steep cut-off present in the soft power law: indeed, \citet{2021Xu} also find a very flat slope when fitting the type-1 AGN ESO 362-G18 with a double \texttt{nthComp} model as compared to modelling with blurred reflection. While these flat slopes may be consistent with the expected photon indices given their low accretion rates of $\sim0.001-0.01$ times the Eddington limit (see top panel of Fig.~\ref{fig:eddse}), \citet{2022Laurenti} demonstrate that this correlation may not necessarily hold, and that sources accreting at approximately their Eddington limit display a broad range of photon indices. These findings suggest that the model parameters obtained in this work are acceptable for this analysis. 

One limitation to modelling the warm corona using the method outlined in this work is that it fails to characterise the effect one corona has on the emission from the other. In a more realistic scenario, photons from the warm corona would be incident on the hot corona, and vice versa. It could even be plausible that the hard and soft components are part of the same, multi-temperature and multi-density cloud of electrons. Therefore, it is interesting to investigate evidence for interplay between the two coronae. One way to do this is by understanding the soft excess strength, that is the ratio of the fluxes emitted by each of the two coronae in the same energy band. Examining the distribution of soft excesses (Fig.~\ref{fig:softex}), most of the soft excesses span a relatively small parameter space of factor $\sim10$, with few extreme values. Interestingly, using a similar definition for the soft excess strength and modelling broad-line Seyfert 1 galaxies observed with \textit{Suzaku}, \citet{2020Waddell} found that soft excess strengths span about factor $\sim10$. By contrast, the observed soft excess strengths in NLS1 galaxies spanned a much larger parameter space (factor $\sim 100$), extending to very strong soft excess strengths ($\sim 10$), similar to the extreme values found for some sources in this work. This may suggest some coupling between the two coronae which only permits certain ratios of fluxes between the two components, while this restriction does not exist for blurred reflection dominated sources, where the flux ratio between components is more dependant on the height of the corona above the accretion disc.

\subsection{Dense absorption in eFEDS may be warm and complex}
When modelling eFEDS sources with the baseline absorbed power law model, many sources appear to show evidence for Compton-thin absorption with column densities $>10^{22}\pscm$. However, examining these sources more closely, a majority (19/26) show evidence for complex absorption (18 warm absorbers and one partial covering absorber), as seen in Fig.~\ref{fig:sespace}. This includes all eight sources with column densities above $10^{23}\pscm$, as measured by the baseline model, which are all best fit by a warm absorber model. This raises an intriguing possibility that some, if not most, of the apparently Compton-thin absorbers observed in the Universe are associated with ionised gas hosted in disc winds, and not necessarily in a more distant obscurer such as the torus. 

\begin{figure}
   \centering
    \includegraphics[width=0.95\columnwidth]{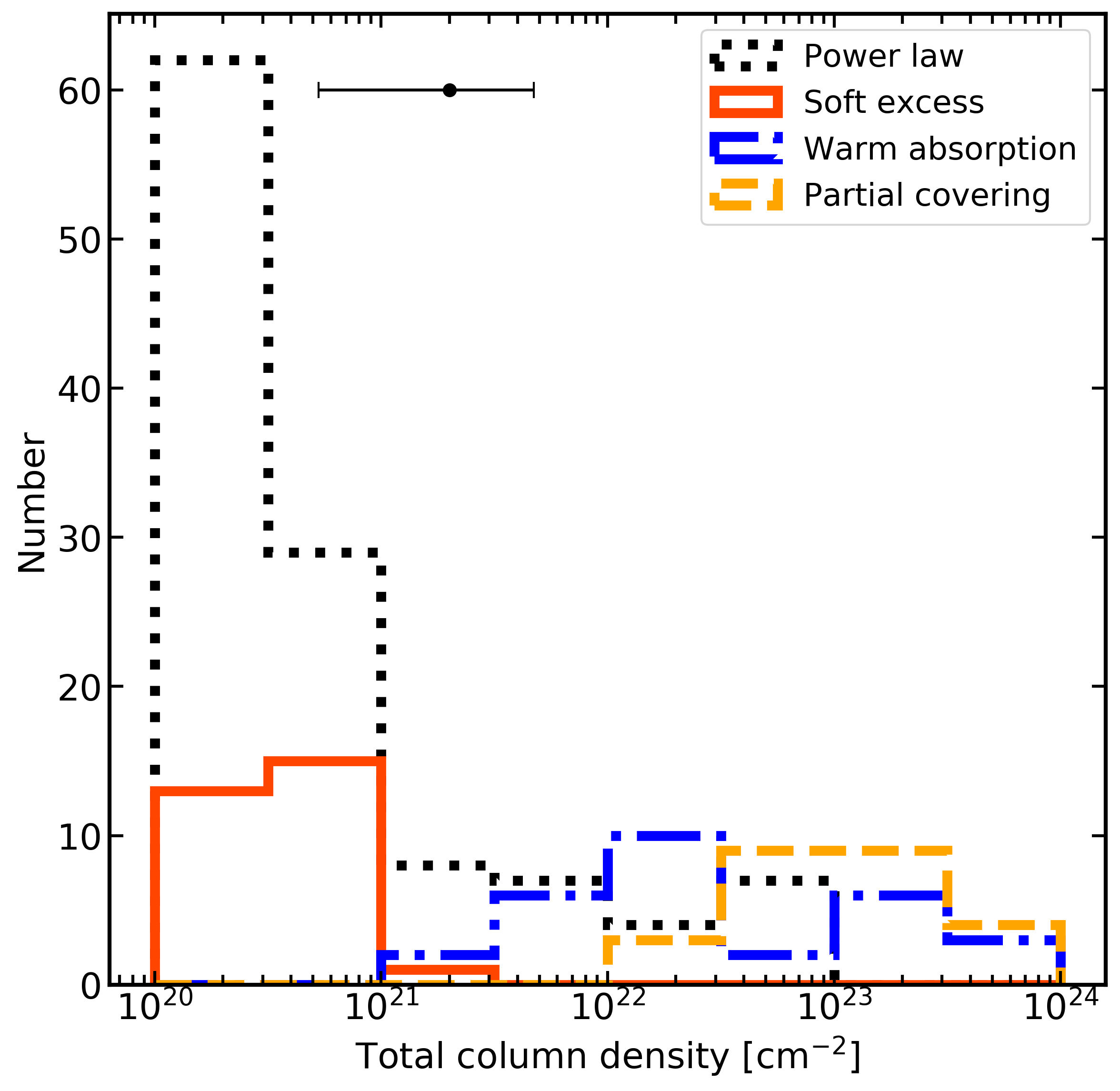}
    \caption{Distribution of total absorbing column density for each model. For sources best fit with a single power law or soft excess, this corresponds to the column density measured in the \texttt{ztbabs} component. For sources best fit with a warm absorber, this corresponds to the sum of the column densities from the \texttt{ztbabs} model and the XSTAR grid, and for sources best fit with neutral partial covering, this corresponds to the sum of the column densities measured from the \texttt{ztbabs} and zpcfabs components. Sources best fit with soft excesses are shown in red, sources best fit with warm absorbers are shown as blue dash-dot lines, and sources best fit with partial covering are shown as orange dashed lines. The typical error bar is shown in black. }
    \label{fig:nhcomp}

\end{figure}

In sources with complex absorbers, while the cold and warm absorption column densities are not individually well constrained, the sums of the simple (\texttt{ztbabs}) and complex (cwa18 or \texttt{zpcfabs}) absorbers are constrained (see also Fig.~\ref{fig:wacorner}, Fig.~\ref{fig:pcfcorner}, Fig.~\ref{fig:secorner} and Fig.~\ref{fig:abscorner}.) Fig.~\ref{fig:nhcomp} shows the distributions of total column density for each model. Contrary to Fig.~\ref{fig:sespace}, the definition of the column density changes depending on the best-fit model for each source; for sources best fit with a single power law or soft excess, this corresponds to the column density measured in the \texttt{ztbabs} component. For sources best fit with a warm absorber, this corresponds to the sum of the column densities from the \texttt{ztbabs} model and the cwa18 grid, and for sources best fit with neutral partial covering, this corresponds to the sum of the column densities measured from the \texttt{ztbabs} and \texttt{zpcfabs} components. A majority of sources best fit with a power law have low column densities ($<10^{21}\pscm$). This can be associated with absorption in the host galaxy, with a small minority showing evidence for some form of Compton-thin obscuration. However, the majority of absorbed sources are obscured by a partial covering absorber or a warm absorber, and thus not a distant torus. Indeed, no significant difference is found between soft excess and a power law, demonstrating that most of the absorption originates in the partial covering of warm absorbers. 

To investigate this in more detail, the 19 apparently obscured sources best fit with complex absorption are re-fit with a more physical torus model, \texttt{UXClumpy} \citep{2019Buchner}, which describes the X-ray spectrum of a clumpy absorber illuminated by the corona. The Galactic column density is also included in the model, as before. Free parameters include the coronal photon index (given a uniform prior between one and three, as in all other models in this work), the line-of-sight column density (given a log-uniform prior between $10^{20}\pscm$ and $10^{25}\pscm$), and the torus inclination, vertical extent of the clouds, and covering factor, all of which were given uniform priors between the minimum and maximum values allowed by the model and none of which were well constrained in spectral fits. The energy cut-off is fixed at $400\kev$, but does not influence the spectral shape for large cut-off values given the limited hard X-ray coverage of eROSITA. The Bayes factor is then computed for each source to ease comparison with other models. 

Of these 19 sources, only six are better fit with \texttt{UXClumpy}. In all of these cases the column density in the torus is $\sim10^{23}\pscm$, and other torus parameters cannot be constrained. Furthermore, very steep photon indices of $\Gamma \sim 2-3$, with many in agreement with the upper limit of $\Gamma = 3$, are required to explain the spectral shape. Four of these sources have clear evidence for optical broad lines and thus have type-1 optical spectra. This seems highly unlikely, as type-1 AGN do not typically show evidence for absorption with column densities $>10^{22}\pscm$ \citep[e.g.][]{2018Shimizu}, and this is at odds with the AGN unification model wherein type-1 AGN offer a direct view of the central region and only type-2 AGN are viewed through the torus \citep[e.g.][]{1993Antonucci,1995Urry}. Furthermore, the high photon indices are likely un-physical and are atypical for obscured AGN studied with high quality spectra \citep[e.g.][]{2017Ricci}. Therefore, the warm absorber model, which yields more reasonable photon indices of $\Gamma \sim 1.8-2$ is preferred for these sources. Future modelling with higher signal to noise spectra or expanded energy coverage (e.g. with \nustar) would be required to confirm this result (Waddell \et in prep.). 

If indeed the majority of Compton-thin obscuration in this sample is caused by warm absorption, this may change our view of the way obscuration is considered and treated in AGN. If these warm absorbers are indeed physically produced by winds launched from the accretion disc, these winds are believed to play a significant role in the evolution of the system. In a comprehensive work on absorption in AGN, \citet{2015Buchner} studied the evolution of AGN which appear unobscured (column densities $<10^{22}\pscm$), Compton-thin (column densities $10^{22}-10^{24}\pscm$), and Compton-thick (column densities $>10^{24}\pscm$), and found some evidence that Compton-thin AGN are evolving faster at redshifts $z = 0.5-4$ (specifically, their space density rises more rapidly and peaks earlier than unobscured and Compton-thick AGN). This is most apparent for sources with column densities $10^{22}-10^{23.5}\pscm$, where a majority of the sources in this work (15/23) show evidence for warm absorption. This might suggest that disc winds are playing a role in the more rapid space density evolution of apparently Compton-thin sources in this redshift range. A more in-depth analysis incorporating more sources spanning a larger redshift range would be required to more fully understand this result.

\subsection{The Eddington ratio distinction between spectral models}
Perhaps the most important result in this work is that sources with warm absorbers tend to have low Eddington ratios with a few having higher values of $\lambda_{\rm Edd} \sim 0.1$, while sources with soft excesses have higher Eddington ratios and an increasing fraction of sources with higher Eddington ratios have soft excesses. The result is robust to various tests of selection effects and bias. This suggests that this is an intrinsic property of the objects. It is therefore of interest to investigate which physical mechanisms may be responsible for these observed differences.

With very few sources having both warm absorbers and Eddington ratios available (due to incomplete optical coverage or lack of a well-defined broad line), it is difficult to identify possible correlations here between parameters. However, it appears that sources with higher Eddington ratios have higher warm absorber ionisations than those without. There are not enough sources to confirm this, but such a correlation may well be expected if the warm absorber is indeed associated with a disc wind. There are three primary launch mechanisms for these low-velocity disc winds; a thermal driven wind in which the wind is formed when the disc loses upper layers due to irradiation of the outer disc, a radiation pressure driven wind wherein the wind is launched via radiation pressure in the disc, or by magnetic fields which give rise to magneto-rotational instabilities in the disc \citep[e.g.][]{1994Lubow, 2017Fukumura, 2021Mizumoto}. Thermal winds are likely less important in many systems, given that radiation pressure driven winds can begin to dominate the wind in systems as cool as $10^5$ K. While it is well known that magnetic fields are important in the context of AGN, it is easiest to understand disc winds in the context of radiation pressure. It is possible that the sources with higher accretion rates have higher radiation pressures in the disc, resulting in the launching of ionised winds. However, if the radiation pressure becomes too high, the wind may become over-ionised such that the velocity of the wind is not sufficient to escape the system, resulting in a failed wind \citep{2006Schurch,2017Parker, 2018Parker, 2018Pinto, 2019Gallo, 2019Giustini, 2021Giustini, 2021Boller}. This suggests that highly ionised winds must be launched close to the accretion disc and must not become over-ionised in order to be detected, whereas low ionisation winds can be launched at larger disc radii and should be easier to detect \citep[e.g.][]{2017Fukumura}. A tentative positive correlation is found between the Eddington ratio and warm absorber ionisation for sources in this work, supporting this interpretation. 

Of particular interest here is the proposition of \citet{2006Schurch,2019Giustini, 2021Giustini} that these failed winds may be associated with a physical component; the broad-line region (BLR), an X-ray obscurer on BLR scales, or a warm corona. This would nicely explain the observed distributions of fractions of sources in each accretion rate bin presented in Fig.~\ref{fig:eddfrac}. For sources with lower $\lambda_{\rm Edd}$, the disc exerts less radiation pressure and thus produces less failed winds, so a warm absorber is more frequently detected in the spectrum (provided the wind intersects the line of sight). At intermediate $\lambda_{\rm Edd}$, there is sufficient radiation pressure exerted near the inner disc for some winds to become over-ionised and the velocity is insufficient for the wind to escape. It remains in the system, producing an X-ray obscurer which is detected as a partial covering absorber. Finally, at larger $\lambda_{\rm Edd}$, a larger fraction of winds will have insufficient velocity to escape the system and will become failed winds. These failed winds may remain closer to the disc, forming the warm corona and resulting in a large fraction of sources with soft excesses also featuring larger $\lambda_{\rm Edd}$. In this framework, sources best fit with a power law only, do not have strong enough winds intercepting the line of sight to be detected, or have not had failed winds with the correct conditions to form an absorber or soft excess.

\begin{figure*}
   \centering
    \includegraphics[width=0.32\textwidth]{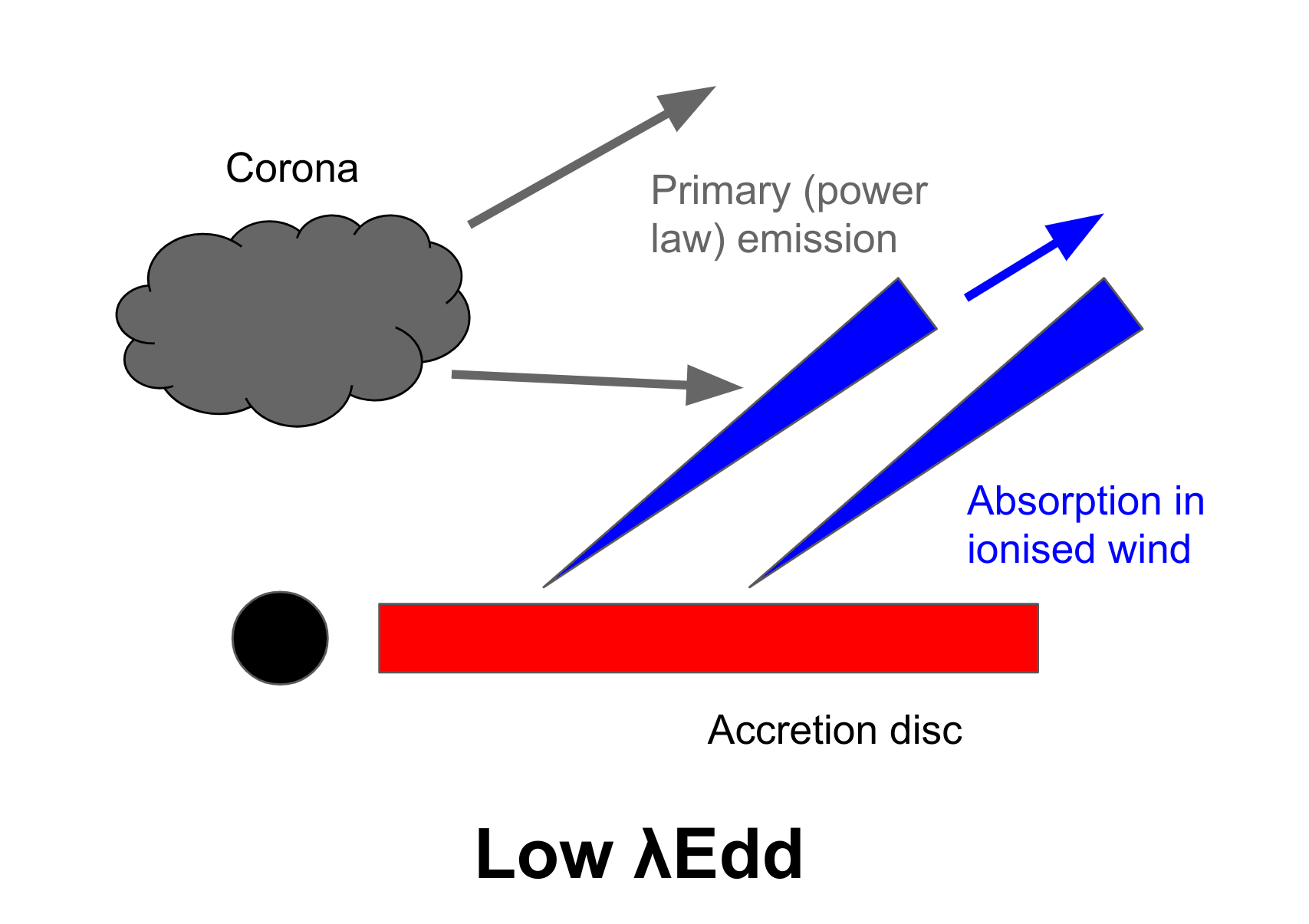}
    \includegraphics[width=0.32\textwidth]{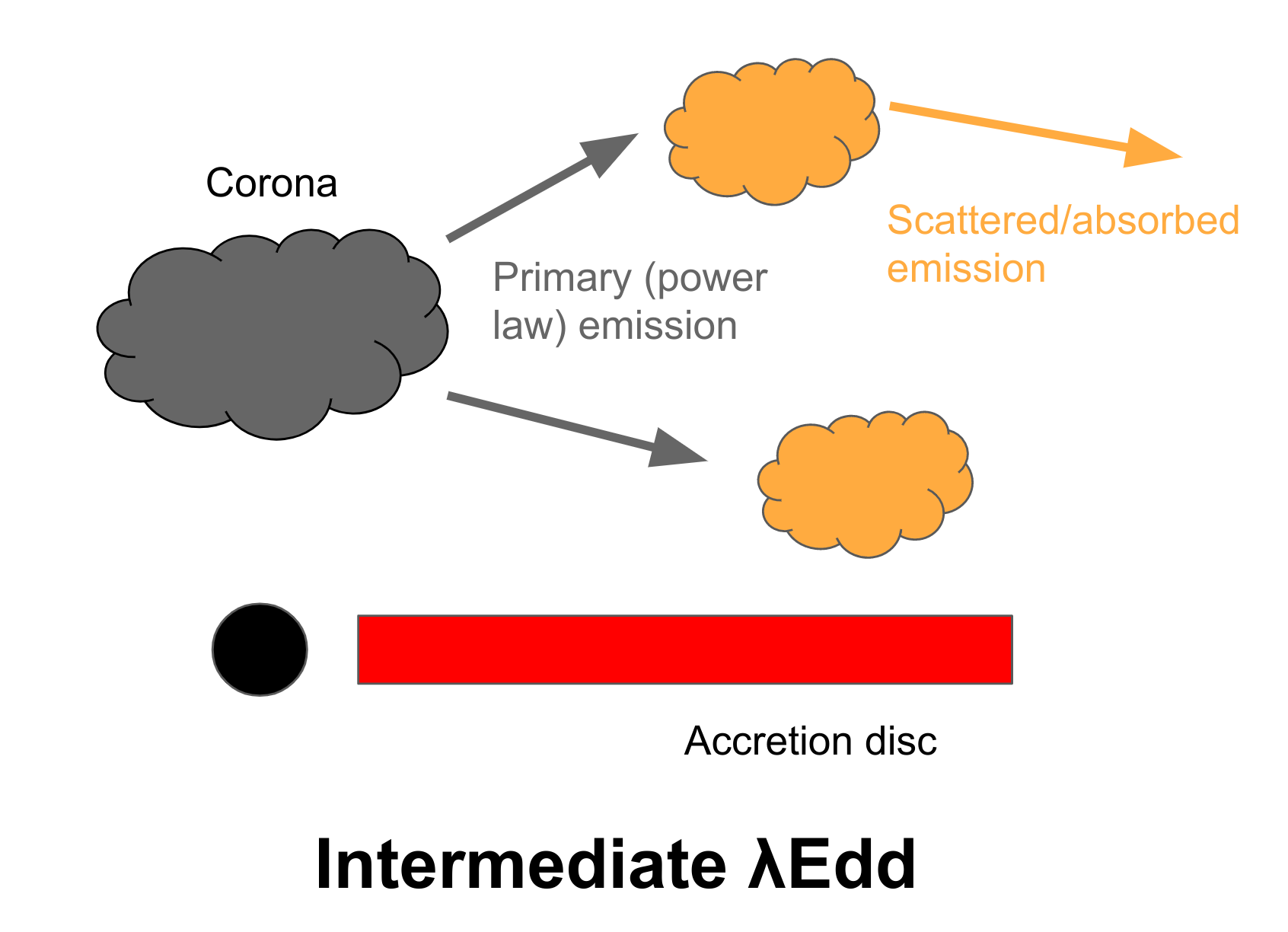}
    \includegraphics[width=0.33\textwidth]{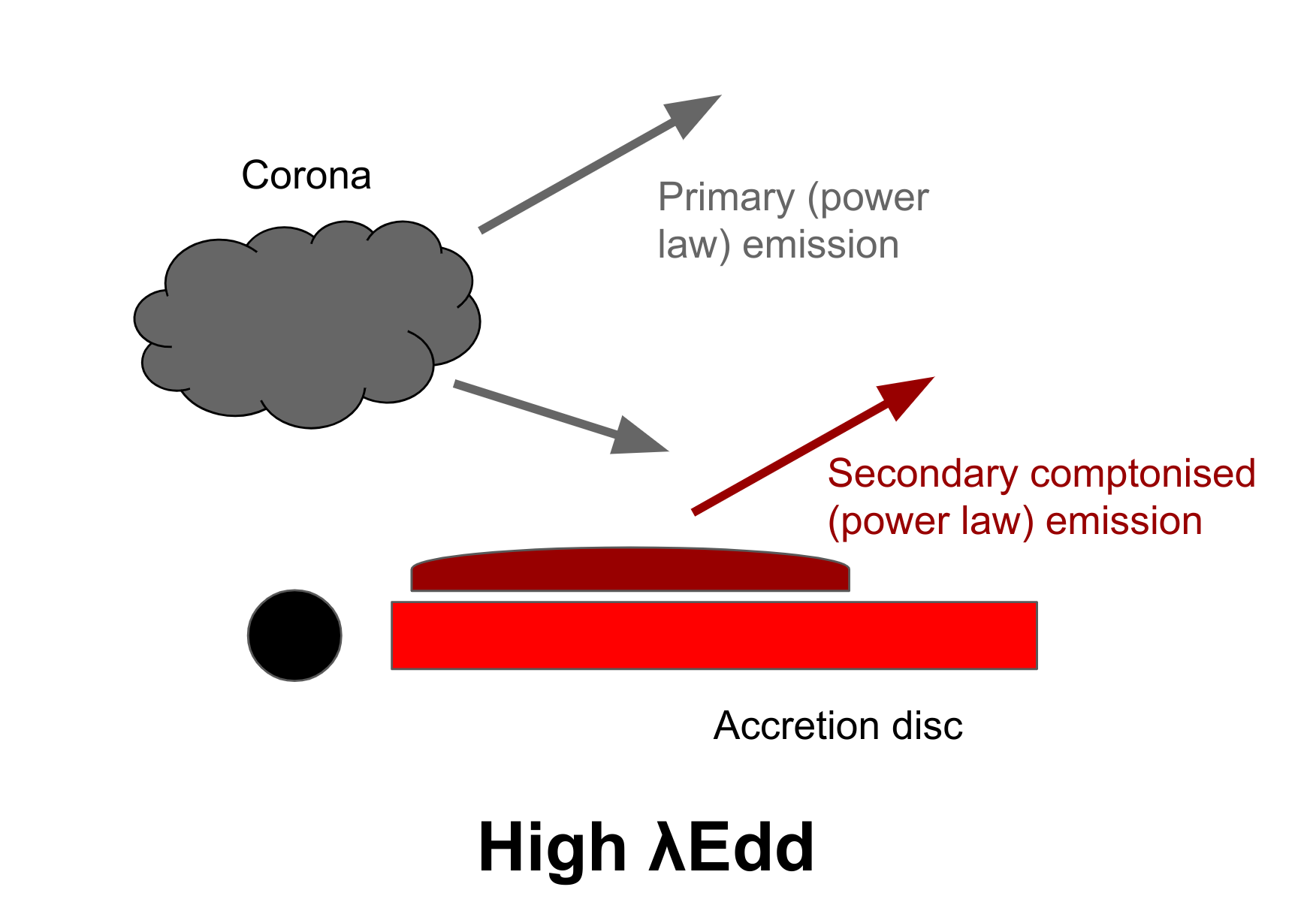}
    \caption{Cartoon schematic showing how the emission may vary with Eddington ratio. Left: Sources with relatively lower Eddington ratios (few $\times 10^{-3}$), there is sufficient radiation pressure to launch line driven winds, but not so much that the winds become over-ionised, so more warm absorbers are detected. Middle: Sources with intermediate Eddington ratios, (few $\times 10^{-2}$) some winds will become over-ionised and their velocities will not exceed the escape velocity and will fail, but will remain somewhat bound to the system, creating patchy partial covering absorbers. Right: Sources accreting at $0.1-1$ Eddington, high radiation pressure results in too many ionising photons and the wind is destroyed, remaining tightly bound to the system and forming the warm corona.}
    \label{fig:cartoon}

\end{figure*}

All these physical scenarios are illustrated in Fig.~\ref{fig:cartoon}, with one panel showing each of the three Eddington ratio regimes. This interpretation is consistent with the tentative correlation found between the warm absorber ionisation and the Eddington ratio, and may also explain why there are no correlations found between the Eddington ratio and the properties of the warm corona, as it suggests that a warm corona is more likely to form for sources with higher $\lambda_{\rm Edd}$, but is not necessarily hotter, denser or stronger. It should also be stressed here that this work does not attempt to model the ultra-fast outflows \citep[UFOs; e.g.][]{2010Tombesiufo, 2020Igo, 2023Matzeu} which typically have higher column densities ($\sim10^{23}$), higher ionisations ($10^3-10^6$ erg s$^{-1}$) and large outflow velocities (v$\sim0.033$c - $0.3$c). Such outflows are often detected based on absorption features in the $7-10\kev$ range, at the edge of or outside of the eROSITA bandpass. UFOs are often found in high Eddington ratio sources, but the typical ionisations and large outflow velocities are mostly inconsistent with the ionisation range of ($10^{-4}-10^4$ erg s$^{-1}$) and the zero outflow velocity used in this work. More in-depth modelling with a broader parameter space and likely including data from \xmm\ or \nustar\ would be required to attempt to identify signatures of UFOs in this sample.

One caveat to this analysis is that this work does not attempt to characterise properties of potential warm absorbers and disc winds simultaneously. To address this, all sources which showed evidence for a soft excess or warm absorber were also fit with a model including both a disc wind (warmabs) and a secondary power law component (in XSPEC, this corresponds to \texttt{tbabs} $\times$ \texttt{ztbabs} $\times$ cwa18 $\times$ (\texttt{powerlaw} + \texttt{powerlaw})). The same priors as in Sect. 3 were used, and Bayes factors were again calculated for each source. For most sources, the inclusion of both components did not result in a better fit to the spectrum, and many spectral parameters could not be well constrained. Therefore, it appears that for the eFEDS analysis, the components can only be treated separately, which may lead to some modelling errors. \citet{2013Laha} discuss the complications of simultaneously modelling a soft excess and warm absorption component in NLS1 IRAS 13349+2438, and demonstrate that different ionising continua produce different ionisation structures in the warm absorption component, whereas including a soft excess tends to decrease the ionisation while the column density remains consistent. More recently, \citet{2022Parker} discuss the degeneracies between X-ray winds and relativistic blurred reflection, where systematic biases are found in the derived outflow parameters of the wind when the emission from the disc is not properly characterised. These are particularly crucial in the case of broad features, where even microcalorimeter resolution does not help to break some degeneracies \citep{2022Parker}. Care should therefore be taken in interpreting the ionisation parameters in this work, and further analysis should be done to better understand any potential superposition between these components.

\section{Conclusions}
In this work, the 200 sources classified as AGN from the eFEDS hard X-ray--selected sample are modelled with a variety of phenomenological and physically motivated models in order to investigate the nature of the X-ray soft excess. X-ray spectra are fit using BXA so that the Bayesian evidence can be compared to select the best fitting models. This work demonstrates that eROSITA can be used to identify signatures of both complex absorption and soft excesses, and using simulations, the significance of these features can be evaluated. This analysis identifies a total of 29 sources that have warm absorbers ($14.5\%$ of the sample), 25 sources that have neutral partial covering absorbers ($12.5\%$ of the sample), and 29 sources ($14.5\%$ of the sample) with soft excesses (all with 97.5\% purity), which clearly shows that soft excesses and complex absorbers are key features for understanding the properties of large samples of AGN. 

It is shown that most sources with true soft excesses are best explained by a warm corona model as opposed to a relativistic blurred reflection scenario. Follow-up observations of these sources with better sensitivity in the hard X-ray (e.g. with \xmm\ or \nustar) can help to search for the presence of a broad iron line or Compton hump which are strong signatures of blurred reflection \citep[e.g.][]{1989Fabian,2005Ross}, and a timing analysis incorporating reverberation mapping \citep[e.g. following the prescription of][]{2014Uttley} can also help to distinguish between soft excess and absorption models.

Several interesting results were also found in studying the properties of the warm absorbers and soft excesses, including that warm absorbers are likely the true nature of the absorption in many apparently Compton-thin AGN, and that sources with soft excesses are found in higher Eddington ratio sources while sources with warm absorbers are found in lower Eddington ratios. This Eddington ratio division may be explained in the context of winds which escape the system and intercept the line of sight at low $\lambda_{\rm Edd}$, but which become over-ionised failed winds resulting in the formation of a partial covering absorber at intermediate $\lambda_{\rm Edd}$ or the warm corona at high $\lambda_{\rm Edd}$. However, it is difficult to confirm these findings or to completely explain their physical interpretation using eFEDS alone. Using the results from this work and repeating this analysis using the all-sky survey (e.g. eRASS:1, eRASS:4) will provide a large sample of $\sim1000$s of AGN for enhanced analysis. Furthermore, using data with very high spectral resolution (e.g. XRISM and Athena) will help to confirm these results, and to create a more complete picture of the nature of warm absorbers and soft excesses in AGN in the local Universe.

\begin{acknowledgements}
We thank the referee for their careful reading of this manuscript and for their very helpful comments and suggestions which improved this work.
\\
This work is based on data from eROSITA, the soft X-ray instrument onboard SRG, a joint Russian-German science mission supported by the Russian Space Agency (Roskosmos), in the interests of the Russian Academy of Sciences represented by its Space Research Institute (IKI), and the Deutsches Zentrum f\"ur Luft- und Raumfahrt (DLR). The SRG spacecraft was built by Lavochkin Association (NPOL) and its subcontractors, and is operated by NPOL with support from the Max-Planck Institute for Extraterrestrial Physics (MPE). The development and construction of the eROSITA X-ray instrument was led by MPE, with contributions from the Dr. Karl Remeis Observatory Bamberg \& ECAP (FAU Erlangen-Nuernberg), the University of Hamburg Observatory, the Leibniz Institute for Astrophysics Potsdam (AIP), and the Institute for Astronomy and Astrophysics of the University of T\"ubingen, with the support of DLR and the Max Planck Society. The Argelander Institute for Astronomy of the University of Bonn and the Ludwig Maximilians Universit\"at Munich also participated in the science preparation for eROSITA. The eROSITA data shown here were processed using the eSASS/NRTA software system developed by the German eROSITA consortium.
\\
The Hyper Suprime-Cam (HSC) collaboration includes the astronomical communities of Japan and Taiwan, and Princeton University.  The HSC instrumentation and software were developed by the National Astronomical Observatory of Japan (NAOJ), the Kavli Institute for the Physics and Mathematics of the Universe (Kavli IPMU), the University of Tokyo, the High Energy Accelerator Research Organization (KEK), the Academia Sinica Institute for Astronomy and Astrophysics in Taiwan (ASIAA), and Princeton University.  Funding was contributed by the FIRST program from Japanese Cabinet Office, the Ministry of Education, Culture, Sports, Science and Technology (MEXT), the Japan Society for the Promotion of Science (JSPS), Japan Science and Technology Agency (JST),the Toray Science Foundation, NAOJ, Kavli IPMU, KEK,ASIAA, and Princeton University.
\\
Funding for the Sloan Digital Sky Survey IV has been provided by the Alfred P. Sloan Foundation, the U.S. Department of Energy Office of Science, and the Participating Institutions. SDSS acknowledges support and resources from the Center for High-Performance Computing at the University of Utah. The SDSS web site is www.sdss.org.

SDSS is managed by the Astrophysical Research Consortium for the Participating Institutions of the SDSS Collaboration including the Brazilian Participation Group, the Carnegie Institution for Science, Carnegie Mellon University, Center for Astrophysics | Harvard \& Smithsonian (CfA), the Chilean Participation Group, the French Participation Group, Instituto de Astrofísica de Canarias, The Johns Hopkins University, Kavli Institute for the Physics and Mathematics of the Universe (IPMU) / University of Tokyo, the Korean Participation Group, Lawrence Berkeley National Laboratory, Leibniz Institut für Astrophysik Potsdam (AIP), Max-Planck-Institut für Astronomie (MPIA Heidelberg), Max-Planck-Institut für Astrophysik (MPA Garching), Max-Planck-Institut für Extraterrestrische Physik (MPE), National Astronomical Observatories of China, New Mexico State University, New York University, University of Notre Dame, Observatório Nacional / MCTI, The Ohio State University, Pennsylvania State University, Shanghai Astronomical Observatory, United Kingdom Participation Group, Universidad Nacional Autónoma de México, University of Arizona, University of Colorado Boulder, University of Oxford, University of Portsmouth, University of Utah, University of Virginia, University of Washington, University of Wisconsin, Vanderbilt University, and Yale University.
\\
Funding for the Sloan Digital Sky Survey (SDSS) has been provided by the Alfred P. Sloan Foundation, the Participating Institutions, the National Aeronautics and Space Administration, the National Science Foundation, the US Department of Energy, the Japanese Monbukagakusho, and the Max Planck Society. The SDSS Web site is http://www.sdss.org/. The SDSS is managed by the Astrophysical Research Consortium (ARC) for the Participating Institutions. The Participating Institutions are The University of Chicago, Fermilab, the Institute for Advanced Study, the Japan Participation Group, The Johns Hopkins University, Los Alamos National Laboratory, the Max-Planck-Institute for Astronomy (MPIA), the Max-Planck-Institute for Astrophysics (MPA), New Mexico State University, University of Pittsburgh, Princeton University, the United States Naval Observatory, and the University of Washington.
\\
MB acknowledges support from the European Innovative Training Network (ITN) “BiD4BEST” funded by the Marie Sklodowska-Curie Actions in Horizon 2020 (GA 860744).
\end{acknowledgements}

%
\bibliographystyle{aa} 
\bibliography{efeds_sewa} 
%

\appendix

\section{Simulations to assess Bayes factor thresholds}
In order to determine which values of the Bayes factors for each model correspond to a significant improvement to the fit, simulations are performed. Here, 1000 spectra are generated using the fakeit command in XSPEC. Spectral simulations are based on an absorbed power law model (\texttt{tbabs} $\times$ \texttt{ztbabs} $\times$ \texttt{powerlaw}) using the average spectral properties from the absorbed power law (PL) model, with a photon index of $\Gamma \sim2$ and a host galaxy column density of $\sim2\times10^{20}\pscm$. In this way, the simulated spectra are representative of the average eFEDS spectrum and can be used to calibrate the Bayes factor threshholds for the entire sample.

\begin{figure}
   \centering
    \includegraphics[width=44mm]{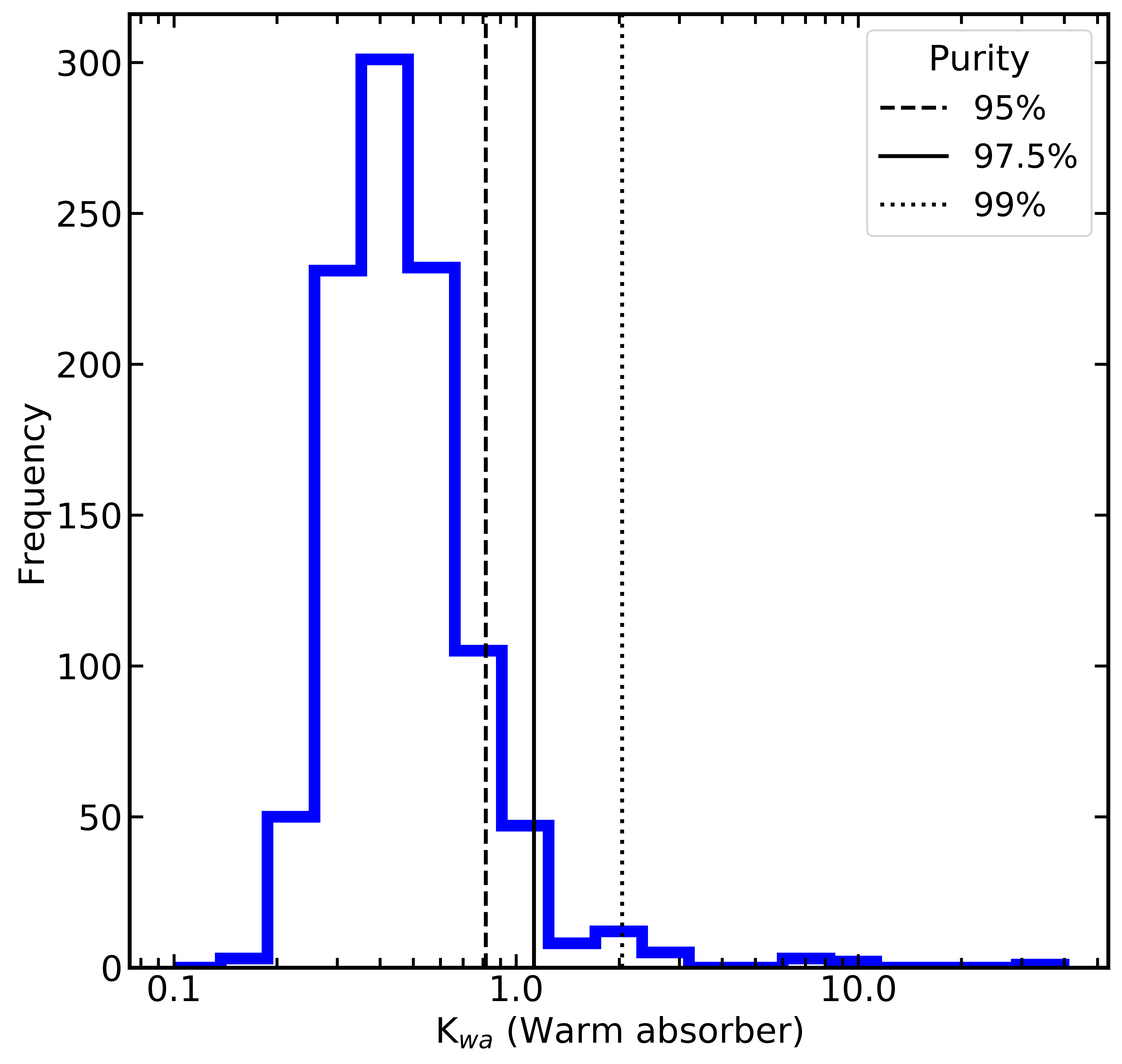}
    \includegraphics[width=44mm]{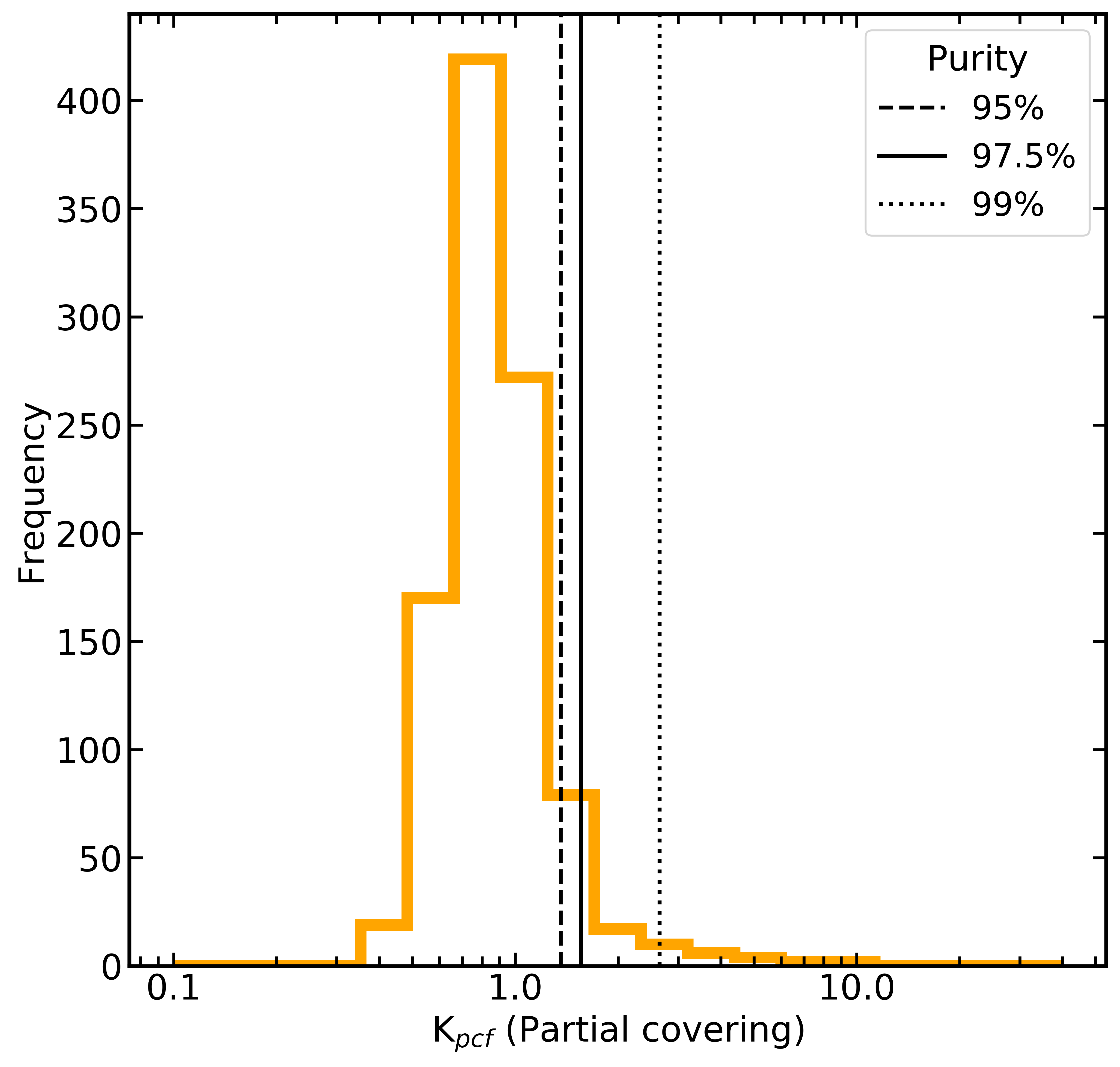}
    \caption{Distributions of Bayes factors obtained from simulations. Left: Computed warm absorber Bayes factors (K$_{wa}$). Right: computed partial covering Bayes factors (K$_{pcf}$). Confidence levels of 95\%, 97.5\% and 99\% are indicated using vertical lines of different linestyles.  }
    \label{fig:warmabs_k}

\end{figure}

First, to determine the significance of the absorber models, each simulated spectrum is then fit with the baseline (PL), warm absorber (PL+WA), and partial covering (PL+PCF) models, and the Bayes factor is computed for each model using equation~\ref{eqn:kbayes}. Formulae for the Bayes factor for all models are given in Table~\ref{tab:bayes}. The resulting histograms are shown in Fig.~\ref{fig:warmabs_k}, with results for warm absorption modelling shown in the left-hand panel and results for neutral partial covering shown on the right. Purity thresholds (the confidence per object that the absorption component is real) levels of 95\%, 97.5\% and 99\% are indicated using vertical lines with dashed, solid and dotted linestyles, respectively. These Bayes factor purity thresholds are selected such that cases where if the model both has the lowest Bayesian evidence of all models, and the Bayes factor exceeds the threshold, it is considered a false detection. 

\begin{table}
\caption{Summary of detection thresholds for complex absorbers. Column (1) gives the percentile of sources, and column (2) gives the corresponding Bayes factor purity threshhold. Column (3) gives the resulting assumed number of false detections, obtained from multiplying the purity by the total number of sources (199). Column (4) gives the number of sources in the sample with K$_{wa}$ values greater than the value given in column (2) and where the model provides the best fit to the data. Column (5) gives the assumed number of true detections in the eFEDS hard X-ray--selected sample of 200 sources.}
\label{tab:warmabs}
\resizebox{\columnwidth}{!}{%
\begin{tabular}{lcccc}
\hline
(1) & (2) & (3) & (4) & (5) \\
Percentile & K & Expected & Expected & True detections \\
 &  & false detections & Sources &  \\
\hline
\multicolumn{5}{c}{Warm absorption}  \\
\hline
95\% & 0.815 & 10 & 36 & 26 \\
97.5\% & 1.126 & 5 & 29 & 24 \\
99\% & 2.040 & 2 & 15 & 13 \\
\hline
\multicolumn{5}{c}{Partial covering}  \\
\hline
95\% & 1.360 & 10 & 28 & 18 \\
97.5\% & 1.555 & 5 & 25 & 20 \\
99\% & 2.646 & 2 & 12 & 10\\
\hline
\hline
\end{tabular}
}
\end{table}

By applying these purity thresholds to the eFEDS hard X-ray sample, the number of expected false detections of an absorption component at each confidence level can be predicted. These results are summarised in Table~\ref{tab:warmabs}. The percentile gives the percentage of simulated sources with Bayes factors below the quoted value, and the false detection rate gives the corresponding number of false detections. The number of sources is the number of real sources with Bayes factors above the stated value and where the given model provides the best fit, and the number of true detections is calculated by subtracting the expected number of false detections in a sample of 200 sources. Since evidence comparison is used to determine which of the complex models provides the best fit to the data, scenarios where a source appears in multiple sub-samples are excluded. The 97.5\% purity thresholds are taken as the samples of sources with warm absorbers or partial covering components respectively.

\begin{figure}
   \centering
    \includegraphics[width=0.95\columnwidth]{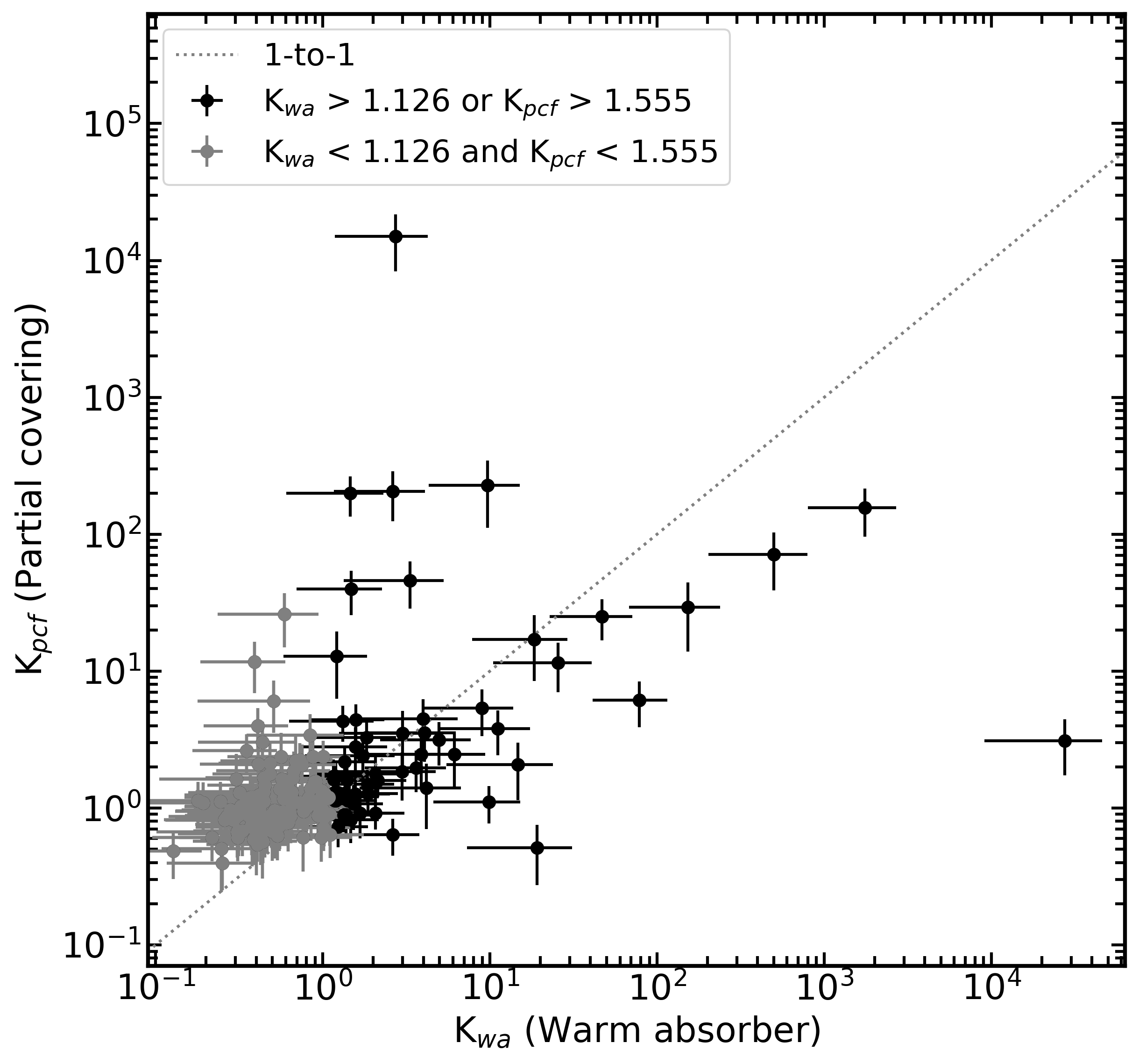}
    \caption{Comparison of Bayes factors between warm absorber models. The Bayes factor for the warm absorption model is shown on the horizontal axis, while the Bayes factor for the neutral partial covering model is shown on the vertical axis. The dashed grey line shows the one-to-one relation, with sources lying below the line being better fit with the PL+WA model, and sources above the line being better fit with the PL+PCF model. Sources which do not show evidence for any complex absorption are shown in grey.The values of K$_{wa}$ = 1.126 and K$_{pcf}$ = 1.555 correspond to purities of 97.5\%.  }
    \label{fig:wavspcf}

\end{figure}

The Bayes factors obtained from fitting the eFEDS spectra with the two complex absorption models are compared in Fig.~\ref{fig:wavspcf}, with error bars are shown for each source. One source, ID 00011, is excluded as the K$_{wa}$ and K$_{pcf}$ values are many orders of magnitude larger than the rest of the sample (see tables in the appendix). There is clear evidence both for sources which are very well fit with neutral partial covering and with ionised warm absorption, confirming that both models should be included in the analysis in order to properly characterise the shape of the observed X-ray spectra. 

\begin{figure}
   \centering
    \includegraphics[width=0.95\columnwidth]{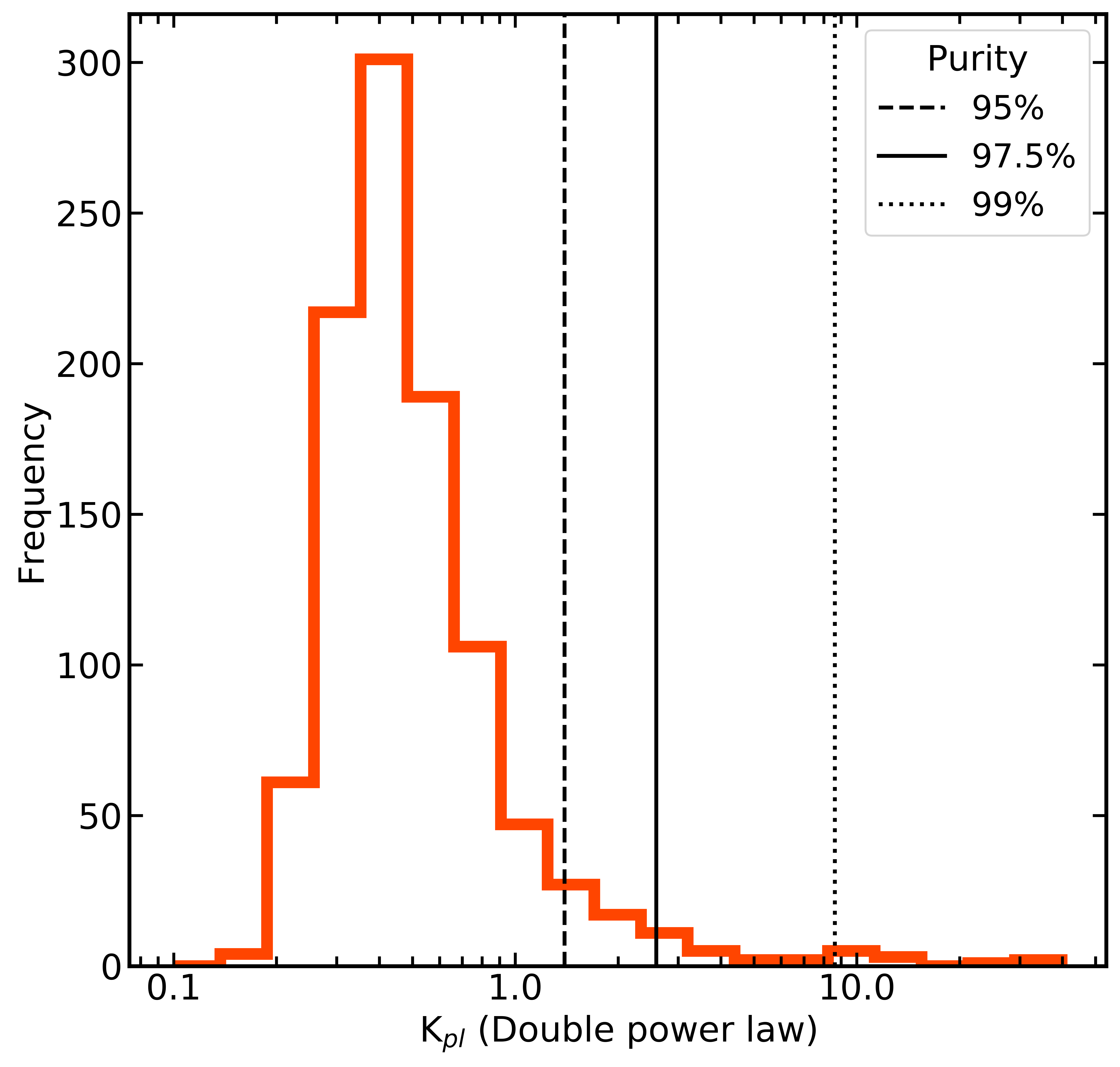}
    \caption{Distribution of computed K$_{pl}$ values obtained from simulations. Confidence levels of 95\%, 97.5\% and 99\% are indicated using vertical lines of different linestyles.  }
    \label{fig:plpl_k}

\end{figure}

From the computed PL+PL Bayes factors, to determine which sources have significant soft excesses, the same method employed to determine the significance of warm absorbers and partial covering components is used. The same 1000 simulated spectra are used, but are now fit with the PL+PL model, and Bayes factors are again computed. The resulting histogram of K$_{pl}$ values is shown in Fig.~\ref{fig:plpl_k}. Confidence levels of 95\%, 97.5\% and 99\% are indicated using vertical lines with solid, dashed, dotted and dash-dot linestyles, respectively. In general, a higher threshold is required to distinguish a soft excess than to distinguish complex absorption. The false detection rates are again calculated and are summarised in Table~\ref{tab:plpl}.

\begin{table}
\caption{Summary of detection thresholds for soft excesses. Column (1) gives the percentile of sources, and column (2) gives the corresponding K$_{pl}$ values. Column (3) gives expected number of false detections. Column (4) gives the number of sources in the sample with K$_{pl}$ values greater than the value given in column (2) and where the model provides the best fit to the data. Column (5) gives the assumed number of true detections in the eFEDS hard X-ray--selected sample of 200 sources.}
\label{tab:plpl}
\resizebox{\columnwidth}{!}{%
\begin{tabular}{lcccc}
\hline
(1) & (2) & (3) & (4) & (5) \\
Percentile & K & Expected & Expected & True detections \\
 &  & false detections & Sources &  \\
 \hline
95\% & 1.392 & 10 & 40 & 30 \\
97.5\% & 2.586 & 5 & 29 & 24 \\
99\% & 8.613 & 2 & 20 & 18 \\
\hline
\hline
\end{tabular}
}
\end{table}

\section{Spectral fit parameters}
The tables below give the spectral fit parameters for the models presented in this work. table.~\ref{tab:apppl} gives parameters for the baseline (PL) fit for all sources. table.~\ref{tab:appwa} gives parameters for the warm absorber (PL+WA) model for sources with $>95\%$ significance. table.~\ref{tab:apppcf} gives  parameters for the partial covering absorber (PL+PCF) model for sources with $>95\%$ significance. table.~\ref{tab:appse} gives  parameters for the soft excess (PL+PL) model for sources with $>95\%$ significance. Spectral fit parameters for soft excess sources best fit with a warm corona are given in table.~\ref{tab:appnth}, and parameters for sources best fit with blurred reflection are given in Table~\ref{tab:apprel}.

\newpage

\renewcommand{\arraystretch}{1.3}
\begin{table*}[]
\caption{Summary of warm absorber parameters for sources with significant warm absorbers ($>95\%$). Column (1) gives the eROSITA name, column (2) gives the source ID (ID\_SRC in the catalog, used throughout this work for source identification), column (3) gives the warm absorber column density, column (4) gives the warm absorber ionisation, column (5) gives the photon index, column (6) gives the logarithm of the $2-10\kev$ flux measured from the absorbed power law model in cgs units, and column (7) gives the best-fit model (where PL is the absorbed power law model, SE is the soft excess model, PCF is the partial covering absorber model, and WA is the warm absorber model). }
\label{tab:apppl}
\centering
\begin{tabular}{ccccccc}
\hline
(1) & (2) & (3) & (4) & (5) & (6) & (7) \\
Name & ID & nH$_{gal}$ & log(NH$_z$) & $\Gamma$ & log(F) & Model \\
& & ($\times10^{20}\pscm$) & & & \fluxcgs & \\
\hline
eFEDS J093701.0+010545 & 00001  &  4.95  &  20.03 $_{- 0.02 } ^{+ 0.05 }$ &  2.69 $_{- 0.03 } ^{+ 0.03 }$ & -11.67 & SE \\
eFEDS J085617.9-013806 & 00003  &  2.08  &  20.02 $_{- 0.02 } ^{+ 0.03 }$ &  2.92 $_{- 0.03 } ^{+ 0.03 }$ & -12.14 & SE \\
eFEDS J084025.5+033303 & 00004  &  3.18  &  20.02 $_{- 0.02 } ^{+ 0.03 }$ &  2.33 $_{- 0.03 } ^{+ 0.03 }$ & -11.77 & SE \\
eFEDS J091702.3-004416 & 00007  &  3.53  &  20.06 $_{- 0.04 } ^{+ 0.08 }$ &  2.35 $_{- 0.04 } ^{+ 0.04 }$ & -12.10 & SE \\
eFEDS J091401.8+050750 & 00008  &  4.06  &  20.14 $_{- 0.10 } ^{+ 0.15 }$ &  2.10 $_{- 0.04 } ^{+ 0.05 }$ & -11.78 & PL \\
eFEDS J084303.0+030218 & 00011  &  3.71  &  20.04 $_{- 0.03 } ^{+ 0.05 }$ &  2.99 $_{- 0.01 } ^{+ 0.01 }$ & -12.64 & SE \\
eFEDS J091420.1+013748 & 00014  &  2.66  &  20.08 $_{- 0.06 } ^{+ 0.11 }$ &  2.27 $_{- 0.06 } ^{+ 0.05 }$ & -12.15 & PCF \\
eFEDS J083356.4-010113 & 00016  &  3.91  &  20.07 $_{- 0.05 } ^{+ 0.10 }$ &  1.87 $_{- 0.06 } ^{+ 0.05 }$ & -11.90 & WA \\
eFEDS J092414.7+030860 & 00023  &  3.81  &  20.11 $_{- 0.08 } ^{+ 0.13 }$ &  2.06 $_{- 0.06 } ^{+ 0.06 }$ & -12.13 & PCF \\
eFEDS J090910.2+012136 & 00025  &  3.30  &  20.37 $_{- 0.25 } ^{+ 0.32 }$ &  1.77 $_{- 0.07 } ^{+ 0.07 }$ & -11.98 & PL \\
eFEDS J093117.8+033121 & 00028  &  3.20  &  20.05 $_{- 0.04 } ^{+ 0.08 }$ &  1.92 $_{- 0.06 } ^{+ 0.07 }$ & -12.11 & WA \\
eFEDS J083702.1+024536 & 00029  &  3.83  &  20.06 $_{- 0.05 } ^{+ 0.09 }$ &  2.56 $_{- 0.07 } ^{+ 0.07 }$ & -12.56 & SE \\
eFEDS J091333.6-004250 & 00030  &  3.11  &  20.12 $_{- 0.08 } ^{+ 0.16 }$ &  2.22 $_{- 0.07 } ^{+ 0.07 }$ & -12.34 & PCF \\
eFEDS J093518.2+020415 & 00032  &  4.10  &  20.26 $_{- 0.19 } ^{+ 0.28 }$ &  2.01 $_{- 0.07 } ^{+ 0.08 }$ & -12.19 & PL \\
eFEDS J090446.0+020843 & 00033  &  4.12  &  20.90 $_{- 0.40 } ^{+ 0.23 }$ &  2.01 $_{- 0.11 } ^{+ 0.14 }$ & -12.25 & PL \\
eFEDS J090352.1-005353 & 00034  &  2.82  &  20.06 $_{- 0.04 } ^{+ 0.08 }$ &  2.31 $_{- 0.07 } ^{+ 0.07 }$ & -12.42 & SE \\
eFEDS J084110.8+022951 & 00035  &  4.60  &  20.13 $_{- 0.09 } ^{+ 0.16 }$ &  2.48 $_{- 0.08 } ^{+ 0.08 }$ & -12.52 & SE \\
eFEDS J085624.3-003644 & 00037  &  3.22  &  20.11 $_{- 0.08 } ^{+ 0.14 }$ &  2.08 $_{- 0.07 } ^{+ 0.07 }$ & -12.26 & WA \\
eFEDS J083834.1-002440 & 00038  &  4.05  &  20.12 $_{- 0.09 } ^{+ 0.15 }$ &  2.16 $_{- 0.08 } ^{+ 0.08 }$ & -12.34 & SE \\
eFEDS J093714.1+031536 & 00039  &  3.01  &  20.07 $_{- 0.05 } ^{+ 0.10 }$ &  2.97 $_{- 0.04 } ^{+ 0.02 }$ & -12.94 & SE \\
eFEDS J093303.4+045236 & 00041  &  3.41  &  20.28 $_{- 0.19 } ^{+ 0.24 }$ &  2.24 $_{- 0.08 } ^{+ 0.09 }$ & -12.44 & PL \\
eFEDS J085900.1-020028 & 00042  &  2.01  &  20.15 $_{- 0.11 } ^{+ 0.17 }$ &  2.14 $_{- 0.08 } ^{+ 0.08 }$ & -12.56 & PL \\
eFEDS J085240.1+051216 & 00043  &  4.35  &  20.15 $_{- 0.10 } ^{+ 0.18 }$ &  2.60 $_{- 0.09 } ^{+ 0.09 }$ & -12.58 & PCF \\ 
eFEDS J091250.7+035221 & 00044  &  3.56  &  20.16 $_{- 0.12 } ^{+ 0.20 }$ &  2.00 $_{- 0.08 } ^{+ 0.08 }$ & -12.28 & PL \\
eFEDS J091846.9-005929 & 00045  &  3.73  &  20.10 $_{- 0.08 } ^{+ 0.13 }$ &  2.37 $_{- 0.08 } ^{+ 0.08 }$ & -12.63 & SE \\
eFEDS J090743.4+013327 & 00048  &  3.61  &  20.92 $_{- 0.26 } ^{+ 0.19 }$ &  1.93 $_{- 0.17 } ^{+ 0.19 }$ & -12.18 & PL \\
eFEDS J085915.7+011801 & 00050  &  4.59  &  20.24 $_{- 0.17 } ^{+ 0.25 }$ &  2.56 $_{- 0.10 } ^{+ 0.11 }$ & -12.71 & PL \\
eFEDS J085904.0+020503 & 00051  &  4.18  &  20.16 $_{- 0.11 } ^{+ 0.18 }$ &  2.53 $_{- 0.09 } ^{+ 0.10 }$ & -12.69 & PL \\
eFEDS J091459.6+023512 & 00052  &  2.42  &  20.14 $_{- 0.10 } ^{+ 0.15 }$ &  2.61 $_{- 0.08 } ^{+ 0.09 }$ & -12.80 & PL \\
eFEDS J084625.1+044110 & 00054  &  4.02  &  20.10 $_{- 0.07 } ^{+ 0.12 }$ &  2.51 $_{- 0.09 } ^{+ 0.09 }$ & -12.51 & SE \\
eFEDS J091437.9+024558 & 00055  &  2.50  &  20.47 $_{- 0.31 } ^{+ 0.28 }$ &  2.00 $_{- 0.10 } ^{+ 0.12 }$ & -12.34 & PL \\
eFEDS J093240.5+023333 & 00056  &  3.96  &  20.18 $_{- 0.13 } ^{+ 0.18 }$ &  1.87 $_{- 0.10 } ^{+ 0.09 }$ & -12.25 & PL \\
eFEDS J093434.0+030645 & 00057  &  3.06  &  20.16 $_{- 0.12 } ^{+ 0.18 }$ &  2.06 $_{- 0.09 } ^{+ 0.09 }$ & -12.38 & SE \\
eFEDS J083535.5-014020 & 00060  &  3.30  &  21.71 $_{- 0.08 } ^{+ 0.07 }$ &  1.90 $_{- 0.21 } ^{+ 0.23 }$ & -12.04 & SE \\
eFEDS J084622.7+031324 & 00064  &  3.55  &  20.15 $_{- 0.11 } ^{+ 0.17 }$ &  2.09 $_{- 0.09 } ^{+ 0.10 }$ & -12.37 & PL \\
eFEDS J091232.9+001654 & 00067  &  2.63  &  20.24 $_{- 0.17 } ^{+ 0.25 }$ &  2.11 $_{- 0.09 } ^{+ 0.10 }$ & -12.45 & PL \\
eFEDS J091247.5-012125 & 00070  &  3.19  &  20.18 $_{- 0.13 } ^{+ 0.20 }$ &  2.47 $_{- 0.09 } ^{+ 0.10 }$ & -12.74 & SE \\
eFEDS J092544.6+001529 & 00071  &  3.15  &  20.22 $_{- 0.15 } ^{+ 0.22 }$ &  2.13 $_{- 0.10 } ^{+ 0.11 }$ & -12.50 & PCF \\
eFEDS J090821.0+045059 & 00075  &  3.68  &  20.66 $_{- 0.43 } ^{+ 0.33 }$ &  1.57 $_{- 0.11 } ^{+ 0.14 }$ & -12.06 & WA \\
eFEDS J092851.8+041629 & 00076  &  3.37  &  20.20 $_{- 0.14 } ^{+ 0.23 }$ &  2.38 $_{- 0.09 } ^{+ 0.10 }$ & -12.63 & SE \\
eFEDS J090915.9+035442 & 00078  &  3.23  &  21.17 $_{- 0.74 } ^{+ 0.72 }$ &  1.19 $_{- 0.10 } ^{+ 0.11 }$ & -11.86 & WA \\
eFEDS J083110.5-010006 & 00080  &  4.19  &  20.29 $_{- 0.20 } ^{+ 0.25 }$ &  2.33 $_{- 0.11 } ^{+ 0.12 }$ & -12.63 & PL \\
\hline
\hline
\end{tabular}
\twocolumn
\end{table*}

\renewcommand{\arraystretch}{1.3}
\begin{table*}[]
\centering
\begin{tabular}{ccccccc}
\hline
(1) & (2) & (3) & (4) & (5) & (6) & (7) \\
Name & ID & nH$_{gal}$ & log(NH$_z$) & $\Gamma$ & log(F) & Model \\
& & ($\times10^{20}\pscm$) & & & \fluxcgs & \\
\hline
eFEDS J092918.3+041933 & 00082  &  3.39  &  20.26 $_{- 0.18 } ^{+ 0.24 }$ &  2.23 $_{- 0.10 } ^{+ 0.11 }$ & -12.54 & PL \\
eFEDS J083402.6-014846 & 00101  &  3.37  &  20.50 $_{- 0.34 } ^{+ 0.34 }$ &  1.91 $_{- 0.11 } ^{+ 0.14 }$ & -12.59 & PL \\
eFEDS J085259.2+031322 & 00105  &  3.68  &  20.21 $_{- 0.15 } ^{+ 0.22 }$ &  2.51 $_{- 0.11 } ^{+ 0.12 }$ & -12.72 & WA \\
eFEDS J083310.4+041036 & 00107  &  2.77  &  20.22 $_{- 0.15 } ^{+ 0.23 }$ &  1.60 $_{- 0.11 } ^{+ 0.11 }$ & -12.18 & WA \\
eFEDS J091609.5+000019 & 00108  &  2.86  &  20.26 $_{- 0.18 } ^{+ 0.28 }$ &  2.41 $_{- 0.11 } ^{+ 0.11 }$ & -12.80 & PCF \\
eFEDS J092656.7+041613 & 00109  &  3.47  &  20.36 $_{- 0.26 } ^{+ 0.36 }$ &  2.10 $_{- 0.11 } ^{+ 0.11 }$ & -12.55 & PL \\
eFEDS J090826.8+002136 & 00110  &  3.18  &  20.23 $_{- 0.16 } ^{+ 0.24 }$ &  2.02 $_{- 0.10 } ^{+ 0.11 }$ & -12.49 & PL \\
eFEDS J083949.6+010428 & 00112  &  4.96  &  20.57 $_{- 0.40 } ^{+ 0.43 }$ &  1.60 $_{- 0.11 } ^{+ 0.11 }$ & -12.20 & PL \\
eFEDS J093814.9+020023 & 00113  &  4.06  &  20.32 $_{- 0.23 } ^{+ 0.24 }$ &  2.62 $_{- 0.13 } ^{+ 0.14 }$ & -12.72 & PL \\
eFEDS J085451.7+012611 & 00114  &  4.07  &  20.37 $_{- 0.27 } ^{+ 0.32 }$ &  2.14 $_{- 0.11 } ^{+ 0.13 }$ & -12.57 & PL \\
eFEDS J091405.7-000009 & 00117  &  2.68  &  20.18 $_{- 0.13 } ^{+ 0.19 }$ &  2.53 $_{- 0.11 } ^{+ 0.11 }$ & -12.90 & SE \\
eFEDS J091437.4+012622 & 00118  &  2.69  &  20.31 $_{- 0.21 } ^{+ 0.27 }$ &  2.05 $_{- 0.12 } ^{+ 0.13 }$ & -12.56 & PL \\
eFEDS J085554.3+005111 & 00120  &  4.89  &  20.26 $_{- 0.18 } ^{+ 0.27 }$ &  1.30 $_{- 0.11 } ^{+ 0.11 }$ & -12.00 & PCF \\
eFEDS J090457.9+010423 & 00121  &  4.01  &  20.15 $_{- 0.11 } ^{+ 0.18 }$ &  2.41 $_{- 0.11 } ^{+ 0.12 }$ & -12.83 & SE \\
eFEDS J084138.5+014221 & 00122  &  5.60  &  20.29 $_{- 0.20 } ^{+ 0.29 }$ &  2.36 $_{- 0.11 } ^{+ 0.12 }$ & -12.73 & SE \\
eFEDS J091412.6-021348 & 00123  &  3.84  &  20.37 $_{- 0.26 } ^{+ 0.32 }$ &  2.20 $_{- 0.11 } ^{+ 0.12 }$ & -12.92 & PL \\
eFEDS J091459.0+012630 & 00125  &  2.69  &  20.47 $_{- 0.31 } ^{+ 0.29 }$ &  1.95 $_{- 0.13 } ^{+ 0.15 }$ & -12.48 & PL \\
eFEDS J090915.8-011632 & 00130  &  2.70  &  20.39 $_{- 0.26 } ^{+ 0.31 }$ &  1.95 $_{- 0.12 } ^{+ 0.13 }$ & -12.50 & PL \\
eFEDS J085907.6+044436 & 00132  &  4.02  &  20.33 $_{- 0.23 } ^{+ 0.28 }$ &  2.37 $_{- 0.14 } ^{+ 0.14 }$ & -12.83 & WA \\
eFEDS J093506.3+034514 & 00133  &  3.06  &  20.53 $_{- 0.35 } ^{+ 0.36 }$ &  1.91 $_{- 0.13 } ^{+ 0.17 }$ & -12.50 & PL \\
eFEDS J091459.2+050243 & 00135  &  4.04  &  20.50 $_{- 0.32 } ^{+ 0.30 }$ &  1.68 $_{- 0.12 } ^{+ 0.16 }$ & -12.23 & PL \\
eFEDS J092008.1+032245 & 00141  &  3.63  &  20.25 $_{- 0.18 } ^{+ 0.26 }$ &  2.19 $_{- 0.12 } ^{+ 0.13 }$ & -12.68 & PL \\
eFEDS J083029.0+024304 & 00142  &  4.25  &  20.34 $_{- 0.23 } ^{+ 0.31 }$ &  2.55 $_{- 0.15 } ^{+ 0.16 }$ & -12.70 & PL \\
eFEDS J083911.2-020229 & 00143  &  2.94  &  20.37 $_{- 0.24 } ^{+ 0.28 }$ &  2.03 $_{- 0.11 } ^{+ 0.13 }$ & -12.54 & PL \\
eFEDS J084257.6+030841 & 00145  &  3.54  &  20.14 $_{- 0.10 } ^{+ 0.16 }$ &  2.03 $_{- 0.11 } ^{+ 0.11 }$ & -12.56 & SE \\
eFEDS J085301.3-015050 & 00148  &  2.03  &  20.39 $_{- 0.28 } ^{+ 0.36 }$ &  1.47 $_{- 0.11 } ^{+ 0.11 }$ & -12.39 & SE \\
eFEDS J090816.6+052012 & 00151  &  3.84  &  20.31 $_{- 0.22 } ^{+ 0.29 }$ &  1.74 $_{- 0.13 } ^{+ 0.14 }$ & -12.38 & PL \\
eFEDS J090921.6+003608 & 00153  &  3.05  &  20.15 $_{- 0.11 } ^{+ 0.18 }$ &  2.26 $_{- 0.12 } ^{+ 0.12 }$ & -12.76 & SE \\
eFEDS J092725.2+024348 & 00156  &  4.40  &  20.79 $_{- 0.51 } ^{+ 0.40 }$ &  2.11 $_{- 0.19 } ^{+ 0.24 }$ & -12.75 & PL \\
eFEDS J085944.7-015645 & 00167  &  2.07  &  20.20 $_{- 0.15 } ^{+ 0.22 }$ &  1.83 $_{- 0.12 } ^{+ 0.12 }$ & -12.49 & WA \\
eFEDS J083723.1+001724 & 00168  &  4.63  &  20.69 $_{- 0.46 } ^{+ 0.44 }$ &  1.81 $_{- 0.14 } ^{+ 0.16 }$ & -12.46 & PL \\
eFEDS J084535.4+001621 & 00169  &  3.35  &  20.42 $_{- 0.29 } ^{+ 0.33 }$ &  1.90 $_{- 0.13 } ^{+ 0.15 }$ & -12.51 & PL \\
eFEDS J092759.5-002905 & 00175  &  3.05  &  20.23 $_{- 0.17 } ^{+ 0.25 }$ &  2.25 $_{- 0.13 } ^{+ 0.13 }$ & -12.81 & PL \\
eFEDS J093912.7+033956 & 00176  &  3.33  &  20.26 $_{- 0.18 } ^{+ 0.29 }$ &  2.14 $_{- 0.13 } ^{+ 0.13 }$ & -12.72 & SE \\
eFEDS J093826.3+022037 & 00184  &  3.60  &  20.29 $_{- 0.20 } ^{+ 0.29 }$ &  1.94 $_{- 0.12 } ^{+ 0.13 }$ & -12.43 & PCF \\
eFEDS J091551.6+044732 & 00192  &  4.04  &  20.33 $_{- 0.23 } ^{+ 0.32 }$ &  1.92 $_{- 0.14 } ^{+ 0.16 }$ & -12.60 & PL \\
eFEDS J084833.9+025312 & 00193  &  3.29  &  20.47 $_{- 0.32 } ^{+ 0.40 }$ &  1.81 $_{- 0.12 } ^{+ 0.15 }$ & -12.52 & PL \\
eFEDS J090610.1-004553 & 00194  &  2.89  &  20.24 $_{- 0.17 } ^{+ 0.25 }$ &  1.98 $_{- 0.12 } ^{+ 0.14 }$ & -12.62 & PCF \\
eFEDS J092723.6-014121 & 00195  &  2.94  &  20.36 $_{- 0.25 } ^{+ 0.32 }$ &  2.35 $_{- 0.15 } ^{+ 0.16 }$ & -13.05 & PL \\
eFEDS J084759.9-003333 & 00200  &  2.79  &  20.19 $_{- 0.14 } ^{+ 0.22 }$ &  2.22 $_{- 0.13 } ^{+ 0.14 }$ & -12.69 & SE \\
eFEDS J091034.3+031329 & 00204  &  3.11  &  20.19 $_{- 0.14 } ^{+ 0.22 }$ &  2.10 $_{- 0.13 } ^{+ 0.15 }$ & -12.72 & SE \\
eFEDS J091624.8+040944 & 00206  &  3.72  &  20.25 $_{- 0.18 } ^{+ 0.25 }$ &  2.19 $_{- 0.14 } ^{+ 0.14 }$ & -12.81 & WA \\
eFEDS J085439.4+004143 & 00213  &  4.86  &  20.32 $_{- 0.23 } ^{+ 0.32 }$ &  1.79 $_{- 0.13 } ^{+ 0.14 }$ & -12.48 & PL \\
eFEDS J090831.9+045105 & 00214  &  3.70  &  20.24 $_{- 0.18 } ^{+ 0.26 }$ &  1.76 $_{- 0.14 } ^{+ 0.13 }$ & -12.49 & PCF \\
eFEDS J084131.6-005217 & 00216  &  3.24  &  20.16 $_{- 0.12 } ^{+ 0.18 }$ &  2.33 $_{- 0.13 } ^{+ 0.15 }$ & -12.87 & SE \\
eFEDS J085004.0+032642 & 00217  &  3.56  &  20.89 $_{- 0.44 } ^{+ 0.30 }$ &  1.67 $_{- 0.23 } ^{+ 0.32 }$ & -12.38 & PL \\
\hline
\hline
\end{tabular}
\twocolumn
\end{table*}

\renewcommand{\arraystretch}{1.3}
\begin{table*}[]
\centering
\begin{tabular}{ccccccc}
\hline
(1) & (2) & (3) & (4) & (5) & (6) & (7) \\
Name & ID & nH$_{gal}$ & log(NH$_z$) & $\Gamma$ & log(F) & Model \\
& & ($\times10^{20}\pscm$) & & & \fluxcgs & \\
\hline
eFEDS J093000.5-011807 & 00223  &  3.05  &  20.38 $_{- 0.27 } ^{+ 0.37 }$ &  1.56 $_{- 0.13 } ^{+ 0.14 }$ & -12.39 & SE \\
eFEDS J085643.1-020804 & 00230  &  1.89  &  20.91 $_{- 0.49 } ^{+ 0.28 }$ &  2.14 $_{- 0.19 } ^{+ 0.25 }$ & -12.95 & PL \\
eFEDS J085224.3+020139 & 00233  &  3.30  &  20.36 $_{- 0.24 } ^{+ 0.34 }$ &  2.13 $_{- 0.16 } ^{+ 0.15 }$ & -12.72 & WA \\
eFEDS J083346.6-011202 & 00235  &  3.82  &  20.30 $_{- 0.21 } ^{+ 0.29 }$ &  2.21 $_{- 0.14 } ^{+ 0.15 }$ & -12.83 & PL \\
eFEDS J093329.6+030811 & 00237  &  3.18  &  20.18 $_{- 0.14 } ^{+ 0.22 }$ &  1.98 $_{- 0.14 } ^{+ 0.14 }$ & -12.71 & SE \\
eFEDS J085559.0+005745 & 00238  &  4.79  &  20.35 $_{- 0.25 } ^{+ 0.33 }$ &  2.19 $_{- 0.15 } ^{+ 0.18 }$ & -12.83 & SE \\
eFEDS J084039.3-011433 & 00245  &  3.28  &  20.73 $_{- 0.48 } ^{+ 0.40 }$ &  1.84 $_{- 0.17 } ^{+ 0.23 }$ & -12.60 & PL \\
eFEDS J093509.2+005729 & 00252  &  4.55  &  20.26 $_{- 0.18 } ^{+ 0.24 }$ &  2.31 $_{- 0.15 } ^{+ 0.16 }$ & -12.91 & PL \\
eFEDS J092814.6+005631 & 00256  &  4.28  &  20.34 $_{- 0.24 } ^{+ 0.32 }$ &  2.38 $_{- 0.16 } ^{+ 0.16 }$ & -12.99 & PL \\
eFEDS J085812.4+010644 & 00263  &  4.53  &  20.60 $_{- 0.36 } ^{+ 0.32 }$ &  1.85 $_{- 0.18 } ^{+ 0.21 }$ & -12.59 & PL \\
eFEDS J090612.0+014412 & 00264  &  3.93  &  20.58 $_{- 0.39 } ^{+ 0.44 }$ &  2.23 $_{- 0.19 } ^{+ 0.22 }$ & -12.94 & PL \\
eFEDS J090933.7-014231 & 00267  &  2.60  &  20.45 $_{- 0.31 } ^{+ 0.36 }$ &  2.42 $_{- 0.15 } ^{+ 0.17 }$ & -13.21 & PL \\
eFEDS J083143.5+013952 & 00268  &  4.59  &  20.81 $_{- 0.51 } ^{+ 0.35 }$ &  2.09 $_{- 0.22 } ^{+ 0.29 }$ & -12.78 & PL \\
eFEDS J092237.4-004751 & 00272  &  3.24  &  20.36 $_{- 0.25 } ^{+ 0.35 }$ &  2.04 $_{- 0.14 } ^{+ 0.15 }$ & -12.74 & PL \\
eFEDS J092738.1+014430 & 00288  &  5.21  &  20.21 $_{- 0.15 } ^{+ 0.24 }$ &  2.27 $_{- 0.15 } ^{+ 0.16 }$ & -12.88 & SE \\
eFEDS J092739.7+053004 & 00297  &  4.29  &  20.60 $_{- 0.39 } ^{+ 0.42 }$ &  2.38 $_{- 0.19 } ^{+ 0.20 }$ & -12.91 & PL \\
eFEDS J090543.3+020716 & 00303  &  3.90  &  20.28 $_{- 0.19 } ^{+ 0.29 }$ &  1.89 $_{- 0.15 } ^{+ 0.17 }$ & -12.73 & PCF \\
eFEDS J092120.9+003628 & 00322  &  2.99  &  20.66 $_{- 0.45 } ^{+ 0.46 }$ &  2.07 $_{- 0.16 } ^{+ 0.18 }$ & -12.84 & PL \\
eFEDS J084310.5-013425 & 00323  &  3.23  &  20.37 $_{- 0.25 } ^{+ 0.36 }$ &  2.11 $_{- 0.15 } ^{+ 0.14 }$ & -12.95 & PCF \\
eFEDS J093259.6+040505 & 00329  &  3.12  &  20.19 $_{- 0.14 } ^{+ 0.22 }$ &  1.26 $_{- 0.14 } ^{+ 0.13 }$ & -12.28 & PCF \\
eFEDS J085536.0-004544 & 00330  &  2.95  &  20.34 $_{- 0.24 } ^{+ 0.31 }$ &  1.91 $_{- 0.15 } ^{+ 0.18 }$ & -12.71 & PL \\
eFEDS J091406.1-021119 & 00337  &  3.82  &  21.89 $_{- 0.14 } ^{+ 0.13 }$ &  1.28 $_{- 0.19 } ^{+ 0.31 }$ & -12.23 & WA \\
eFEDS J085512.8-000733 & 00340  &  3.98  &  20.29 $_{- 0.21 } ^{+ 0.26 }$ &  2.51 $_{- 0.16 } ^{+ 0.18 }$ & -13.14 & SE \\
eFEDS J085552.5+051039 & 00358  &  4.32  &  20.24 $_{- 0.17 } ^{+ 0.26 }$ &  2.22 $_{- 0.18 } ^{+ 0.18 }$ & -12.81 & SE \\
eFEDS J085130.8+044221 & 00364  &  4.56  &  20.43 $_{- 0.31 } ^{+ 0.38 }$ &  2.70 $_{- 0.18 } ^{+ 0.17 }$ & -13.21 & PL \\
eFEDS J083520.2+012517 & 00367  &  4.88  &  20.64 $_{- 0.41 } ^{+ 0.41 }$ &  1.82 $_{- 0.21 } ^{+ 0.23 }$ & -12.65 & PL \\
eFEDS J085814.1+045453 & 00373  &  4.16  &  21.00 $_{- 0.63 } ^{+ 0.49 }$ &  1.80 $_{- 0.18 } ^{+ 0.24 }$ & -12.62 & PL \\
eFEDS J091420.9+031121 & 00378  &  2.82  &  20.43 $_{- 0.29 } ^{+ 0.38 }$ &  1.90 $_{- 0.16 } ^{+ 0.16 }$ & -12.70 & PL \\
eFEDS J084555.5-001255 & 00384  &  3.05  &  20.35 $_{- 0.24 } ^{+ 0.29 }$ &  2.56 $_{- 0.18 } ^{+ 0.19 }$ & -13.18 & PL \\
eFEDS J092331.0-000243 & 00385  &  3.01  &  20.38 $_{- 0.27 } ^{+ 0.37 }$ &  2.17 $_{- 0.17 } ^{+ 0.19 }$ & -12.91 & PL \\
eFEDS J093708.6+012543 & 00389  &  4.92  &  20.30 $_{- 0.21 } ^{+ 0.29 }$ &  2.29 $_{- 0.18 } ^{+ 0.18 }$ & -12.96 & PL \\
eFEDS J092946.4+020352 & 00395  &  5.55  &  21.23 $_{- 0.31 } ^{+ 0.23 }$ &  1.44 $_{- 0.28 } ^{+ 0.42 }$ & -12.31 & PL \\
eFEDS J092107.5+001320 & 00400  &  3.04  &  21.15 $_{- 0.56 } ^{+ 0.36 }$ &  1.18 $_{- 0.13 } ^{+ 0.24 }$ & -12.27 & WA \\
eFEDS J083236.3+044507 & 00420  &  2.78  &  20.94 $_{- 0.46 } ^{+ 0.29 }$ &  1.25 $_{- 0.16 } ^{+ 0.22 }$ & -12.05 & PCF \\
eFEDS J090215.4+032408 & 00426  &  4.30  &  20.20 $_{- 0.15 } ^{+ 0.24 }$ &  2.01 $_{- 0.17 } ^{+ 0.18 }$ & -12.86 & SE \\
eFEDS J083838.6+000552 & 00434  &  4.43  &  21.00 $_{- 0.45 } ^{+ 0.27 }$ &  2.26 $_{- 0.31 } ^{+ 0.40 }$ & -12.98 & PL \\
eFEDS J085447.4-003959 & 00436  &  3.08  &  20.41 $_{- 0.27 } ^{+ 0.33 }$ &  2.34 $_{- 0.19 } ^{+ 0.21 }$ & -13.09 & PL \\
eFEDS J091157.5+014327 & 00439  &  2.88  &  22.42 $_{- 0.13 } ^{+ 0.11 }$ &  2.22 $_{- 0.43 } ^{+ 0.44 }$ & -12.44 & WA \\
eFEDS J083647.5+042214 & 00450  &  3.06  &  20.23 $_{- 0.16 } ^{+ 0.27 }$ &  2.12 $_{- 0.18 } ^{+ 0.19 }$ & -12.97 & SE \\
eFEDS J092507.9+001914 & 00457  &  3.10  &  21.15 $_{- 0.74 } ^{+ 0.55 }$ &  1.52 $_{- 0.18 } ^{+ 0.24 }$ & -12.48 & PL \\
eFEDS J085318.2-010309 & 00461  &  2.48  &  21.98 $_{- 0.11 } ^{+ 0.11 }$ &  1.37 $_{- 0.25 } ^{+ 0.33 }$ & -12.09 & PL \\
eFEDS J082934.1-003310 & 00462  &  4.71  &  21.81 $_{- 1.01 } ^{+ 0.42 }$ &  1.84 $_{- 0.29 } ^{+ 0.39 }$ & -12.67 & PL \\
eFEDS J092435.2+012846 & 00528  &  3.62  &  20.35 $_{- 0.24 } ^{+ 0.34 }$ &  2.27 $_{- 0.19 } ^{+ 0.19 }$ & -13.11 & SE \\
eFEDS J090306.4+030449 & 00534  &  4.25  &  20.51 $_{- 0.34 } ^{+ 0.44 }$ &  2.18 $_{- 0.20 } ^{+ 0.24 }$ & -12.97 & PL \\
eFEDS J084528.7-002724 & 00540  &  2.97  &  20.60 $_{- 0.38 } ^{+ 0.37 }$ &  2.10 $_{- 0.24 } ^{+ 0.33 }$ & -12.91 & PCF \\
eFEDS J090604.1+011114 & 00545  &  3.87  &  20.60 $_{- 0.40 } ^{+ 0.46 }$ &  2.06 $_{- 0.21 } ^{+ 0.22 }$ & -12.94 & PL \\
\hline
\hline
\end{tabular}
\twocolumn
\end{table*}

\renewcommand{\arraystretch}{1.3}
\begin{table*}[]
\centering
\begin{tabular}{ccccccc}
\hline
(1) & (2) & (3) & (4) & (5) & (6) & (7) \\
Name & ID & nH$_{gal}$ & log(NH$_z$) & $\Gamma$ & log(F) & Model \\
& & ($\times10^{20}\pscm$) & & & \fluxcgs & \\
\hline
eFEDS J092017.3-004659 & 00558  &  3.45  &  21.52 $_{- 0.26 } ^{+ 0.18 }$ &  1.95 $_{- 0.44 } ^{+ 0.46 }$ & -12.71 & PL \\
eFEDS J090326.2-005527 & 00618  &  2.80  &  20.54 $_{- 0.37 } ^{+ 0.49 }$ &  2.13 $_{- 0.19 } ^{+ 0.18 }$ & -13.07 & WA \\
eFEDS J091119.2+031150 & 00626  &  3.15  &  20.39 $_{- 0.27 } ^{+ 0.39 }$ &  1.41 $_{- 0.19 } ^{+ 0.20 }$ & -12.56 & PL \\
eFEDS J093123.9+051344 & 00632  &  3.79  &  21.95 $_{- 0.19 } ^{+ 0.13 }$ &  2.36 $_{- 0.51 } ^{+ 0.42 }$ & -12.52 & WA \\
eFEDS J092938.4+021627 & 00667  &  5.24  &  22.18 $_{- 0.51 } ^{+ 0.28 }$ &  1.36 $_{- 0.24 } ^{+ 0.34 }$ & -12.41 & WA \\
eFEDS J092227.5+030909 & 00713  &  3.91  &  20.26 $_{- 0.19 } ^{+ 0.29 }$ &  1.39 $_{- 0.18 } ^{+ 0.18 }$ & -12.57 & PCF \\
eFEDS J091517.8-001720 & 00758  &  3.01  &  21.71 $_{- 0.49 } ^{+ 0.26 }$ &  1.14 $_{- 0.10 } ^{+ 0.20 }$ & -12.38 & PL \\
eFEDS J092200.7-012209 & 00760  &  3.60  &  20.29 $_{- 0.20 } ^{+ 0.31 }$ &  1.87 $_{- 0.22 } ^{+ 0.21 }$ & -13.06 & SE \\
eFEDS J093255.9+034230 & 00774  &  3.09  &  21.03 $_{- 0.66 } ^{+ 0.53 }$ &  1.54 $_{- 0.22 } ^{+ 0.30 }$ & -12.63 & PL \\
eFEDS J083141.8+014556 & 00781  &  4.62  &  21.50 $_{- 0.34 } ^{+ 0.21 }$ &  1.30 $_{- 0.21 } ^{+ 0.34 }$ & -12.41 & PL \\
eFEDS J085043.9-013638 & 00784  &  2.24  &  20.33 $_{- 0.23 } ^{+ 0.33 }$ &  2.42 $_{- 0.22 } ^{+ 0.22 }$ & -13.53 & SE \\
eFEDS J090852.2-013741 & 00881  &  2.54  &  21.08 $_{- 0.57 } ^{+ 0.39 }$ &  1.08 $_{- 0.06 } ^{+ 0.14 }$ & -12.51 & WA \\
eFEDS J090623.7+031745 & 01000  &  3.17  &  20.56 $_{- 0.39 } ^{+ 0.44 }$ &  1.88 $_{- 0.23 } ^{+ 0.27 }$ & -13.09 & PL \\
eFEDS J090022.7+013116 & 01007  &  4.57  &  22.26 $_{- 0.17 } ^{+ 0.17 }$ &  1.66 $_{- 0.41 } ^{+ 0.59 }$ & -12.35 & WA \\
eFEDS J093841.3+032422 & 01029  &  3.19  &  20.48 $_{- 0.34 } ^{+ 0.39 }$ &  2.05 $_{- 0.27 } ^{+ 0.29 }$ & -13.15 & PL \\
eFEDS J091443.7+035410 & 01058  &  3.56  &  21.56 $_{- 0.35 } ^{+ 0.28 }$ &  1.31 $_{- 0.22 } ^{+ 0.41 }$ & -12.46 & WA \\
eFEDS J090909.1-013201 & 01091  &  2.61  &  20.75 $_{- 0.53 } ^{+ 0.61 }$ &  1.93 $_{- 0.25 } ^{+ 0.26 }$ & -13.20 & PL \\
eFEDS J093201.6+031859 & 01099  &  3.22  &  20.67 $_{- 0.45 } ^{+ 0.53 }$ &  1.89 $_{- 0.28 } ^{+ 0.35 }$ & -13.02 & PL \\
eFEDS J090111.8+044900 & 01116  &  3.80  &  20.97 $_{- 0.65 } ^{+ 0.64 }$ &  1.84 $_{- 0.25 } ^{+ 0.28 }$ & -12.94 & PL \\
eFEDS J084221.3+000459 & 01124  &  3.60  &  21.49 $_{- 0.46 } ^{+ 0.30 }$ &  1.26 $_{- 0.19 } ^{+ 0.36 }$ & -12.51 & PL \\
eFEDS J083857.6+041426 & 01136  &  2.96  &  20.40 $_{- 0.28 } ^{+ 0.39 }$ &  1.89 $_{- 0.29 } ^{+ 0.28 }$ & -13.11 & SE \\
eFEDS J091320.6+034013 & 01137  &  3.40  &  20.42 $_{- 0.30 } ^{+ 0.37 }$ &  1.64 $_{- 0.25 } ^{+ 0.27 }$ & -12.83 & PL \\
eFEDS J091821.2+002954 & 01153  &  2.94  &  20.53 $_{- 0.38 } ^{+ 0.50 }$ &  1.31 $_{- 0.18 } ^{+ 0.21 }$ & -12.66 & PL \\
eFEDS J091755.4+003060 & 01161  &  2.91  &  21.19 $_{- 0.69 } ^{+ 0.37 }$ &  2.06 $_{- 0.37 } ^{+ 0.52 }$ & -13.14 & PCF \\
eFEDS J085135.5+044126 & 01165  &  4.55  &  21.71 $_{- 0.18 } ^{+ 0.17 }$ &  1.22 $_{- 0.17 } ^{+ 0.32 }$ & -12.23 & PL \\
eFEDS J083002.0+005059 & 01277  &  4.48  &  22.85 $_{- 0.18 } ^{+ 0.17 }$ &  2.21 $_{- 0.76 } ^{+ 0.57 }$ & -12.63 & WA \\
eFEDS J084324.6+023121 & 01283  &  4.82  &  20.74 $_{- 0.51 } ^{+ 0.58 }$ &  1.69 $_{- 0.23 } ^{+ 0.27 }$ & -12.85 & PL \\
eFEDS J083718.2+050938 & 01492  &  3.06  &  20.96 $_{- 0.62 } ^{+ 0.41 }$ &  2.19 $_{- 0.35 } ^{+ 0.41 }$ & -13.28 & PL \\
eFEDS J093441.8+034511 & 01536  &  3.05  &  22.13 $_{- 0.21 } ^{+ 0.19 }$ &  1.28 $_{- 0.21 } ^{+ 0.47 }$ & -12.38 & WA \\
eFEDS J084157.8-022617 & 01543  &  3.04  &  22.37 $_{- 0.47 } ^{+ 0.21 }$ &  1.56 $_{- 0.45 } ^{+ 0.66 }$ & -12.66 & WA \\
eFEDS J083956.9+035338 & 01637  &  3.04  &  22.65 $_{- 0.15 } ^{+ 0.12 }$ &  2.08 $_{- 0.60 } ^{+ 0.56 }$ & -12.51 & WA \\
eFEDS J083013.5-013558 & 01736  &  3.82  &  20.22 $_{- 0.16 } ^{+ 0.25 }$ &  2.24 $_{- 0.30 } ^{+ 0.32 }$ & -13.47 & SE \\
eFEDS J085324.2-011357 & 01911  &  2.39  &  20.72 $_{- 0.50 } ^{+ 0.56 }$ &  1.96 $_{- 0.35 } ^{+ 0.39 }$ & -13.30 & PL \\
eFEDS J091800.1+042506 & 02010  &  3.78  &  20.33 $_{- 0.23 } ^{+ 0.31 }$ &  2.67 $_{- 0.33 } ^{+ 0.23 }$ & -13.83 & PCF \\
eFEDS J090313.1+032927 & 02242  &  4.09  &  20.94 $_{- 0.58 } ^{+ 0.43 }$ &  1.92 $_{- 0.43 } ^{+ 0.57 }$ & -13.19 & PL \\
eFEDS J082841.3-004219 & 02507  &  4.80  &  20.50 $_{- 0.36 } ^{+ 0.48 }$ &  1.50 $_{- 0.27 } ^{+ 0.27 }$ & -12.85 & PCF \\
eFEDS J085409.3-005029 & 02634  &  2.74  &  21.90 $_{- 0.27 } ^{+ 0.22 }$ &  1.24 $_{- 0.18 } ^{+ 0.40 }$ & -12.68 & PCF \\
eFEDS J090328.2+030908 & 02660  &  4.13  &  20.87 $_{- 0.58 } ^{+ 0.57 }$ &  1.25 $_{- 0.17 } ^{+ 0.27 }$ & -12.68 & PCF \\
eFEDS J085548.0+004738 & 03205  &  4.93  &  23.32 $_{- 0.13 } ^{+ 0.12 }$ &  1.98 $_{- 0.68 } ^{+ 0.69 }$ & -12.07 & WA \\
eFEDS J091928.9+043847 & 03376  &  3.79  &  21.22 $_{- 0.69 } ^{+ 0.47 }$ &  1.26 $_{- 0.19 } ^{+ 0.38 }$ & -12.83 & PCF \\
eFEDS J085552.8-013855 & 03447  &  2.08  &  22.65 $_{- 0.19 } ^{+ 0.16 }$ &  1.91 $_{- 0.55 } ^{+ 0.65 }$ & -12.81 & PL \\
eFEDS J093518.1+003235 & 03798  &  4.05  &  20.60 $_{- 0.44 } ^{+ 0.54 }$ &  1.36 $_{- 0.25 } ^{+ 0.42 }$ & -13.04 & SE \\
eFEDS J083601.6-002460 & 04215  &  4.24  &  22.29 $_{- 0.34 } ^{+ 0.24 }$ &  1.34 $_{- 0.25 } ^{+ 0.59 }$ & -12.79 & PL \\
eFEDS J083848.2+040735 & 04444  &  2.96  &  20.36 $_{- 0.24 } ^{+ 0.32 }$ &  2.66 $_{- 0.33 } ^{+ 0.24 }$ & -13.94 & WA \\
eFEDS J083404.6-015103 & 04871  &  3.34  &  22.63 $_{- 0.56 } ^{+ 0.22 }$ &  1.42 $_{- 0.33 } ^{+ 0.72 }$ & -12.72 & PCF \\
eFEDS J083553.3-000346 & 05196  &  4.47  &  22.49 $_{- 0.23 } ^{+ 0.18 }$ &  1.45 $_{- 0.32 } ^{+ 0.65 }$ & -12.78 & WA \\
\hline
\hline
\end{tabular}
\twocolumn
\end{table*}

\renewcommand{\arraystretch}{1.3}
\begin{table*}[]
\centering
\begin{tabular}{ccccccc}
\hline
(1) & (2) & (3) & (4) & (5) & (6) & (7) \\
Name & ID & nH$_{gal}$ & log(NH$_z$) & $\Gamma$ & log(F) & Model \\
& & ($\times10^{20}\pscm$) & & & \fluxcgs & \\
\hline
eFEDS J094032.7+043452 & 05278  &  3.87  &  20.51 $_{- 0.36 } ^{+ 0.49 }$ &  1.45 $_{- 0.30 } ^{+ 0.40 }$ & -12.90 & PCF \\
eFEDS J083253.7+011622 & 05546  &  4.52  &  20.69 $_{- 0.48 } ^{+ 0.54 }$ &  1.79 $_{- 0.42 } ^{+ 0.46 }$ & -13.37 & PCF \\
eFEDS J083247.1+001340 & 05842  &  4.52  &  21.16 $_{- 0.65 } ^{+ 0.41 }$ &  1.13 $_{- 0.10 } ^{+ 0.25 }$ & -12.88 & PCF \\
eFEDS J083322.6-000851 & 06275  &  4.43  &  22.24 $_{- 0.25 } ^{+ 0.20 }$ &  1.50 $_{- 0.38 } ^{+ 0.75 }$ & -12.86 & PL \\
eFEDS J085506.8-001147 & 06549  &  3.82  &  21.09 $_{- 0.75 } ^{+ 0.68 }$ &  1.44 $_{- 0.32 } ^{+ 0.56 }$ & -13.27 & PL \\
eFEDS J083624.1+044724 & 06693  &  3.12  &  22.99 $_{- 0.16 } ^{+ 0.14 }$ &  1.94 $_{- 0.65 } ^{+ 0.74 }$ & -12.55 & WA \\
eFEDS J090852.0+053713 & 06932  &  3.91  &  22.52 $_{- 0.19 } ^{+ 0.17 }$ &  1.57 $_{- 0.43 } ^{+ 0.75 }$ & -12.47 & PL \\
eFEDS J083548.4+014325 & 08243  &  4.88  &  20.61 $_{- 0.41 } ^{+ 0.46 }$ &  1.53 $_{- 0.37 } ^{+ 0.62 }$ & -13.27 & PCF \\
eFEDS J085240.9-013923 & 08968  &  2.12  &  22.94 $_{- 0.30 } ^{+ 0.29 }$ &  1.49 $_{- 0.37 } ^{+ 0.74 }$ & -12.86 & WA \\
eFEDS J092511.4-010623 & 09040  &  3.13  &  22.50 $_{- 0.34 } ^{+ 0.32 }$ &  1.39 $_{- 0.30 } ^{+ 0.64 }$ & -12.69 & PL \\
eFEDS J093432.2+002831 & 09905  &  3.79  &  22.70 $_{- 0.19 } ^{+ 0.16 }$ &  1.87 $_{- 0.58 } ^{+ 0.71 }$ & -12.84 & PL \\
eFEDS J093100.9+021125 & 10285  &  5.11  &  23.17 $_{- 0.21 } ^{+ 0.17 }$ &  2.13 $_{- 0.74 } ^{+ 0.63 }$ & -12.56 & WA \\
eFEDS J085801.6-021650 & 10831  &  1.85  &  23.20 $_{- 0.16 } ^{+ 0.15 }$ &  2.07 $_{- 0.68 } ^{+ 0.64 }$ & -12.94 & WA \\
eFEDS J093144.2-022851 & 11430  &  3.39  &  20.71 $_{ -0.48 } ^{ +0.62 }$ &  1.66 $_{ -0.48 } ^{ +0.72 }$ & -13.58 & PCF  \\
eFEDS J090735.4-013734 & 11437  &  2.47  &  23.41 $_{- 0.32 } ^{+ 0.19 }$ &  1.89 $_{- 0.62 } ^{+ 0.72 }$ & -13.04 & WA \\
eFEDS J092820.0+004356 & 13817  &  3.93  &  22.79 $_{- 0.36 } ^{+ 0.29 }$ &  1.57 $_{- 0.42 } ^{+ 0.76 }$ & -12.79 & PL \\
eFEDS J093904.1+034206 & 15909  &  3.33  &  23.65 $_{- 0.20 } ^{+ 0.18 }$ &  2.04 $_{- 0.67 } ^{+ 0.65 }$ & -12.16 & WA \\
eFEDS J092203.3-004445 & 16185  &  3.27  &  23.20 $_{- 0.35 } ^{+ 0.27 }$ &  1.89 $_{- 0.64 } ^{+ 0.73 }$ & -12.75 & WA \\
eFEDS J085819.9+031048 & 18184  &  4.23  &  23.57 $_{- 0.21 } ^{+ 0.18 }$ &  2.12 $_{- 0.72 } ^{+ 0.62 }$ & -12.61 & WA \\
eFEDS J093252.4+015246 & 27045  &  4.96  &  24.11 $_{- 1.16 } ^{+ 0.25 }$ &  1.93 $_{- 0.63 } ^{+ 0.71 }$ & -12.67 & WA \\
\hline
\hline
\end{tabular}
\twocolumn
\end{table*}

\renewcommand{\arraystretch}{1.3}
\begin{table}[h]
\caption{Summary of warm absorber parameters for sources with significant warm absorbers ($>95\%$). Column (1) gives the eROSITA ID, column (2) gives the warm absorber column density, column (3) gives the warm absorber ionisation, column (4) gives the photon index, and column (5) gives the Bayes factor K$_{wa}$. }
\label{tab:appwa}
\resizebox{\columnwidth}{!}{%
\begin{tabular}{lcccc}
\hline
(1) & (2) & (3) & (4) & (5) \\
eROID & log(nH) & log($\xi$) & $\Gamma_{wa}$ & K$_{wa}$ \\
\hline
00016 & 22.22 $_{ -0.13 } ^{ +0.13 }$ & 1.86 $_{ -0.11 } ^{ +0.10 }$ & 1.95 $_{ -0.06 } ^{ +0.07 }$ & 2.98$\times$10$^8$ \\
00028 & 22.26 $_{ -0.25 } ^{ +0.13 }$ & 1.84 $_{ -0.30 } ^{ +0.12 }$ & 1.94 $_{ -0.07 } ^{ +0.10 }$ & 5.66$\times$10$^{10}$ \\
00037 & 21.47 $_{ -0.36 } ^{ +0.38 }$ & 1.30 $_{ -3.68 } ^{ +0.51 }$ & 2.20 $_{ -0.12 } ^{ +0.14 }$ & 18.42 \\
00075 & 21.14 $_{ -0.58 } ^{ +0.31 }$ & -0.90 $_{ -2.12 } ^{ +3.30 }$ & 1.82 $_{ -0.25 } ^{ +0.32 }$ & 1.11 \\
00078 & 21.47 $_{ -0.97 } ^{ +1.33 }$ & 1.62 $_{ -3.47 } ^{ +1.29 }$ & 1.27 $_{ -0.13 } ^{ +0.19 }$ & 1.18 \\
00105 & 22.46 $_{ -0.43 } ^{ +0.32 }$ & 2.25 $_{ -0.19 } ^{ +0.25 }$ & 2.43 $_{ -0.14 } ^{ +0.13 }$ & 9.89 \\
00107 & 21.27 $_{ -0.23 } ^{ +0.62 }$ & -1.99 $_{ -1.47 } ^{ +3.87 }$ & 1.91 $_{ -0.27 } ^{ +0.31 }$ & 8.95 \\
00132 & 21.18 $_{ -0.62 } ^{ +1.15 }$ & 1.83 $_{ -4.11 } ^{ +0.86 }$ & 2.43 $_{ -0.18 } ^{ +0.23 }$ & 1.10 \\
00167 & 21.78 $_{ -0.56 } ^{ +0.83 }$ & 1.71 $_{ -0.91 } ^{ +0.88 }$ & 1.90 $_{ -0.17 } ^{ +0.24 }$ & 4.18 \\
00206 & 22.23 $_{ -0.45 } ^{ +0.31 }$ & 1.88 $_{ -0.33 } ^{ +0.26 }$ & 2.27 $_{ -0.19 } ^{ +0.22 }$ & 11.18 \\
00233 & 22.45 $_{ -1.77 } ^{ +0.98 }$ & 2.61 $_{ -2.15 } ^{ +0.82 }$ & 2.12 $_{ -0.19 } ^{ +0.23 }$ & 0.98 \\
00337 & 21.74 $_{ -0.21 } ^{ +0.17 }$ & -1.77 $_{ -1.55 } ^{ +1.24 }$ & 1.49 $_{ -0.27 } ^{ +0.36 }$ & 14.67 \\
00400 & 21.25 $_{ -0.70 } ^{ +0.59 }$ & 0.11 $_{ -2.86 } ^{ +2.45 }$ & 1.31 $_{ -0.22 } ^{ +0.38 }$ & 1.37 \\
00439 & 22.06 $_{ -1.18 } ^{ +0.26 }$ & -1.09 $_{ -1.82 } ^{ +3.06 }$ & 2.35 $_{ -0.46 } ^{ +0.38 }$ & 1.67 \\
00618 & 22.24 $_{ -1.33 } ^{ +0.29 }$ & 1.67 $_{ -0.75 } ^{ +0.36 }$ & 2.68 $_{ -0.49 } ^{ +0.23 }$ & 1.45 \\
00632 & 21.69 $_{ -0.25 } ^{ +0.19 }$ & -1.74 $_{ -1.69 } ^{ +1.36 }$ & 2.56 $_{ -0.47 } ^{ +0.30 }$ & 6.14 \\
00667 & 21.76 $_{ -1.16 } ^{ +0.63 }$ & 0.00 $_{ -2.85 } ^{ +2.72 }$ & 1.44 $_{ -0.26 } ^{ +0.41 }$ & 1.00 \\
00881 & 22.23 $_{ -0.18 } ^{ +0.11 }$ & 0.96 $_{ -0.15 } ^{ +0.14 }$ & 2.20 $_{ -0.65 } ^{ +0.54 }$ & 1748.16 \\
01007 & 21.96 $_{ -0.97 } ^{ +0.33 }$ & -0.81 $_{ -2.10 } ^{ +2.41 }$ & 1.81 $_{ -0.50 } ^{ +0.63 }$ & 1.38 \\
01058 & 21.66 $_{ -0.72 } ^{ +0.50 }$ & 0.06 $_{ -2.72 } ^{ +1.67 }$ & 1.64 $_{ -0.45 } ^{ +0.67 }$ & 2.14 \\
01277 & 22.02 $_{ -1.39 } ^{ +0.69 }$ & -0.18 $_{ -2.59 } ^{ +2.57 }$ & 2.21 $_{ -0.74 } ^{ +0.56 }$ & 0.94 \\
01536 & 21.98 $_{ -0.43 } ^{ +0.33 }$ & -0.89 $_{ -1.94 } ^{ +2.12 }$ & 1.40 $_{ -0.29 } ^{ +0.58 }$ & 4.06 \\
01543 & 22.53 $_{ -0.21 } ^{ +0.18 }$ & 1.52 $_{ -0.27 } ^{ +0.28 }$ & 1.73 $_{ -0.43 } ^{ +0.58 }$ & 46.91 \\
01637 & 21.91 $_{ -1.25 } ^{+ 0.57 }$ & -0.24 $_{ -2.58 } ^{ +3.06 }$ & 2.14 $_{ -0.63 } ^{ +0.54 }$ & 1.01 \\
03205 & 22.43 $_{ -1.66 } ^{ +0.66 }$ & -0.61 $_{ -2.27 } ^{ +2.85 }$ & 2.02 $_{ -0.72 } ^{ +0.65 }$ & 1.20 \\
04444 & 23.77 $_{ -0.21 } ^{ +0.09 }$ & 3.35 $_{ -0.56 } ^{ +0.57 }$ & 1.18 $_{ -0.13 } ^{ +0.26 }$ & 78.3 \\
05196 & 21.87 $_{ -1.21 } ^{ +0.57 }$ & -0.39 $_{ -2.40 } ^{ +2.92 }$ & 1.53 $_{ -0.4 } ^{ +0.68 }$ & 1.53 \\
06693 & 22.23 $_{ -1.45 } ^{ +0.62 }$ & -0.44 $_{ -2.46 } ^{ +3.02 }$ & 2.02 $_{ -0.64 } ^{ +0.68 }$ & 1.02 \\
08968 & 22.60 $_{ -1.43 } ^{ +0.45 }$ & -0.05 $_{ -2.75 } ^{ +2.43 }$ & 1.61 $_{ -0.45 } ^{ +0.79 }$ & 1.87 \\
10285 & 22.29 $_{ -1.52 } ^{ +0.68 }$ & -0.05 $_{ -2.64 } ^{ +2.37 }$ & 2.17 $_{ -0.73 } ^{ +0.60 }$ & 1.37 \\
10831 & 22.40 $_{ -1.57 } ^{ +0.62 }$ & -0.28 $_{ -2.55 } ^{ +2.70 }$ & 2.14 $_{ -0.69 } ^{ +0.60 }$ & 1.47 \\
11437 & 22.53 $_{ -1.64 } ^{ +0.70 }$ & -0.39 $_{ -2.50 } ^{ +2.57 }$ & 1.90 $_{ -0.66 } ^{ +0.78 }$ & 1.42 \\
15909 & 22.50 $_{ -1.63 } ^{ +0.89 }$ & -0.05 $_{ -2.66 } ^{ +2.69 }$ & 2.09 $_{ -0.74 } ^{ +0.65 }$ & 1.23 \\
16185 & 22.55 $_{ -1.67 } ^{ +0.59 }$ & 0.08 $_{ -2.69 } ^{ +2.41 }$ & 1.88 $_{ -0.60 } ^{ +0.77 }$ & 1.87 \\
18184 & 22.53 $_{ -1.77 } ^{ +0.79 }$ & -0.09 $_{ -2.67 } ^{ +2.62 }$ & 2.09 $_{ -0.69 } ^{ +0.59 }$ & 1.34 \\
27045 & 23.59 $_{ -2.16 } ^{ +0.30 }$ & -0.80 $_{ -2.24 } ^{ +2.73 }$ & 1.96 $_{ -0.63 } ^{ +0.67 }$ & 2.06 \\
\hline
\hline
\end{tabular}
}
\end{table}

\renewcommand{\arraystretch}{1.3}
\begin{table}[h]
\caption{Summary of partial covering absorber parameters for sources with significant partial coverers ($>95\%$). Column (1) gives the eROSITA ID, column (2) gives the partial covering absorber column density, column (3) gives the partial covering absorber ionisation, column (4) gives the photon index, and column (5) gives the Bayes factor K$_{pcf}$. }
\label{tab:apppcf}
\resizebox{\columnwidth}{!}{%
\begin{tabular}{lcccc}
\hline
(1) & (2) & (3) & (4) & (5) \\
eROID & log(nH) & C & $\Gamma_{pcf}$ & K$_{pcf}$ \\
\hline
00014  &  22.88 $_{- 0.29 } ^{+ 0.52 }$ &  0.53 $_{- 0.17 } ^{+ 0.14 }$ & 2.37 $_{- 0.09 } ^{+ 0.11 }$ & 6.03 \\
00023  &  23.44 $_{- 0.25 } ^{+ 0.27 }$ &  0.73 $_{- 0.2 } ^{+ 0.11 }$ & 2.13 $_{- 0.07 } ^{+ 0.08 }$ & 11.67 \\
00030  &  22.63 $_{- 0.14 } ^{+ 0.14 }$ &  0.62 $_{- 0.13 } ^{+ 0.09 }$ & 2.61 $_{- 0.18 } ^{+ 0.17 }$ & 26.00 \\
00043  &  23.22 $_{- 0.67 } ^{+ 0.81 }$ &  0.51 $_{- 0.31 } ^{+ 0.25 }$ & 2.66 $_{- 0.10 } ^{+ 0.11 }$ & 1.63 \\
00071  &  23.59 $_{- 0.46 } ^{+ 0.39 }$ &  0.77 $_{- 0.34 } ^{+ 0.13 }$ & 2.21 $_{- 0.11 } ^{+ 0.13 }$ & 3.03 \\
00108  &  23.97 $_{- 0.46 } ^{+ 0.24 }$ &  0.87 $_{- 0.39 } ^{+ 0.08 }$ & 2.46 $_{- 0.11 } ^{+ 0.13 }$ & 3.98 \\
00120  &  22.85 $_{- 1.01 } ^{+ 1.31 }$ &  0.44 $_{- 0.28 } ^{+ 0.25 }$ & 1.41 $_{- 0.16 } ^{+ 0.27 }$ & 1.56 \\
00184  &  22.93 $_{- 1.41 } ^{+ 1.19 }$ &  0.45 $_{- 0.29 } ^{+ 0.26 }$ & 2.07 $_{- 0.18 } ^{+ 0.25 }$ & 1.47 \\
00194  &  22.89 $_{- 0.29 } ^{+ 0.27 }$ &  0.66 $_{- 0.32 } ^{+ 0.15 }$ & 2.24 $_{- 0.24 } ^{+ 0.28 }$ & 2.62 \\
00214  &  22.62 $_{- 0.81 } ^{+ 1.52 }$ &  0.45 $_{- 0.29 } ^{+ 0.26 }$ & 1.89 $_{- 0.19 } ^{+ 0.32 }$ & 1.60 \\
00303  &  22.98 $_{- 0.45 } ^{+ 0.76 }$ &  0.62 $_{- 0.34 } ^{+ 0.18 }$ & 2.11 $_{- 0.24 } ^{+ 0.32 }$ & 2.08 \\
00323  &  23.4 $_{- 1.72 } ^{+ 0.81 }$ &  0.52 $_{- 0.34 } ^{+ 0.28 }$ & 2.2 $_{- 0.18 } ^{+ 0.22 }$ & 1.77 \\
00329  &  22.39 $_{- 0.10 } ^{+ 0.09 }$ &  0.91 $_{- 0.07 } ^{+ 0.03 }$ & 2.61 $_{- 0.36 } ^{+ 0.25 }$ & 227.95 \\
00420  &  22.17 $_{- 1.27 } ^{+ 1.63 }$ &  0.55 $_{- 0.37 } ^{+ 0.26 }$ & 1.48 $_{- 0.31 } ^{+ 0.64 }$ & 1.71 \\
00540  &  23.4 $_{- 0.96 } ^{+ 0.53 }$ &  0.78 $_{- 0.48 } ^{+ 0.16 }$ & 2.23 $_{- 0.28 } ^{+ 0.34 }$ & 2.09 \\
00713  &  22.38 $_{- 0.25 } ^{+ 0.39 }$ &  0.73 $_{- 0.36 } ^{+ 0.14 }$ & 2.06 $_{- 0.58 } ^{+ 0.57 }$ & 3.27 \\
01161  &  23.78 $_{- 1.89 } ^{+ 0.33 }$ &  0.86 $_{- 0.48 } ^{+ 0.1 }$ & 2.34 $_{- 0.47 } ^{+ 0.43 }$ & 2.12 \\
02010  &  23.66 $_{- 0.12 } ^{+ 0.12 }$ &  0.98 $_{- 0.05 } ^{+ 0.01 }$ & 2.74 $_{- 0.27 } ^{+ 0.18 }$ & 15005.82 \\
02507  &  23.22 $_{- 0.31 } ^{+ 0.24 }$ &  0.82 $_{- 0.38 } ^{+ 0.1 }$ & 2.28 $_{- 0.66 } ^{+ 0.50 }$ & 4.32 \\
02634  &  22.96 $_{- 1.84 } ^{+ 0.98 }$ &  0.58 $_{- 0.35 } ^{+ 0.26 }$ & 1.3 $_{- 0.22 } ^{+ 0.48 }$ & 1.65 \\
02660  &  23.31 $_{- 2.30 } ^{+ 0.85 }$ &  0.54 $_{- 0.34 } ^{+ 0.28 }$ & 1.37 $_{- 0.25 } ^{+ 0.45 }$ & 1.38 \\
03376  &  22.98 $_{- 2.04 } ^{+ 1.19 }$ &  0.56 $_{- 0.36 } ^{+ 0.29 }$ & 1.34 $_{- 0.25 } ^{+ 0.48 }$ & 1.39 \\
04871  &  22.86 $_{- 0.50 } ^{+ 0.36 }$ &  0.86 $_{- 0.35 } ^{+ 0.09 }$ & 1.42 $_{- 0.32 } ^{+ 0.64 }$ & 4.41 \\
05278  &  23.23 $_{- 1.17 } ^{+ 0.76 }$ &  0.67 $_{- 0.41 } ^{+ 0.22 }$ & 1.71 $_{- 0.45 } ^{+ 0.61 }$ & 2.17 \\
05546  &  22.65 $_{- 1.90 } ^{+ 1.56 }$ &  0.47 $_{- 0.32 } ^{+ 0.34 }$ & 1.89 $_{- 0.46 } ^{+ 0.49 }$ & 1.63 \\
05842  &  23.24 $_{- 0.20 } ^{+ 0.18 }$ &  0.96 $_{- 0.05 } ^{+ 0.02 }$ & 1.36 $_{- 0.27 } ^{+ 0.53 }$ & 206.15 \\
08243  &  23.04 $_{- 1.08 } ^{+ 0.72 }$ &  0.77 $_{- 0.48 } ^{+ 0.18 }$ & 1.65 $_{- 0.45 } ^{+ 0.62 }$ & 1.56 \\
11430  &  23.19 $_{- 0.30 } ^{+ 0.32 }$ &  0.92 $_{- 0.65 } ^{+ 0.06 }$ & 1.76 $_{- 0.52 } ^{+ 0.68 }$ & 2.37 \\
\hline
\hline
\end{tabular}
}
\end{table}

\renewcommand{\arraystretch}{1.3}
\begin{table}[h]
\caption{Summary of soft excess parameters for sources with significant soft excesses measured with a second power law component ($>95\%$). Column (1) gives the eROSITA ID, column (2) gives the hard X-ray photon index, column (3) gives the soft X-ray photon index, column (4) gives the soft excess strength (SE), and column (5) gives the Bayes factor K$_{pl}$. }
\label{tab:appse}
\resizebox{\columnwidth}{!}{%
\begin{tabular}{lcccc}
\hline
(1) & (2) & (3) & (4) & (5) \\
eROID & $\Gamma_h$ & $\Gamma_{s}$ & SE & K$_{pl}$ \\
\hline
00001  &  2.63 $_{- 0.14 } ^{+ 0.05 }$ &  5.83 $_{- 2.54 } ^{+ 1.10 }$ & 0.22 $_{- 0.11 } ^{+ 0.38 }$ & 2.65 \\
00003  &  2.87 $_{- 0.05 } ^{+ 0.05 }$ &  6.60 $_{- 1.69 } ^{+ 0.87 }$ & 0.18 $_{- 0.09 } ^{+ 0.23 }$ & 2.54 \\
00004  &  2.18 $_{- 0.10 } ^{+ 0.06 }$ &  5.68 $_{- 1.33 } ^{+ 1.11 }$ & 0.38 $_{- 0.12 } ^{+ 0.24 }$ & 37405.4 \\
00007  &  2.07 $_{- 0.22 } ^{+ 0.14 }$ &  4.39 $_{- 0.91 } ^{+ 1.26 }$ & 0.91 $_{- 0.39 } ^{+ 0.84 }$ & 840.36 \\
00011  &  2.57 $_{- 0.85 } ^{+ 0.36 }$ &  3.72 $_{- 0.27 } ^{+ 0.77 }$ & 4.89 $_{- 3.69 } ^{+ 30.64 }$ & 122$\times10^7$ \\
00029  &  2.40 $_{- 0.12 } ^{+ 0.11 }$ &  6.71 $_{- 1.58 } ^{+ 0.92 }$ & 0.64 $_{- 0.28 } ^{+ 0.54 }$ & 18.68 \\
00034  &  2.05 $_{- 0.14 } ^{+ 0.12 }$ &  6.51 $_{- 1.38 } ^{+ 1.00 }$ & 0.98 $_{- 0.38 } ^{+ 0.84 }$ & 3723.68 \\
00035  &  2.31 $_{- 0.15 } ^{+ 0.12 }$ &  6.21 $_{- 1.87 } ^{+ 1.22 }$ & 1.08 $_{- 0.53 } ^{+ 1.19 }$ & 11.30 \\
00038  &  1.92 $_{- 0.36 } ^{+ 0.2 }$ &  4.84 $_{- 1.58 } ^{+ 1.94 }$ & 1.43 $_{- 0.89 } ^{+ 2.94 }$ & 9.82 \\
00039  &  1.64 $_{- 0.35 } ^{+ 0.42 }$ &  4.70 $_{- 0.57 } ^{+ 0.74 }$ & 18.00 $_{- 8.11 } ^{+ 11.48 }$ & 129896.89 \\
00045  &  1.78 $_{- 0.42 } ^{+ 0.29 }$ &  4.50 $_{- 0.92 } ^{+ 1.40 }$ & 2.60 $_{- 1.19 } ^{+ 2.66 }$ & 1554.05 \\
00054  &  1.92 $_{- 0.40 } ^{+ 0.30 }$ &  4.68 $_{- 1.02 } ^{+ 1.29 }$ & 2.74 $_{- 1.53 } ^{+ 2.93 }$ & 47.84 \\
00057  &  1.86 $_{- 0.24 } ^{+ 0.19 }$ &  5.19 $_{- 1.66 } ^{+ 1.31 }$ & 1.81 $_{- 1.33 } ^{+ 4.18 }$ & 4.22 \\
00060  &  2.08 $_{- 0.23 } ^{+ 0.25 }$ &  8.49 $_{- 2.59 } ^{+ 1.11 }$ & 37.11 $_{- 35.19 } ^{+ 125.97 }$ & 2.15 \\
00070  &  2.27 $_{- 0.37 } ^{+ 0.19 }$ &  4.72 $_{- 1.60 } ^{+ 1.82 }$ & 0.94 $_{- 0.70 } ^{+ 2.15 }$ & 1.66 \\
00076  &  1.83 $_{- 0.54 } ^{+ 0.44 }$ &  3.79 $_{- 0.83 } ^{+ 1.68 }$ & 3.67 $_{- 2.99 } ^{+ 6.85 }$ & 3.44 \\
00117  &  2.34 $_{- 0.40 } ^{+ 0.21 }$ &  4.58 $_{- 1.59 } ^{+ 2.1 }$ & 0.87 $_{- 0.73 } ^{+ 2.77 }$ & 2.06 \\
00121  &  1.96 $_{- 0.43 } ^{+ 0.29 }$ &  4.81 $_{- 1.22 } ^{+ 1.55 }$ & 1.96 $_{- 1.11 } ^{+ 3.39 }$ & 16.54 \\
00122  &  2.33 $_{- 0.23 } ^{+ 0.27 }$ &  7.25 $_{- 2.55 } ^{+ 1.83 }$ & 8.24 $_{- 7.33 } ^{+ 86.15 }$ & 2.85 \\
00145  &  1.80 $_{- 0.31 } ^{+ 0.21 }$ &  5.06 $_{- 1.75 } ^{+ 1.82 }$ & 0.97 $_{- 0.69 } ^{+ 1.63 }$ & 2.35 \\
00148  &  1.32 $_{- 0.16 } ^{+ 0.16 }$ &  4.67 $_{- 1.48 } ^{+ 1.61 }$ & 2.04 $_{- 1.65 } ^{+ 8.27 }$ & 1.71 \\
00153  &  1.84 $_{- 0.35 } ^{+ 0.26 }$ &  4.85 $_{- 1.18 } ^{+ 1.51 }$ & 1.83 $_{- 1.06 } ^{+ 2.44 }$ & 11.96 \\
00176  &  1.95 $_{- 0.22 } ^{+ 0.20 }$ &  6.21 $_{- 2.20 } ^{+ 1.35 }$ & 2.09 $_{- 1.40 } ^{+ 4.33 }$ & 2.70 \\
00200  &  1.89 $_{- 0.33 } ^{+ 0.27 }$ &  6.03 $_{- 1.88 } ^{+ 2.09 }$ & 5.19 $_{- 3.85 } ^{+ 23.39 }$ & 11.70 \\
00204  &  1.45 $_{- 0.30 } ^{+ 0.39 }$ &  4.86 $_{- 1.22 } ^{+ 1.85 }$ & 5.46 $_{- 3.21 } ^{+ 10.66 }$ & 24.08 \\
00216  &  1.40 $_{- 0.28 } ^{+ 0.42 }$ &  4.31 $_{- 0.79 } ^{+ 1.15 }$ & 5.08 $_{- 2.87 } ^{+ 4.90 }$ & 46.87 \\
00223  &  1.31 $_{- 0.21 } ^{+ 0.21 }$ &  4.45 $_{- 1.22 } ^{+ 1.62 }$ & 3.21 $_{- 2.39 } ^{+ 5.38 }$ & 2.28 \\
00237  &  1.62 $_{- 0.19 } ^{+ 0.20 }$ &  6.74 $_{- 1.45 } ^{+ 1.10 }$ & 3.74 $_{- 1.84 } ^{+ 5.33 }$ & 109.52 \\
00238  &  1.93 $_{- 0.43 } ^{+ 0.33 }$ &  4.86 $_{- 1.72 } ^{+ 2.24 }$ & 4.21 $_{- 3.89 } ^{+ 30.73 }$ & 1.39 \\
00288  &  1.84 $_{- 0.25 } ^{+ 0.24 }$ &  7.14 $_{- 1.56 } ^{+ 1.00 }$ & 5.98 $_{- 3.01 } ^{+ 6.92 }$ & 118.59 \\
00340  &  1.47 $_{- 0.34 } ^{+ 0.6 }$ &  4.59 $_{- 1.07 } ^{+ 2.09 }$ & 14.10 $_{- 10.92 } ^{+ 119.11 }$ & 6.16 \\
00358  &  1.76 $_{- 0.26 } ^{+ 0.27 }$ &  7.16 $_{- 1.34 } ^{+ 1.24 }$ & 10.51 $_{- 5.91 } ^{+ 26.74 }$ & 336.1 \\
00426  &  1.28 $_{- 0.18 } ^{+ 0.27 }$ &  6.01 $_{- 1.08 } ^{+ 1.82 }$ & 13.67 $_{- 8.19 } ^{+ 137.51 }$ & 1152.91 \\
00450  &  1.86 $_{- 0.32 } ^{+ 0.27 }$ &  5.27 $_{- 2.07 } ^{+ 1.86 }$ & 1.03 $_{- 0.83 } ^{+ 1.96 }$ & 1.61 \\
00528  &  1.96 $_{- 0.43 } ^{+ 0.33 }$ &  4.98 $_{- 1.69 } ^{+ 1.97 }$ & 3.06 $_{- 2.67 } ^{+ 11.30 }$ & 1.98 \\
00760  &  1.53 $_{- 0.31 } ^{+ 0.31 }$ &  6.48 $_{- 1.86 } ^{+ 2.06 }$ & 7.94 $_{- 6.09 } ^{+ 173.10 }$ & 6.35 \\
00784  &  2.08 $_{- 0.42 } ^{+ 0.39 }$ &  6.20 $_{- 2.22 } ^{+ 2.44 }$ & 7.08 $_{- 6.03 } ^{+ 105.53 }$ & 3.49 \\
01136  &  1.39 $_{- 0.28 } ^{+ 0.39 }$ &  5.17 $_{- 1.62 } ^{+ 1.75 }$ & 5.46 $_{- 4.14 } ^{+ 23.77 }$ & 2.85 \\
01736  &  1.46 $_{- 0.31 } ^{+ 0.39 }$ &  6.56 $_{- 1.55 } ^{+ 1.62 }$ & 6.95 $_{- 4.00 } ^{+ 14.30 }$ & 23.66 \\
03798  &  1.24 $_{- 0.18 } ^{+ 0.31 }$ &  6.42 $_{- 2.28 } ^{+ 2.18 }$ & 21.34 $_{- 19.82 } ^{+ 586.14 }$ & 1.70 \\
\hline
\hline
\end{tabular}
}
\end{table}

\renewcommand{\arraystretch}{1.3}
\begin{table}[h]
\caption{Summary of soft excess parameters for sources with significant soft excesses which are best fit with the warm corona model as opposed to a blurred reflection model. Column (1) gives the eROSITA ID, column (2) gives the hard X-ray photon index, column (3) gives the soft X-ray photon index, column (4) gives the warm corona temperature (kT$_e$), and column (5) gives the Bayes factor K$_{pl}$. }
\label{tab:appnth}
\resizebox{\columnwidth}{!}{%
\begin{tabular}{lcccc}
\hline
(1) & (2) & (3) & (4) & (5) \\
eROID & $\Gamma_h$ & $\Gamma_{s}$ & kT$_e$ & K$_{pl}$ \\
\hline
00001  &  2.40 $_{- 0.47 } ^{+ 0.26 }$ &  3.04 $_{- 0.30 } ^{+ 0.31 }$ & 0.61 $_{- 0.31 } ^{+ 0.27 }$ & 1.48 \\
00004  &  1.91 $_{- 0.16 } ^{+ 0.13 }$ &  3.26 $_{- 0.28 } ^{+ 0.18 }$ & 0.61 $_{- 0.32 } ^{+ 0.27 }$ & 8354.43 \\
00007  &  1.92 $_{- 0.18 } ^{+ 0.16 }$ &  3.24 $_{- 0.29 } ^{+ 0.19 }$ & 0.45 $_{- 0.28 } ^{+ 0.38 }$ & 321.73 \\
00011  &  2.01 $_{- 0.63 } ^{+ 0.61 }$ &  3.30 $_{- 0.15 } ^{+ 0.13 }$ & 0.71 $_{- 0.26 } ^{+ 0.20 }$ & 6.95$\times10^6$ \\
00029  &  2.29 $_{- 0.37 } ^{+ 0.27 }$ &  3.10 $_{- 0.54 } ^{+ 0.29 }$ & 0.61 $_{- 0.37 } ^{+ 0.28 }$ & 1.17 \\
00034  &  1.67 $_{- 0.25 } ^{+ 0.24 }$ &  3.27 $_{- 0.31 } ^{+ 0.17 }$ & 0.55 $_{- 0.36 } ^{+ 0.31 }$ & 163.32 \\
00038  &  1.61 $_{- 0.33 } ^{+ 0.32 }$ &  2.99 $_{- 0.40 } ^{+ 0.34 }$ & 0.59 $_{- 0.30 } ^{+ 0.28 }$ & 6.86 \\
00039  &  1.83 $_{- 0.41 } ^{+ 0.33 }$ &  3.15 $_{- 0.53 } ^{+ 0.27 }$ & 0.18 $_{- 0.05 } ^{+ 0.15 }$ & 74560.61 \\
00045  &  1.42 $_{- 0.25 } ^{+ 0.30 }$ &  3.22 $_{- 0.28 } ^{+ 0.20 }$ & 0.58 $_{- 0.30 } ^{+ 0.29 }$ & 1069.04 \\
00054  &  1.67 $_{- 0.33 } ^{+ 0.31 }$ &  3.23 $_{- 0.32 } ^{+ 0.19 }$ & 0.45 $_{- 0.30 } ^{+ 0.38 }$ & 34.23 \\
00076  &  1.58 $_{- 0.39 } ^{+ 0.52 }$ &  2.97 $_{- 0.35 } ^{+ 0.34 }$ & 0.61 $_{- 0.31 } ^{+ 0.26 }$ & 5.74 \\
00121  &  1.64 $_{- 0.36 } ^{+ 0.41 }$ &  3.15 $_{- 0.40 } ^{+ 0.25 }$ & 0.51 $_{- 0.32 } ^{+ 0.32 }$ & 9.09 \\
00153  &  1.59 $_{- 0.29 } ^{+ 0.33 }$ &  3.15 $_{- 0.44 } ^{+ 0.25 }$ & 0.48 $_{- 0.31 } ^{+ 0.36 }$ & 8.98 \\
00200  &  1.66 $_{- 0.33 } ^{+ 0.37 }$ &  3.10 $_{- 0.48 } ^{+ 0.29 }$ & 0.45 $_{- 0.30 } ^{+ 0.38 }$ & 3.99 \\
00216  &  1.3 $_{- 0.22 } ^{+ 0.31 }$ &  3.13 $_{- 0.46 } ^{+ 0.26 }$ & 0.33 $_{- 0.17 } ^{+ 0.43 }$ & 71.42 \\
00237  &  1.44 $_{- 0.25 } ^{+ 0.28 }$ &  3.17 $_{- 0.43 } ^{+ 0.25 }$ & 0.41 $_{- 0.26 } ^{+ 0.38 }$ & 7.54 \\
00288  &  1.56 $_{- 0.32 } ^{+ 0.40 }$ &  3.18 $_{- 0.45 } ^{+ 0.23 }$ & 0.4 $_{- 0.27 } ^{+ 0.42 }$ & 7.05 \\
00340  &  1.35 $_{- 0.25 } ^{+ 0.50 }$ &  3.02 $_{- 0.56 } ^{+ 0.34 }$ & 0.31 $_{- 0.14 } ^{+ 0.42 }$ & 12.52 \\
00358  &  1.55 $_{- 0.29 } ^{+ 0.32 }$ &  3.17 $_{- 0.46 } ^{+ 0.24 }$ & 0.19 $_{- 0.08 } ^{+ 0.47 }$ & 13.10 \\
00426  &  1.21 $_{- 0.15 } ^{+ 0.20 }$ &  3.11 $_{- 0.54 } ^{+ 0.3 }$ & 0.14 $_{- 0.03 } ^{+ 0.15 }$ & 372.76 \\
00784  &  2.03 $_{- 0.60 } ^{+ 0.44 }$ &  2.93 $_{- 0.56 } ^{+ 0.41 }$ & 0.55 $_{- 0.34 } ^{+ 0.29 }$ & 2.31 \\
01136  &  1.33 $_{- 0.25 } ^{+ 0.46 }$ &  2.95 $_{- 0.54 } ^{+ 0.39 }$ & 0.47 $_{- 0.27 } ^{+ 0.34 }$ & 2.93 \\
01736  &  1.35 $_{- 0.26 } ^{+ 0.56 }$ &  3.13 $_{- 0.53 } ^{+ 0.28 }$ & 0.35 $_{- 0.22 } ^{+ 0.43 }$ & 3.79 \\
\hline
\hline
\end{tabular}
}
\end{table}

\renewcommand{\arraystretch}{1.3}
\begin{table*}
\centering
\caption{Summary of blurred reflection parameters for all sources which are best fit with blurred reflection. Column (1) gives the eROSITA ID, column (2) gives the emissivity index, column (3) gives the disc inclination, column (4) gives the photon index, column (5) gives the disc ionisation and column (6) gives the reflection fraction. }
\label{tab:apprel}
\resizebox{1.5\columnwidth}{!}{%
\begin{tabular}{lccccccc}
\hline
(1) & (2) & (3) & (4) & (5) & (6) & (7) \\
eROID & Inner emissivity (q$_{1}$) & Inclination ($^o$) & $\Gamma_{rel}$ & log($\xi$) & log($R$) & K$_{rel}$\\
 \hline
00035  &  7.1 $_{- 2.3 } ^{+ 2.0 }$ &  36 $_{- 18 } ^{+ 21 }$ & 2.45 $_{- 0.11 } ^{+ 0.10 }$ & 1.8 $_{- 1.2 } ^{+ 1.4 }$ & 0.4 $_{- 0.8 } ^{+ 0.5 }$ & 2.73 \\
00057  &  6.4 $_{- 2.0 } ^{+ 2.4 }$ &  39 $_{- 19 } ^{+ 19 }$ & 2.00 $_{- 0.15 } ^{+ 0.13 }$ & 1.8 $_{- 1.0 } ^{+ 1.2 }$ & 0.6 $_{- 0.7 } ^{+ 0.3 }$ & 3.53 \\
00122  &  6.5 $_{- 2.3 } ^{+ 2.3 }$ &  41 $_{- 20 } ^{+ 22 }$ & 2.34 $_{- 0.13 } ^{+ 0.13 }$ & 2.0 $_{- 1.3 } ^{+ 1.4 }$ & 0.2 $_{- 0.8 } ^{+ 0.5 }$ & 1.13 \\
00176  &  6.9 $_{- 2.4 } ^{+ 2.1 }$ &  39 $_{- 20 } ^{+ 24 }$ & 2.12 $_{- 0.15 } ^{+ 0.15 }$ & 2.0 $_{- 1.3 } ^{+ 1.4 }$ & 0.2 $_{- 0.8 } ^{+ 0.6 }$ & 1.56 \\
00204  &  4.3 $_{- 1.0 } ^{+ 2.6 }$ &  43 $_{- 21 } ^{+ 19 }$ & 1.86 $_{- 0.25 } ^{+ 0.22 }$ & 2.0 $_{- 0.9 } ^{+ 0.8 }$ & 0.8 $_{- 0.3 } ^{+ 0.1 }$ & 27.36 \\
00760  &  5.8 $_{- 1.9 } ^{+ 2.7 }$ &  40 $_{- 21 } ^{+ 22 }$ & 1.75 $_{- 0.37 } ^{+ 0.26 }$ & 2.4 $_{- 1.5 } ^{+ 0.8 }$ & 0.4 $_{- 0.8 } ^{+ 0.4 }$ & 3.14 \\
\hline
\hline
\end{tabular}
}
\end{table*}

\end{document}